%% file: main.tex
\documentclass[aps,prd,twocolumn,10pt,nofootinbib,superscriptaddress,preprintnumbers,floatfix]{revtex4-1}
\pdfoutput=1 % for arXiv to force using pdflatex, must be in first 5 lines
\usepackage[utf8]{inputenc} % for umlauts etc
\usepackage{amsmath} % ams stuff for maths
\usepackage{amsfonts}
\usepackage{amsthm}
\usepackage{amssymb}
\usepackage{graphicx} % for pics
\usepackage{color} % for coloured comments
\usepackage[normalem]{ulem} % for \stout (strike out)
\usepackage{slashed} % for feynman slash
\usepackage{physics} % for mathematical notation (e.g. partial derivatives and integral measures)
\usepackage{mathrsfs} % Lagrangian (\mathscr{L})
\usepackage{tikz}  % Scale hierarchy figure
\usepackage[compat=1.1.0]{tikz-feynman} % Feynman diagrams
\usepackage{hyperref} % for hyperlinks
\usepackage{verbatim} % comment blocks

\graphicspath{{./}{./figs/}}

\include{feynman-diagrams}

\definecolor{AVcolour}{rgb}{0.8,0.6,0.0}

\definecolor{JHcolour}{rgb}{0.96,0.0,0.63}
\newcommand{\jh}[1]{\textcolor{JHcolour}{[JH: #1]}}

\definecolor{WRIcolor}{rgb}{0,0.9,0.3}
\newcommand{\jw}[1]{\textcolor{WRIcolor}{#1}}

\definecolor{OLIcolour}{rgb}{0.0,0.78,0.61}

\definecolor{ghost}{rgb}{0.38,0.38,0.38}

\newcommand{\rmi}[1]{{\mbox{\scriptsize #1}}}
\newcommand{\rmii}[1]{{\mbox{\tiny\rm{#1}}}}

\newcommand{\Sef}{S_{\rmi{eff}}}
\newcommand{\Seu}{S_{\mathrm{E}}}

\newcommand\sumint[1]{\hbox{$\sum$}\!\!\!\!\!\!\!\int_{#1}}
\newcommand{\cou}[1]{\delta_{#1}}
\newcommand{\MSbar}{\overline{\text{MS}}}
\newcommand{\gE}{\gamma_{\text{E}}}
\newcommand{\heavy}{\Phi^{\rmii{heavy}}}
\newcommand{\light}{\Phi^{\rmii{light}}}
\newcommand{\effphi}{\Phi^{\rmii{eff}}}
\newcommand{\ir}{\Phi^{\rmii{IR}}}
\newcommand{\uv}{\Phi^{\rmii{UV}}}
\newcommand{\lightlow}{\light_{\rmii{IR}}}
\newcommand{\lighthigh}{\light_{\rmii{UV}}}

\newcommand{\Chi}{\mathrm{X}}
\newcommand{\parttitle}[1]{ \begin{center} {\bf \emph{#1}} \end{center} }
\newcommand{\meff}{m_{\text{eff}}}
\newcommand{\irqe}[1]{\left. #1 \;\right\rvert_{\substack{\text{IRq}\\\text{exp.}}}}
\newcommand{\kOrd}[2]{\left. #1 \;\right\rvert_{\qty(\spat{k}^2)^{#2}}}
\newcommand{\kmOrd}[1]{\left. #1 \;\right\rvert_{\qty(m_3^2)^{0},\qty(\spat{k}^2)^{0}}}
\newcommand{\EFT}[1]{\left. #1 \;\right\rvert_{\text{EFT}}}
\newcommand{\massless}[1]{\left. #1 \;\right\rvert_{\substack{\text{mass-}\\\text{less}}}}
\newcommand{\thermalscale}[1]{\left. #1 \;\right\rvert_{\substack{\text{thermal}\\\text{scale}}}}
\newcommand{\spat}[1]{{\mathbf{#1}}}

\newcommand{\oeps}{\order{\epsilon}}
\newcommand{\phiphys}{\Phi_{\text{eig}}}
\newcommand{\chiphys}{\Chi_{\text{eig}}}
\newcommand{\mixm}{\mathfrak{M}}

\newcommand{\Helsinki}{\affiliation{
    Department of Physics and Helsinki Institute of Physics,
    P.O.\ Box 64,
    FI-00014 University of Helsinki,
    Finland
}}

\begin{document}

% opening
\title{Intuitive method for constructing effective field theories}
%\title{Intuitive method for constructing high-temperature effective descriptions}
\date{May 5, 2022}

\author{Joonas Hirvonen}
\email{joonas.o.hirvonen@helsinki.fi}
\Helsinki

\begin{abstract}
%\av{``Flesh out and intuitive method'' is way too murky. Use ``Here, we derive a novel method designed for A, B and C'' or something equivalent.}
%\av{Not much of motivation.}
%\av{So called 1PI effective action}
%The construction of effective thermodynamic descriptions at high temperatures is an integral part of studying cosmological phase transitions. 
We derive a novel method for constructing effective field theories.
%, intuitive method designed to construct high-temperature effective field theories. It is computationally nearly identical to the perturbative construction of a 1PI effective action,
% and in intuition close to integrating out heavy fields.
Physically, the method is very close to the intuition behind effective field theories: One can integrate out the heavier scale directly from the path integral. We give a detailed recipe for the effective field theory construction,
%built upon already known techniques related to effective field theories. The recipe
which is nearly identical to the construction of the so-called 1PI effective action.
%, leading to simple computations.
%This leads to computational simplifications as well. The method is also nearly identical to the construction of the so-called 1PI effective action, leading to eased transfer of computational knowledge.
We also demonstrate the equivalence of the novel method to the commonly used procedure of matching Green's functions.
%in the high-temperature literature.
\end{abstract}

\preprint{HIP-2022-6/TH}
\maketitle

%\tableofcontents

%%%%%%%%%%%%%%%%%%%%%%%%%%%%%%%%%%%%%
%%%%%%%%%%% INTRODUCTION %%%%%%%%%%%% 
%%%%%%%%%%%%%%%%%%%%%%%%%%%%%%%%%%%%%

\section{Introduction} \label{sec:introduction}

%\color{blue}

Effective field theories (EFTs) are an efficient way to handle quantum field theories (QFTs) with scale hierarchies. EFTs also come with a physical picture: One creates an effective description, i.e.\ an EFT, for the longer length (or equivalently for the lower energy) scale. This scale is called the infrared (IR) scale.
%This scale is often referred to as the IR scale.
The EFT implicitly takes into account the effects of the shorter length, i.e.\ the higher energy scale, called the ultraviolet (UV) scale.
%This physical picture is credited to Wilson according to Refs.~\cite{WILSON197475,PhysRev.140.B445}.

In this article, we derive a method for EFT construction, which is in line with the intuitive picture of EFTs, often credited to Wilson~\cite{PhysRevB.4.3174,WILSON197475,POLCHINSKI1984269}: We show how one can simply integrate out the UV scale from a path integral to produce the EFT for the IR scale.

In addition to the construction being physically intuitive, the end result, found in Sec.~\ref{sec:method}, is remarkably simple: The IR-scale effective description contains the low-energy fields, $\lightlow$, which have both IR-scale masses and momenta. The effective action for these fields is directly given by
\begin{align}\label{eq:intro}
    S_{\text{eff}}[\lightlow] %&= -\ln \int\mathcal{D}\uv\,e^{-S_{\mathrm{E}}[\ir+\uv]}\\
    =\Gamma^{\text{1LPI}}[\lightlow]\eval_{\substack{\text{IRq}\\\text{exp.}}}\,.
\end{align}
Here, $\Gamma^{\text{1LPI}}$ refers to the action, which contains all of the one-light-particle-irreducible (1LPI) diagrams (\textit{cf}.\ Fig.~\ref{fig:reduciblereprep}), i.e.\ to the 1LPI action. The diagrams are expanded in the IR-scale masses and external momenta before evaluating the integrals, which is denoted by IRq exp. (shortened from IR quantity expansion). The regulator at use is dimensional regularization~\cite{tHooftDimReg}.

%\jh{One-loop stuff}

This topic has received some attention recently. Reference~\cite{Cohen:2022tir}, which refers to our Eq.~\eqref{eq:intro} as functional matching, reviewed the progress regarding the one-loop functional matching. The basic one-loop result has been known since a long time~\cite{Georgi:1991ch,Georgi:1992xg,Georgi:1993mps}. Our method for EFT construction can be viewed as the generalization of the one-loop result. (Note, that we choose to word the method in more physical terms as integrating out a scale.)
The generalization has already been used in practice in high-temperature QFTs in Refs.~\cite{Gould:2021ccf,Hirvonen:2021zej,Hirvonen:2020jud}. Here, we put the method on a solid foundation by giving the method a complete, physical discussion and derivation.

%Finally, we want to acknowledge a very recent article of Ref.~\cite{Branchina:2022jqc}, which is highly related to the topic at hand but still very distinct from our work. It discusses the physical equivalence between the Wilsonian EFTs containing no scale hierarchies and the corresponding computations using dimensional regularization. This yields physical insight into dimensional regularization. Note however that we are especially interested in scale hierarchies and EFT construction using these hierarchies. We hope to provide similar physical insight into EFT construction in dimensional regularization.

%\jh{Two interesting cases.}

EFTs are widely applicable.
There are two rather distinct motivations for this article in particular: cosmological phase transitions and the Standard Model effective field theory (SMEFT).
%This article will be more rich in discussion on the phase transitions.
Apart from the section explaining the novel method, Sec.~\ref{sec:method}, there is an emphasis on high-temperature QFTs and cosmological phase transitions.
However, the application onto SMEFT can be understood clearly.

In the SMEFT paradigm~\cite{Buchmuller:1985jz,Grzadkowski:2010es,Li:2020gnx,Murphy:2020rsh}, one characterizes the physics of an unknown, higher energy scale, a UV scale, via an effective Lagrangian describing the field content of the Standard Model. One can then find the coefficients of the effective Lagrangian terms experimentally, e.g.\ by collider experiments.
%such as the Large Hadron Collider.\jh{Which ones?}

To study specific extensions of the Standard Model, one needs to derive the corresponding SMEFT starting from the extensions. Here, our novel EFT construction method becomes a handy tool, leading to streamlined computations, which are also physically transparent. For instance, the effective action is dictated by a direct computation, Eq.~\eqref{eq:intro}, so the form of the action does not need to be specified beforehand. The motivations for functional matching in the context of SMEFT are discussed in more detail in Ref.~\cite{Cohen:2022tir}. Noting again the difference of wording functional matching in more physical terms as integrating out a scale.
%The novel method can be understood as the generalization for the so-called functional matching procedure discussed in Ref.~\cite{Cohen:2022tir}. Here, we choose to word the method in more physical terms as integrating out the shorter scale. (Note, that this kind of a functional matching procedure has already been utilized at high-temperature calculations in Refs.~\cite{Gould:2021ccf,Hirvonen:2021zej,Hirvonen:2020jud}.) 

%This article is more interested in thermal field theory applications.
%Thus, the method is presented in the context of Euclidean QFTs. However, the method also applies straightforwardly to zero temperature, collider applications due to the Wick rotation, as emphasized in the seminal article of Ref.~\cite{Appelquist:1974tg}.%
%\footnote{
%Here, we would like to note, that this article gives a generalization to the so-called functional matching procedure discussed in the vacuum context in Ref.~\cite{Cohen:2022tir}. This kind of a functional matching procedure has already been utilized at high-temperature calculations in Refs.~\cite{Gould:2021ccf,Hirvonen:2021zej,Hirvonen:2020jud}. The difference is in word choices. We choose to describe the procedure in more physical terms, as integrating out a scale, rather than as functional matching.
%}

The reason for being interested in cosmological phase transitions, and consequently high-temperature QFTs, is the possibility of a future gravitational wave detection from the early universe~\cite{Arzoumanian:2020vkk,Audley:2017drz,Kawamura:2011zz,Harry:2006fi,Guo:2018npi}. One particular possibility are gravitational waves produced by a first-order phase transition (for reviews see Refs.~\cite{Caprini:2018mtu,Caprini:2019egz,Hindmarsh:2020hop}).

At high temperatures, a QFT forms a thermal medium of plasma. The plasma consists of thermal particles with energies, $E\sim T$.
The medium can affect strongly the physics of long length scales, $L\gg T^{-1}$, through e.g.\ Debye screening. These strong physical effects often render the long length scales the hardest to handle computationally.

As an example, the Debye screening can be viewed computationally as a thermal correction to the mass of a quantum field. If the Debye screening dominates the long-length-scale behavior of the field, the correct thermal mass has to be found before we even attempt to describe the field on these length scales.

Computationally, the need for correct thermal masses results in the divergence of the so-called naive perturbative expansion, which does not attempt to incorporate the correct thermal masses: There are infinite classes of diagrams contributing to the same order. Consequently, the strong effects on the long length scale break down the naive perturbation expansion, requiring resummations of the infinite classes. (See for example Ref.~\cite{Laine:2016hma} Secs.~3.4 and 6.)

In perturbative QFTs, the strong effects on the long length scale are aptly taken care by EFTs~\cite{Kajantie:1995dw,Braaten:1995cm}. This stems from the scale hierarchy between the thermal medium and the long-length IR scale, $L\gg T^{-1}$.

Here, we note that the aforementioned scale hierarchy, $L\gg T^{-1}$, is often in a central role in the cosmological phase transitions of perturbative QFTs~\cite{Gould:2021ccf}. These phase transitions occur as the temperature cools down due to the expansion of the Universe. Thus, the change in temperature has a drastic effect on the physical behavior of the system. In perturbative QFTs, the change in temperature can often only have such drastic effects on the system through the scale hierarchy.

Physically, this can mean for example, that the size of nucleating bubbles is much larger than that of the average thermal fluctuations, $L_{\text{bubble}}\gg T^{-1}$~\cite{Gould:2021ccf}. Thus, the size scale of nucleating bubbles, $L_{\text{bubble}}$, receives strong effects from the thermal medium.% (\textit{cf}.\ Sec.~\ref{sec:exampleII}).

From the above discussion, it is evident that EFTs are an invaluable tool for studying the cosmological phase transitions. The usefulness has been analyzed further, and confirmed, in Refs.~\cite{Croon:2020cgk,Gould:2021oba,Kainulainen:2019kyp}. (See also other recent applications in Refs.~\cite{Niemi:2020hto,Niemi:2021qvp}.) Note, that EFTs also facilitate consistent nucleation rate computations~\cite{Gould:2021ccf,Hirvonen:2021zej}.

Still, the use of high-temperature EFTs is relatively scarce in the literature compared to their usefulness. It is usually replaced with the so-called daisy resummations,
%in the 1-particle-irreducible action (1PI action) or integrating out fields\jh{Latter?} 
often leading to unphysical scale dependence and gauge dependence~\cite{Croon:2020cgk}.

The scarcity of EFTs may be due to the technical appearance of the matching procedure, which is commonly used to construct EFTs. The procedure was introduced to the high-temperature context by the seminal articles~\cite{Kajantie:1995dw,Braaten:1995cm,Farakos:1994kx,Braaten:1995jr}.
It was put in place to circumvent the inconsistencies of integrating out heavy fields~\cite{Kajantie:1995dw}, which had been initially the method for constructing high-temperature effective descriptions~\cite{Ginsparg:1980ef,Appelquist:1981vg,Nadkarni:1982kb,Landsman:1989be}. Thus, the matching procedure provides a consistent way for constructing EFTs.

In this article, our goal is to overcome the technical nature of the matching procedure. We derive a novel method for EFT construction, which is very closely connected to the intuition behind EFTs: One can directly integrate out the UV scale to get the EFT for the IR scale.

%\jh{
%\begin{itemize}
%    \item Matches to the intuition behind EFTs
%    \item Two advantages
%    \begin{itemize}
%        \item Fluent computations due to integrating out
%        \item Coherent physical narrative
%    \end{itemize}
%    \item Almost identical to 1-particle-irreducible action (transition)
%    \item
%\end{itemize}
%}

There are two main advantages to this novel method. The first one is the achieved computational simplification. In the matching procedure, one needs to construct an effective action, which is then matched onto the IR-scale behavior of the full description. Here, we can start from the full description and just integrate over the UV scale.

The computational method, which we will cover in Sec.~\ref{sec:method}, will be almost identical to a widely used action: the 1PI effective action, which contains all the one-particle-irreducible (1PI) diagrams (see e.g.\ Ref.~\cite{Schwartz:2014sze} Sec.~34). This can already be seen from Eq.~\eqref{eq:intro}. There, the effective action is almost directly given by the 1LPI action, which is the 1PI effective action for the light fields. This particular fact facilitates transfer of computational knowledge onto EFTs.

The second advantage is that the method is close to the intuition behind EFTs. Consequently, it allows for a particularly coherent physical narrative. Hopefully, it also enables this article to be a pedagogical entry point to EFTs (including the commonly applied matching procedure discussed in Sec.~\ref{sec:matching}).

The fact that the mathematical method is close to the physical intuition means that we can also have a transparent understanding of EFTs. We will be able to see clearly that an EFT construction is a straightforward reorganization of the perturbation theory. Importantly, it is a very powerful reorganization: It \textit{automatizes} the most general resummations, whose derivative expansion is protected by a scale hierarchy~\cite{Appelquist:1974tg}. This is well known in the EFT literature (see e.g.\ Refs.~\cite{Manohar:2018aog,Laine:2016hma,Schicho:2020xaf}), but it becomes self-evident in the approach taken here.

This article builds upon the literature on EFTs, drawing especially from pedagogical Refs.~\cite{Manohar:2018aog,Laine:2016hma,Braaten:1995cm}. The alignment of the novel EFT construction method with previous literature is stated clearly in Sec.~\ref{sec:matching}, where the equivalence with the matching procedure is shown.

In particular, there are two key elements discussed in Sec.~\ref{sec:method} (in Boxes~\ref{diff:one}, \ref{diff:two}). These have already been known in the EFT literature. The light-mass expansion (Box~\ref{diff:one}) was first used in a high-temperature setting in Ref.~\cite{Braaten:1995cm}. The second key element in Box~\ref{diff:two} was first needed in the high-temperature context in the computation of the quantum chromodynamics (QCD) pressure to $\order{g^6\ln g\,T^4}$ in Ref.~\cite{Kajantie:2002wa}.

Although the effects of the light-mass expansion are mathematically the same in the novel method and in the matching procedure, there is an important difference: In the matching procedure, the purpose of the light-mass expansion is to facilitate computations. This is certainly true here as well. In addition however, its physical interpretation becomes central. As we will discuss, it removes the IR scale from the 1LPI diagrams. Consequently, it allows us to only integrate over the UV scale.% This is discussed in more detail in Sec.~\ref{sec:separatingIRUV}.

We want to explicitly note here, that the novel method applies to perturbative theories with a scale hierarchy between field masses, or masses and temperature. There, the IR-scale degrees of freedom are already manifestly present in the initial QFT. The resulting EFT can actually be strongly coupled (\textit{cf}.\ e.g.\ Refs.~\cite{Kajantie:1995dw,Braaten:1995cm}). However, the integrated UV scale cannot.
As a result, the method cannot be used to construct the chiral effective description for QCD, for example. The UV scale description, QCD, becomes non-perturbative, and the degrees of freedom of the chiral effective description are not manifestly present in the QCD Lagrangian.

%Historically, one of the first EFTs was the chiral effective description~\cite{Weinberg:2021exr}. As noted above, the novel method is not applicable in this particular case. It is quite possible that the method has not historically been derived due to this fact, even though the method is both physically and computationally well motivated and transparent.
% very handy for computations, and understanding EFTs.

This article is, in a sense, split into two parts: Secs.~\ref{sec:method}--\ref{sec:exampleII} and Secs.~\ref{sec:methodlong}--\ref{sec:mixingIRUV}. The novel method is, in principle, very straightforward. Hence, we discuss concisely everything needed for computations in the first part. A short, but complete discussion of the method can be found from Sec.~\ref{sec:method}. Then, we move on to examples in Secs.~\ref{sec:exampleI} and \ref{sec:exampleII}: The first example is a regular computation of free energy, which displays how to handle the thermal scale in different situations. The second example computes the gravitational-wave spectrum from a first-order phase transition as far as possible with pen and paper. It showcases obtaining so-called derivative contributions to an effective action and the execution of resummations in a background field. The first example has not been computed elsewhere in the literature, but the second example has been used in Ref.~\cite{Gould:2021ccf} to illustrate nucleation rates. It was initially computed in Ref.~\cite{Hirvonen:2020jud}.

In the second part, we give more elaborate, and hopefully more pedagogical discussions. Here, the reader may find more complete explanations of the matters handled in the first part. In Sec.~\ref{sec:methodlong}, we discuss through the novel EFT construction method again.
%In Sec.~\ref{sec:thermal}, we discuss general features relating to the thermal scale. We begin with a brief review of imaginary time formalism, and then move on to discussing the handling of the thermal scale in different situations.
In Sec.~\ref{sec:ordereps}, we discuss more thoroughly the appearance of regularization-dependent terms (\textit{cf}.\ Box~\ref{diff:two} in Sec.~\ref{sec:method}). In Sec.~\ref{sec:mixingIRUV}, we discuss the possibility of integrating out the UV scale even if there is some mixing between the light and heavy fields. Finally, we show the equivalence with the matching procedure in Sec.~\ref{sec:matching}, and conclude in Sec.~\ref{sec:discussion}. Notations and conventions can be found from Appendix~\ref{app:notation}.

%%%%%%%%%%%%%%%%%%%%%%%%%%%%%%%%%%%%%
%%%%%%%%%%%% SECTION II %%%%%%%%%%%%% 
%%%%%%%%%%%%%%%%%%%%%%%%%%%%%%%%%%%%%

\section{Integrating out a UV scale:\\an overview}
\label{sec:method}

%\jh{Moved the section \ref{sec:summary} to here. Change wordings and add figs. for self-contained discussion. Maybe some Eqs. as well.}

%We have now completed the method for constructing an EFT by integrating out a UV scale. Here, we will summarize the method by condensing all of the important points found above.

%\jh{Add figs}

Here, we will provide a concise discussion of the novel method for constructing an EFT. There, one directly integrates over a scale in a path integral. A lengthier and hopefully more pedagogical discussion can be found in Sec.~\ref{sec:methodlong}.
%We will reference the

We start with the partition function%
\footnote{Although the partition function is for a Euclidean QFT, the general discussion here applies to collider physics. This is due to the Wick rotation used in computing Feynman diagrams, as emphasized by the seminal article of Ref.~\cite{Appelquist:1974tg}.}
of a theory containing two mass scales, $m^2\ll M^2$:
\begin{align}
    Z&=\int\mathcal{D}\Phi\, e^{-\Seu[\Phi]}\,,\label{eq:reprepone}
\end{align}
where $\Phi$ represent all of the field content of the theory. We keep all of the indices (spin, Lorentz, etc.) implicit.

In order to create an EFT for the IR scale, given by the light mass, $m$, we need to integrate over the UV scale, given by the heavy mass, $M$.

Let us divide the modes of the fields into these two scales:
\begin{equation}\label{eq:fluctdivreprep}
    \mathcal{D}\Phi
    =\mathcal{D}\ir \times \mathcal{D}\uv\,,
\end{equation}
We can now \textit{rewrite} the initial partition function as
\begin{align}
    Z&=\int\mathcal{D}\ir\, e^{-S_{\text{eff}}[\ir]}\,,\label{eq:repreptwo}\\
    S_{\text{eff}}[\ir] &= -\ln \int\mathcal{D}\uv\,e^{-S_{\mathrm{E}}[\ir+\uv]}\,,\label{eq:effactreprep}
\end{align}
where we have denoted $\Phi=\ir+\uv$ in the exponent.
The description given by the partition function in Eq.~\eqref{eq:repreptwo} only contains the IR scale. Therefore, it represents the EFT for the IR scale. Correspondingly, the effective action, $S_{\text{eff}}[\ir]$, is the action governing the EFT.

We will now focus on answering two key questions: What is exactly the split in Eq.~\eqref{eq:fluctdivreprep}? How can one compute the effective action in Eq.~\eqref{eq:effactreprep} in perturbation theory?
% What is the diagrammatic representation for the effective action? How can one compute the resulting diagrammatic expansion?

The UV scale $M$ contains the heavy fields, $\heavy$, with the masses $M$. However, some modes of the light fields, $\light$, also belong to the UV scale, due to a UV-scale momentum, $p^2\sim M^2$. We will label these UV-scale modes of the light field as $\lighthigh$.
%: The light fields, can have a momentum on the UV scale, $p^2\sim M^2$. These UV-momentum modes of the light fields, $\lighthigh$, also belong to the UV scale. Therefore, 
Only the IR-momentum modes of the light fields, $\lightlow$, belong to the IR scale. Thus, we have
\begin{equation}\label{eq:refinedscaleseparationreprep}
    \mathcal{D}\ir = \mathcal{D}\lightlow\,,\qquad\mathcal{D}\uv
    =\mathcal{D}\lighthigh\, \mathcal{D}\heavy\,.
\end{equation}

%In the latter form, the UV-scale modes have been integrated over. These UV modes were found to contain the full heavy field, $\heavy$, and the UV-momentum modes of the light fields, $\lighthigh$. Only the IR-momentum modes of the light fields, $\lightlow$, were left for the EFT.

%The effective action in the latter partition function describes the EFT. It is given by
%\begin{equation}
%    S_{\text{eff}}[\ir] = -\ln \int\mathcal{D}\uv\,e^{-S_{\mathrm{E}}[\ir+\uv]}\,,
%\end{equation}
%where the IR-scale fluctuations are a background for the UV modes.

Let us then discuss the second question, the computation of the effective action in perturbation theory. It is slightly too big of a question. Hence, we split it into two parts: What is the diagrammatic representation for the effective action? How can one compute the resulting diagrammatic expansion?

%We will eventually find, that the diagrammatic representation of the effective action is the same as the that of the 1PI effective action for the light fields, $\light$: Only the diagrams, which are one-particle irreducible in the light fields, i.e.\ the 1LPI diagrams (Fig.~\ref{fig:reduciblereprep}) contribute, and the external legs of the diagrams are only light-field legs. However, the meaning of the diagrams will be slightly different: The internal light-field propagators are $\lighthigh$ propagators, and the external legs are $\lightlow$ legs. 

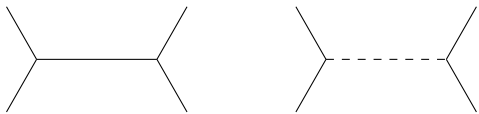
\begin{figure}
    \begin{tikzpicture}
        \node[] at (0,0) {$\feynalignscaled{1}{\notintuitionfourpoint}$};
    \end{tikzpicture}
    \hspace{1cm}
    \begin{tikzpicture}
        \node[] at (0,0) {$\feynalignscaled{1}{\intuitionfourpoint}$};
    \end{tikzpicture}
    \caption{The solid lines represent light fields, $\light$, and the dashed lines heavy fields, $\heavy$. The diagram on the left is not a 1LPI diagram due being reducible in the $\light$ propagator in the middle. The diagram on the right is a 1LPI diagram.}
    \label{fig:reduciblereprep}
\end{figure}

We will eventually find, that the diagrammatic representation of the effective action only contains 1LPI diagrams (Fig.~\ref{fig:reduciblereprep}), whose internal propagators are $\lighthigh$ propagators, and the external legs are $\lightlow$ legs. Thus, the diagrammatic expansion is nearly the same as the 1PI effective action for the light fields, 1LPI action, but the $\light$ propagators and external legs have a special meaning.

We can make three key observations from Eq.~\eqref{eq:effactreprep}:
\begin{itemize}
    \item[1.] There is a logarithm, $\ln$, of the path integral.
    \item[2.] The IR field, $\ir$, acts as a background for $\uv$.
    \item[3.] We only integrate over the UV scale, $\int\mathcal{D}\uv$.
\end{itemize}

The effect of the logarithm is straightforward: The diagrammatic expansion only contains connected diagrams. (See e.g.~\cite{Laine:2016hma}, Sec.~3.)

The background field $\ir$ is the same as $\lightlow$, Eq.~\eqref{eq:refinedscaleseparationreprep}. The background field can be diagrammatically understood as external legs (see e.g.\ Ref.~\cite{Coleman:1973jx}): It is not integrated over. As a consequence, it does not propagate.

The external legs bring the external momenta, $k$, into the diagrams. Since the external legs are the IR-scale background, $\ir$, the external momenta are from the IR scale,
\begin{equation}\label{eq:IRextMomEqreprep}
    k^2\sim m^2\,.
\end{equation}

Finally, we focus on the fact that we only integrate over the UV scale. The UV-scale contains modes from both the heavy and the light fields, Eq.~\eqref{eq:refinedscaleseparationreprep}. Therefore, the diagrams contain propagators from all of the fields.

The heavy fields are integrated over in their entirety. Consequently, their propagators may contain all momenta. However, the light field propagators only contain the high-momentum modes, $\lighthigh$, with $p^2\sim M^2$.

There is an important diagrammatic consequence from the fact that the internal $\light$ propagators are on the UV scale, and the external $\light$ legs are on the IR scale: The effective action of Eq.~\eqref{eq:effactreprep} does not contain diagrams, which are one-particle-reducible in the light fields (\textit{cf}.\ Fig.~\ref{fig:reduciblereprep}). Only the 1LPI diagrams contribute. This follows from the fact that a reducible $\light$ propagator would carry an IR-scale momentum, $p^2\sim m^2$, from the external legs, Eq.~\eqref{eq:IRextMomEqreprep}. Thus, it would be a $\lightlow$ propagator.

To summarize the included diagrams, the diagrammatic expansion of the effective action in Eq.~\eqref{eq:effactreprep} only contains the 1LPI diagrams. In these diagrams, the internal propagators are $\lighthigh$ propagators and the external legs are $\lightlow$ legs.

The question that we are left with is the following: How do we only take into account only the $\lighthigh$ modes in the internal $\light$ propagators, and the $\lightlow$ modes in the $\light$ external legs in dimensional regularization?

%In Sec.~\ref{sec:diagrammaticexp}, we started studying the effective action. We found that only to connected diagrams, which were 1-light-field-irreducible, contribute. Therefore, it has the same diagrammatic representation to the 1PI action for the light fields.

%The method to only compute the UV-scale contributions in dimensional regularization is to expand the 1-light-field-irreducible diagrams in the IR quantities:

The resolution is to expand the 1LPI diagrams in the IR quantities:
\begin{align}\tag*{1}\label{diff:one}
    \fbox{
        \begin{minipage}{7.5cm}
            \vspace{0.1cm}
            Expand the diagrams in IR quantities, i.e.\ the light masses, $m^2$, and the low external momenta, $k$,
            \vspace{-0.2cm}
            $$\frac{1}{(p+k)^2+m^2} \to\frac{1}{p^2}-\frac{2p\cdot k+k^2+m^2}{p^4}+\dots\,,$$
            \vspace{-0.5cm}
            $$\frac{1}{(p+k)^2+M^2} \to\frac{1}{p^2+M^2}-\frac{2p\cdot k+k^2}{(p^2+M^2)^2} +\dots\,,$$
            where $p$ is a loop momentum and $M^2$ is the heavy mass.
            \vspace{0.1cm}
        \end{minipage}
    }
\end{align}

The IR-quantity expansion (the expansion above in Box~\ref{diff:one}) does not affect the UV-scale contributions, where $p^2\sim M^2$ for the light-field propagators. Hence, it does not affect the $\lighthigh$ contributions. However, it expands the IR scale out of the loop integrals. For example,
\begin{align}\label{eq:theRemovalofIRreprep}
    \irqe{\int_p\frac{1}{p^2+m^2}}
    =\int_p\frac{1}{p^2} -m^2\int_p\frac{1}{p^4}+\dots\,,
\end{align}
where IRq exp. refers to the IR-quantity expansion. Thus, the IR-scale contributions of $\lightlow$ from $p^2\sim m^2$ are not taken into account.

Instead of the IR-scale contributions, there are infrared divergences in the momentum region of $p^2\ll M^2$ of the light-field propagators. This may seem alarming. However, the resulting EFT requires renormalization. The aforementioned infrared divergences become the correct counterterms to the effective action, i.e.\ they cancel the ultraviolet divergences of the EFT, Fig.~\ref{fig:regulationreprep}. This is shown in Appendix~\ref{app:IRdivs} along the lines of Ref.~\cite{Manohar:2018aog}.

\begin{figure}
    \centering
    \begin{tikzpicture}[scale=1.0]
        \filldraw[draw=none, fill=black!30, very thick](0,4.5) rectangle (8.5,3.5);
        \filldraw[draw=none, fill=black!20, very thick](0,3.5) rectangle (8.5,1.75);
        \filldraw[draw=none, fill=black!10, very thick](0,1.75) rectangle (8.5,0);
        \draw[dashed] (4.525,4.5) -- (4.525,0);
        \node[] at (0.33,2.625) {$M$};
        \node[] at (0.33,0.875) {$m$};
        \draw[->,thick] (0.75,0) -- (0.75,4.25);
        \node[] at (0.45,4.25) {$E$};
        %\node[] at (0.33,1.75) {$\Lambda$};
        \node[] at (2.625,1.75) {\textbf{Full theory}};
        \draw[->] (2.5,3.9) -- (2.5,3.1);
        \node[] at (1.85,3.75) {$\substack{\text{counter-}\\ \text{terms}}$};
        \draw[->] (2.75,3.1) -- (2.75,3.9);
        \node[] at (3.5,3.25) {$\substack{\text{ultraviolet}\\ \text{divergences}}$};
        \node[] at (6.375,2.625) {\textbf{UV scale}};
        \draw[->] (6.25,3.9) -- (6.25,3.1);
        \node[] at (5.6,3.75) {$\substack{\text{counter-}\\ \text{terms}}$};
        \draw[->] (6.5,3.1) -- (6.5,3.9);
        \node[] at (7.25,3.25) {$\substack{\text{ultraviolet}\\ \text{divergences}}$};
        \draw[->] (6.25,2.15) -- (6.25,1.35);
        \node[] at (5.5,2) {$\substack{\text{infrared}\\ \text{divergences}}$};
        \draw[->] (6.5,1.35) -- (6.5,2.15);
        \node[] at (7.25,1.5) {$\substack{\text{ultraviolet}\\ \text{divergences}}$};
        \node[] at (6.375,0.875) {\textbf{EFT}};
    \end{tikzpicture}
    \caption{The IR-quantity expansion induces infrared divergences to the UV-scale contributions, shown on the right of the dashed line. These infrared divergences cancel identically against the ultraviolet divergences of the EFT. Correspondingly, the infrared divergences are the counterterms for the EFT (\textit{cf}.\ the cancellation of the ultraviolet divergences of the UV scale with the counterterms).}
    \label{fig:regulationreprep}
\end{figure}
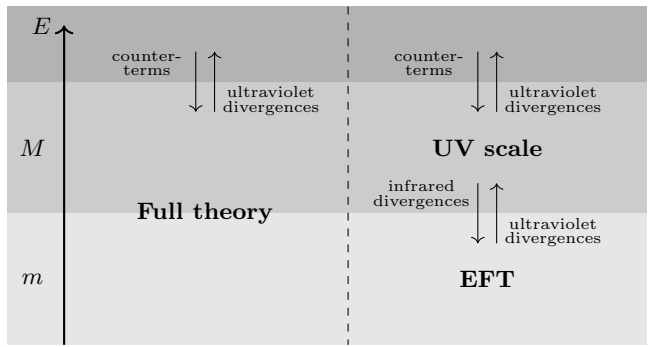

Due to the possible ultraviolet divergences of the EFT, the $\order{\epsilon}$ terms resulting from the 1LPI diagrams cannot be discarded without consideration. (Here, $\epsilon$ is the dimensional modifier, $D=4-2\epsilon$.) The terms can still affect physical quantities by factorizing with the divergences:
%we found in Sec.~\ref{sec:subsecepsilon} that one must keep the UV contributions that are suppressed by powers of $\epsilon$, if they can multiply with the $\epsilon$ poles in the EFT to produce physical contributions of $\order{\epsilon^0}$:
\begin{align}\tag*{2}\label{diff:two}
    \fbox{
        \begin{minipage}{7.5cm}
            \vspace{0.1cm}
            The $\order{\epsilon}$ terms of the UV-scale contributions must be kept if they can multiply with the $\epsilon$ poles of the EFT diagrams to become at least of the physical order of $\epsilon^0$.
            \vspace{0.1cm}
        \end{minipage}
    }
\end{align}
This is discussed in more detail in Sec.~\ref{sec:ordereps}.

Finally, we note that the external legs are naturally from the IR scale, i.e.\ $\lightlow$. The only mass scale present in the EFT is $m^2$. As a consequence, only the IR-scale contributions from $\lightlow$ are taken into account. One can understand this by the fact that the momentum integrals in the EFT only receive ultraviolet divergences from $p^2\gg m^2$. (Slightly non-trivially, the non-existence of the mass scale $M$ in the EFT requires the external-momentum expansion of the $\heavy$ propagator in Box~\ref{diff:one}.)

The EFT construction for the light fields is nearly identical to the construction of the 1PI effective action for them, i.e. the 1LPI action. The diagrammatic representation is the same. The only two differences are given in Boxes above. Thus, we can repackage the effective action in Eq.~\eqref{eq:effactreprep} in terms of the 1LPI action, $\Gamma^{\text{1LPI}}$:
\begin{equation}
    \Sef[\lightlow]=\Gamma^{\text{1LPI}}[\lightlow]\eval_{\substack{\text{IRq}\\\text{exp.}}}\,.\label{eq:finalPackage}
\end{equation}
The right-hand side is an expansion in
\begin{equation}
    \frac{m^2}{M^2}\,,\quad\frac{k^2}{M^2}\,,\quad\epsilon\,,
\end{equation}
where the external-momentum expansion becomes the derivative series in the effective action (\textit{cf}. the example in Sec.~\ref{sec:exampleII}).
All of the terms that contribute to physical quantities to a desired order must be kept and the rest can be truncated.

%\jh{Mixing}

After integrating out the scale according to Eq.~\eqref{eq:finalPackage}, it is possible that there are non-canonical kinetic terms and mixing between the EFT fields, $\lightlow$. These can be removed by field redefinitions. It seems to be true that even mixing between light and heavy fields can be handled within the resulting EFT by field redefinitions. This is explored in Sec.~\ref{sec:mixingIRUV} via an example.

The result for the effective action in Eq.~\eqref{eq:finalPackage} is very close to the one in Refs.~\cite{Georgi:1991ch,Georgi:1992xg,Georgi:1993mps}, that the 1PI action of the EFT reproduces the 1LPI effective action of the initial QFT:
\begin{equation}\label{eq:georgi}
    \Gamma^{\text{1PI}}_{\text{eff}}[\lightlow]=\Gamma^{\text{1LPI}}[\lightlow]\,.
\end{equation}
Note however, that the IR-quantity expansion obtains directly the effective action, $\Sef$, and not the 1PI effective action computed from it.

%\jh{The result resembles that of seminal Refs.~}

By Eqs.~\eqref{eq:reprepone} and \eqref{eq:repreptwo}, the EFT construction is clearly a reorganization of a perturbative computation in scales: one computes the UV scale before the IR scale. This automatizes the most general resummations, which are protected by the scale hierarchy.

Conveniently, it is also a reorganization in the diagrammatic expansion: To compute the effective action in Eq.~\eqref{eq:effactreprep}, one only needs to compute the diagrams that contain at least one heavy-field propagator. If a diagram contains solely light fields, it is zero as a scale-free integral, due to the IR-quantity expansion in Box~\ref{diff:one}. Thus, one only needs to compute the diagrams of purely light fields within the EFT, and the diagrams containing the heavy fields in the EFT construction.

\section{Example 1:\\three-scale system}\label{sec:exampleI}

%\jh{Hand-picked example; there are natural scale hierarchies}

%\jh{Check the mass contribution from vacuum! AND the runnings}

%\jh{This says ``sorry for existing'' too hard. Maybe just say it is easily done and relate to the 1PI discussion by pointing out the diagrams. Also, tell the three different parts, especially the discussion, where the usefulness is discussed.}

In this section, we demonstrate the fluent construction of EFTs by integrating out scales according to Sec.~\ref{sec:method}. 
%is not a burden to the computations, but only serves to streamline them.
The toy model of the section was chosen for being rather minimalistic while still showing how the method presented in the previous section works with respect to a thermal scale.

We start by defining the model and the computational goal for the section. Then, we have three subsections, which focus on a single scale each. Finally, we have a discussion part, where we discuss how the EFT calculation eases computations based on this particular example.

%We apply the method of constructing effective field theories to compute free energy in a model given by the Lagrangian
%\begin{align}
%    \mathscr{L}_{\mathrm{E}}&=\mathscr{L}_{\Phi}+\mathscr{L}_{\Chi}+\mathscr{L}_{\mathrm{I}}\,,\\
%    \mathscr{L}_{\Phi}&=\frac{1}{2}(\partial_\mu\Phi)^2+\frac{m^2}{2}\Phi^2\,,\\
%    \mathscr{L}_{\Chi}&=\frac{1}{2}(\partial_\mu\Chi)^2+\frac{M^2}{2}\Chi^2\,,\\
%    \mathscr{L}_{\mathrm{I}}&=\frac{g}{2}\Chi\Phi^2.
%\end{align}
The example model is given by the Lagrangian
\begin{align}\label{eq:exlagI}
    \mathscr{L}_{\mathrm{E}}=&
    \frac{1}{2}(\partial_\mu\Phi)^2+\frac{m^2+\cou{m^2}}{2}\Phi^2 +\frac{\lambda}{4!}\Phi^4\nonumber\\
    &+\frac{1}{2}(\partial_\mu\Chi)^2+\frac{M^2}{2}\Chi^2+\frac{g}{2}\Chi\Phi^2\,,
\end{align}
where the real scalar field $\Phi$ is much lighter than the real scalar field $\Chi$ with $m^2\ll M^2$ and for perturbativity $g/M\ll 1$, $\lambda\ll 1$.

For power counting, we define a parameter
\begin{equation}
    x\equiv g/M\ll1\,.
\end{equation}
The light mass is set to be
\begin{equation}
    m^2\sim g^2=x^2 M^2\,,
\end{equation}
which is natural because $m^2$ runs at the order $g^2$.
Finally, we set
\begin{equation}
    \lambda\sim x\,.
\end{equation}

\begin{figure}
    \centering
    \begin{tikzpicture}[scale=1.05]
        \draw[->,thick] (-0.25,0) -- (6.25,0) node[anchor=north west] {$E$};
        \draw[thick] (5,-0.1) node[anchor= north] {$\Chi$} -- (5,0.1) node[anchor=south] {$M$};
        \draw[thick] (3,0.1) node[anchor= south] {$\pi T$} -- (3,-0.1) node[anchor=north] {$\Phi_{n\neq0}$};
        \draw[thick] (1,-0.1) node[anchor=north] {$\Phi_0$} -- (1,0.1) node[anchor=south] {$m$};
    \end{tikzpicture}
    \caption{Field degrees of freedom in each scale of the system. The heavy field is at the scale $M$. The non-zero Matsubara modes of $\Phi$, $\Phi_{n\neq0}$, are on the thermal scale (see Appendix.~\ref{app:imtimform}), and the zero Matsubara mode $\Phi_0$ is on the lowest scale $m$.}
    \label{fig:scalesI}
\end{figure}
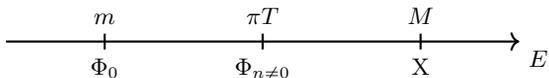

In this section, we will study the system at temperatures of
\begin{equation}
    T\sim x^{1/2}M\,.
\end{equation}
Of course, the temperature can be different physically, but this hand-picked choice is convenient for us. It allows to display different kinds of UV scales:
The heavy $\Chi$ field is near zero temperature,
\begin{equation}
    M\gg T\,,
\end{equation}
and the light $\Phi$ field is at high temperatures,
\begin{equation}
    m\sim x^{1/2}T\ll T\,.
\end{equation}
The scales of the system are shown in Fig.~\ref{fig:scalesI}.

Here, we will compute the free energy density to order $x^{3/2}T^4$. This is again a choice we make to keep the computation illustrative: All of the three scales affect the free energy at this order. Also, the computations do not yet become unnecessarily involved. Still, they display the power of EFTs to render computations rather effortless.

%The task is simplified by leveraging the scales with EFT construction: We can first tackle only the contributions from the scale of $M$ without worrying anything about the lower scales. Then, we can do the same with the scale $\pi T$ and finally with $m$. There is no need to ponder different resummations as every contribution falls onto its place naturally. This is a great advantage with more complicated models, and the generalization of this calculation should be straightforward.

%\jh{Simplify this considerably!}

Before we start computing, we identify UV-scale contributions to the free energy. The total free energy of a system is given by
\begin{equation}\label{eq:thefreeformula}
    F=-T\ln Z.
\end{equation}
%Within this method of constructing effective descriptions, it is clear that the partition function does not change when integrating out a scale (\textit{cf.} Eqs.~\eqref{eq:statdesc}--\eqref{eq:effact}).
%In the following, we identify UV-scale contributions to the free energy.
Let us divide an effective Lagrangian into two pieces,
\begin{equation}
    \mathscr{L}_{\text{eff}}[\ir]=f_{\rmii{UV}}+\mathcal{L}[\ir]\,,
\end{equation}
where $\mathcal{L}$ contains all the field dependence, and $f_{\rmii{UV}}$ is the so-called unit operator of the effective Lagrangian. (It comes from the vacuum-bubble diagrams of a UV scale.) Using the EFT partition function, Eq.~\eqref{eq:effstatdesc}, we obtain:
\begin{align}\label{eq:freenergyfromeft}
    F&=-T \ln \int\mathcal{D}\ir\, e^{-\int_0^\beta\dd \tau\int_{\bf{x}}\left( f_{\rmii{UV}}+\mathcal{L}\right)}\nonumber\\
    &=V f_{\rmii{UV}}-T \ln \int\mathcal{D}\ir\, e^{-\int_0^\beta\dd \tau\int_{\bf{x}}\mathcal{L}}\,,
\end{align}
Thus, the free energy density contribution from a UV scale, $f_{\rmii{UV}}$, is given by the unit term of the effective Lagrangian.

%To show this, we split the effective Lagrangian, $\mathscr{L}_{\rmii{eff}}$, into its unit term, $f_{\rmii{UV}}$, and the IR field dependent part, $\mathcal{L}$.  Then, we insert the partition function of the effective description of Eq.~\eqref{eq:effstatdesc} to the free energy formula,
%\begin{align}\label{eq:freenergyfromeft}
%    F&=-T \ln \int\mathcal{D}\ir\, e^{-\int_0^\beta\dd \tau\int_{\bf{x}}\left( f_{\rmii{UV}}+\mathcal{L}\right)}\nonumber\\
%    &=V f_{\rmii{UV}}-T \ln \int\mathcal{D}\ir\, e^{-\int_0^\beta\dd \tau\int_{\bf{x}}\mathcal{L}}\,,
%\end{align}
%where $V$ is the volume of space. Thus, showing that the unit term is indeed the free energy density contribution from the UV scale, while the IR part in the path integral is still to be computed.

\subsection{Scale $M$}\label{sec:scaleMI}

%\jh{
%\begin{itemize}
%    \item Which is the light field?
%    \item 1$\Phi$I diagrams
%    \item Power counting: Only mass
%    \item What do we mean by the diagrams: Lagrangian coefficient, term or action term?
%    \item Diagrams with only $\Phi$ loops: zero
%    \item IR-quantity expansion directly
%    \item Running, but without fuzz
%    \item No need for the $\order{\epsilon}$ terms
%\end{itemize}
%}

%\jh{Show the split explicitly.}
Due to the scale hierarchy $T\ll M$, the scale $M$ has only exponentially suppressed thermal corrections, $\exp(-M/T)$, which are higher order than our accuracy goal of $\order{x^{3/2}T^4}$. Thus, we can neglect these contributions and deal with the scale as if the system was in zero temperature.

We can divide the field content into the heavy and light fields:
\begin{align*}
    \text{Heavy:}&\quad\Chi\,,\\
    \text{Light:}&\quad\Phi\,.
\end{align*}
Thus, the scales are given by
\begin{align*}
    \text{UV:}&\quad\Chi\,,\;\Phi_{\text{UV}}\,,\\
    \text{IR:}&\quad\Phi_{\text{IR}}\,.
\end{align*}
Here, $\Phi_{\text{UV}}$ are defined as the $\Phi$ fluctuations with UV momenta, $p^2\sim M^2$, and $\Phi_{\text{IR}}$ with $p^2\ll M^2$.

As discussed in Section~\ref{sec:method}, to integrate out the UV scale, we need to compute one-particle-irreducible diagrams in the light fields, $\Phi$, while expanding them in the IR quantities ($m^2$ and external momenta $k$). 

%It turns out that we only need to compute the leading part of 
Below, we will find that we only need to compute the leading part of
\begin{equation}\label{eq:theonlydiag}
    \irqe{\feynalignscaled{0.75}{\intuitionmass}}
\end{equation}
for our accuracy. Other contributions from the scale $M$ will lead to free-energy contributions suppressed beyond the order $x^{3/2}T^4$.
In the diagram, the solid lines refer to the $\Phi$ field, the dashed line to the $\Chi$ field, and IRq exp. refers to the IR-quantity expansion, Box~\ref{diff:one}.

Before examining in detail that the above diagram is the only needed contribution from the scale $M$, we will construct the effective Lagrangian for $\Phi_{\text{IR}}$.

%we can construct an effective field theory for an IR scale by constructing a slightly tweaked 1PI action for the light fields. Here, the light field is the $\Phi$ field, and the heavy field is $\Chi$.

%For the effective action, we only need to compute a potential term,
%\begin{equation}
%    \qquad\feynalignscaled{0.75}{\intuitionmass}\,,
%\end{equation}
%where the solid lines refer to $\Phi$ field and the dashed line to $\Chi$ field. All the higher order contributions, including the derivative expansion of the above diagram, contribute to the total free energy density beyond $\order{x^{3/2}T^4}$.

There are a few ways to interpret the above diagram. The most common one in the literature would probably be to interpret it as a self energy contribution:
\begin{align}\label{eq:diagramMeaning}
    &\irqe{\feynalignscaled{0.75}{\intuitionmass}}\nonumber\\
    =&
    -g^2\int_{p}\frac{1}{(p^2+M^2)p^2}\times\qty(1+\overbrace{\order{\frac{m^2}{M^2},\frac{k^2}{M^2}}}^{\text{IRq exp.}})\nonumber\\[5pt]
    =&-g^2\left(\frac{1}{(4\pi)^2\epsilon} +\frac{1}{(4\pi)^2}\left(1-\ln(\frac{M^2}{\mu^2})\right)\right)\nonumber\\[3pt]
    &\qquad\times\qty(1+\order{\frac{m^2}{M^2},\frac{k^2}{M^2},\epsilon})\,.
\end{align}
Here, we are however constructing an effective action. From this perspective, the self-energy interpretation is the same as interpreting the diagram as a contribution to a Lagrangian coefficient.

We want to emphasize notationally, that we are computing contributions to the effective action. Hence, we interpret the diagram as a contribution to the effective Lagrangian,
\begin{align}\label{eq:theonlylagtermuvI}
    &\irqe{\feynalignscaled{0.75}{\intuitionmass}}\nonumber\\
    =&-\frac{g^2}{2}\left(\frac{1}{(4\pi)^2\epsilon} +\frac{1}{(4\pi)^2}\left(1-\ln(\frac{M^2}{\mu^2})\right)\right)\Phi_{\text{IR}}^2\,,
\end{align}
where $\Phi_{\text{IR}}^2$ comes from the two external IR legs. The action contribution is then just a spatial integral over the Lagrangian contribution.
%(We have dropped the higher-order terms of the expansions.)

The Lagrangian, which describes the thermal scale $\pi T$ is given by
\begin{equation}\label{eq:piTLag}
    \mathscr{L}_{\pi T}=
    \frac{1}{2}(\partial_\mu\Phi_{\text{IR}})^2+\frac{m^2_{\pi T}}{2}\Phi_{\text{IR}}^2+\frac{\lambda}{4!}\Phi_{\text{IR}}^4\,.
\end{equation}
The effective mass,
\begin{align}\label{eq:effpiTmass}
    m^2_{\pi T}&=m^2(\mu)-\frac{g^2}{(4\pi)^2}\left(1-\ln(\frac{M^2}{\mu^2})\right)\,,
\end{align}
contains the UV-scale contribution.
We will power count it as
\begin{equation}
    m^2_{\pi T}\sim xT^2\,.
\end{equation}

Note, that the effective Lagrangian does not contain any counterterms. The $\epsilon$ pole in the UV contribution, Eq.~\eqref{eq:theonlylagtermuvI}, cancels with the initial mass counterterm in Eq.~\eqref{eq:exlagI}.  As a consequence, the Lagrangian coefficients do not run (see e.g.\ Ref.~\cite{Schwartz:2014sze} Sec.~23). Indeed, the same UV contribution cancels the running of the initial mass parameter:
\begin{equation}
    m^2(\mu')=m^2(\mu)+\frac{g^2}{(4\pi)^2}\ln(\frac{{\mu'}^2}{\mu^2})\,.
\end{equation}
All of the physical quantities, obtained from the effective Lagrangian in Eq.~\eqref{eq:piTLag}, will be manifestly independent of the renormalization scale, $\mu$.

Finally, we will discuss, why we only need to take into account the leading part of the computed UV contribution, Eq.~\eqref{eq:theonlylagtermuvI}, and why no other diagrams are needed. We will discuss here the following diagrams:
\begin{align}\label{eq:diaglistign}
    \feynalignscaled{0.5}{\oneloopbubble{dashed}}\,,\quad\feynalignscaled{0.5}{\selfenergyfour{}{}}\,,\quad\feynalignscaled{0.5}{\intuitionlast}\,,\quad \feynalignscaled{0.5}{\intuitionfourpoint}\,.
\end{align}

In the next subsection, Sec.~\ref{sec:scalepitI}, we will discover that the leading contribution to the free energy from the effective mass, $m^2_{\pi T}$, is $m^2_{\pi T}\times T^2$ (\textit{cf}.\ Eq.~\eqref{eq:hidingsunset}). Thus, we still want to know the corrections to the effective mass that are
\begin{equation}
    \Delta m^2_{\pi T}\sim x^{3/2} T^2\,.
\end{equation}
The leading order part, taken into account, is $xT^2$. Higher-order corrections to mass from the IR-quantity expansion are suppressed by the powers of
\begin{equation}
    \frac{m^2}{M^2}\sim x^2\,.
\end{equation}
Consequently, they can be neglected.

By a similar analysis, we can also neglect the derivative terms following from the external-momentum expansion. The leading term would affect the kinetic term $(\partial_\mu\Phi_{\text{IR}})^2/2$. Often, the easiest way to see, if the correction needs to be obtained, is to use field redefinitions to maintain the canonical kinetic term:
\begin{align}
    \frac{Z_{\Phi_{\text{IR}}}}{2}(\partial_\mu\Phi_{\text{IR}})^2&\to\frac{1}{2}(\partial_\mu\Phi_{\text{IR}})^2\,,\\
    \frac{m^2_{\pi T}}{2}\Phi^2_{\text{IR}}&\to\frac{m^2_{\pi T}}{2Z_{\Phi_{\text{IR}}}}\Phi^2_{\text{IR}}\,,\\
    \frac{\lambda}{4!}\Phi^4_{\text{IR}}&\to\frac{\lambda}{4!Z_{\Phi_{\text{IR}}}^2}\Phi^4_{\text{IR}}\,.
\end{align}
This way the correction can be analyzed as part of potential coefficients.

Also, the $\order{\epsilon}$ terms can safely be ignored. Since there are no counterterms, the effective description produces no ultraviolet $\epsilon$ poles.

%The derivative series can be neglected as well. A similar analysis shows that we only need to know the coefficient of the kinetic term, $Z_{\Phi_{\text{IR}}}(\partial_\mu\Phi_{\text{IR}})^2/2$, to
%\begin{equation}
%    \Delta Z_{\Phi_{\text{IR}}}\sim x^{1/2}\,.
%\end{equation}
%The leading derivative correction coming from the IR-quantity expansion is of order
%\begin{equation}
%    \frac{g^2}{M^2}k^2\sim x\, k^2 \to x\,\partial_\mu^2\,.
%\end{equation}
%Thus, negligible.

%We can scale the field normalization away by a field redefinition:
%\begin{align}
%    \Phi^2_{\text{IR}}&\to(1-\Delta Z_{\Phi_{\text{IR}}})\Phi^2_{\text{IR}}\,,\\
%    \frac{1+\Delta Z_{\Phi_{\text{IR}}}}{2}(\partial_\mu\Phi_{\text{IR}})^2&\to\frac{1}{2}(\partial_\mu\Phi_{\text{IR}})^2\,,\\
%    \frac{m^2_{\pi T}}{2}\Phi^2_{\text{IR}}&\to(1-\Delta Z_{\Phi_{\text{IR}}})\frac{m^2_{\pi T}}{2}\Phi^2_{\text{IR}}\,,\\
%    \Phi^2_{\text{IR}}&\to(1-2\Delta Z_{\Phi_{\text{IR}}})\Phi^2_{\text{IR}}\,,\\
%\end{align}

Let us now discuss the diagrams in Eq.~\eqref{eq:diaglistign}.

The one-loop, heavy-field diagram,
\begin{equation}
    \feynalignscaled{0.6}{\oneloopbubble{dashed}}\,,
\end{equation}
only affects the vacuum cosmological constant. Consequently, we can ignore it from the free energy.
%only give exponentially suppressed thermal corrections to the free energy from the scale $M$, which are far beyond $\order{x^{3/2}T^4}$.

%The diagram in Eq.~\eqref{eq:theonlydiag} contains a momentum integral,
%\begin{equation}
%    \int_{p}\frac{1}{(p^2+M^2)(p^2+m^2)}\,,
%\end{equation}
%from its propagator structure. Now, we implement the two steps discovered in the previous section: The step (2) means that we treat the light mass $m^2$ with a strict \jh{IRq} perturbation expansion. Here, we need only the leading term in the strict \jh{IRq} expansion:
%\begin{equation}
%    \int_{p}\frac{1}{(p^2+M^2)p^2}\times\left(1+\order{\frac{m^2}{M^2}}\right)\,.
%\end{equation}
%Then, the step (1) simply states that the integral is evaluated with the renormalization scale set to be the factorization scale $\Lambda$ between $\pi T$ and $M$:
%\begin{align}
%    &\int_{p}\frac{1}{(p^2+M^2)p^2}\nonumber\\
%    =&\frac{1}{(4\pi)^2\epsilon} +\frac{1}{(4\pi)^2}\left(1-\ln(\frac{M^2}{\Lambda^2})\right)+\order{\epsilon}\,.
%\end{align}

%Thus, the full diagram gives a mass contribution to the effective Lagrangian,
%\begin{align}\label{eq:theonlylagtermuvI}
%    &\feynalignscaled{0.75}{\intuitionmass}\nonumber\\
%    =&-\frac{g^2}{2}\left(\frac{1}{(4\pi)^2\epsilon} +\frac{1}{(4\pi)^2}\left(1-\ln(\frac{M^2}{\Lambda^2})\right)\right)\Phi^2,
%\end{align}
%where $\Phi^2$ follows from the external legs and $g^2$ from the vertices.

%The same contribution would appear as the leading radiative contribution to the 1PI potential for the $\Phi$ field.

The two diagrams,
\begin{equation}\label{eq:masscontribNon}
    \irqe{\feynalignscaled{0.6}{\selfenergyfour{}{}}}=0\,,\qquad\irqe{\feynalignscaled{0.6}{\intuitionlast}}=0\,,
\end{equation}
which look like possible possible mass contributions to orders $x^{1/2}T^2$ and $xT^2$ respectively, are identically zero. This is due to the light-field loop, which yields zero under the IR-quantity expansion:
%However, the strict \jh{IRq} perturbation theory in the IR mass, $m^2$, renders the $\Phi$-loop integrals scale free, and thus the diagrams are zero:
%\footnote{The zero contribution of the above diagram might be a surprise. We will discuss the embedding of the diagram into the EFT calculation at the end of this part around Eq.~\eqref{eq:zoomone}. The discussion is not vital to conducting an EFT calculation but it is good to be reassured that we have not made a mistake.}
%\begin{align}
%    \feynalignscaled{0.6}{\selfenergyfour{}{}}&=\frac{\lambda\Phi^2}{4}\int_p\left(\frac{1}{p^2}-\frac{m^2}{p^4}\dots\right)=0\,,\\
%    \feynalignscaled{0.6}{\intuitionlast} &=-\frac{g^2\Phi^2}{4M^2}\int_p\left(\frac{1}{p^2}-\frac{m^2}{p^4}\dots\right)=0\,.\label{eq:roguediag}
%\end{align}
\begin{align}
    \irqe{\feynalignscaled{0.6}{\selfenergyfour{}{}}}&=\frac{\lambda\Phi^2}{4}\int_p\left(\frac{1}{p^2}-\frac{m^2}{p^4}\dots\right)=0\,.%\label{eq:roguediag}
\end{align}

Finally, the four-point diagram,
\begin{equation}\label{eq:fourpointeff}
    \irqe{\feynalignscaled{0.6}{\intuitionfourpoint}} = -\frac{g^2}{8M^2}\Phi_{\text{IR}}^4+\dots\,,
\end{equation}
would contribute to the effective self-coupling:
%Here, $g^2$ follows from the vertices, $M^{-2}$ from the $\Chi$ propagator and $\Phi^4$ from the external legs. It would be an addition to the self-coupling in the effective Lagrangian in Eq.~\eqref{eq:piTLag},
\begin{equation}\label{eq:modifiedselfcoupling}
    \lambda_{\pi T}=\lambda-\frac{3g^2}{M^2}\,.
\end{equation}
%which wouldn't alter the handling of the effective description compared to the next parts of the computation below.
However, the leading contribution from the coupling is $\lambda\times T^4$ (\textit{cf}.\ Eq.~\eqref{eq:twoloopthermcomp}). Thus, this correction is again negligible.

\subsection{Scale $\pi T$}\label{sec:scalepitI}

In the previous part, we calculated the effects of the scale $M$ by integrating it out. This yielded the effective Lagrangian for the scale $\pi T$ in Eq.~\eqref{eq:piTLag}. Here, we will use the effective Lagrangian to obtain the contributions to the free energy density by integrating out the scale $\pi T$. For notational convenience, we will drop the subscript IR from now on,
\begin{equation}\label{eq:piTLagrep}
    \mathscr{L}_{\pi T}=
    \frac{1}{2}(\partial_\mu\Phi)^2+\frac{m^2_{\pi T}}{2}\Phi^2+\frac{\lambda}{4!}\Phi^4\,.
\end{equation}

%With the scale $M$, we separated the contributions into vacuum and thermal parts. The thermal parts were exponentially suppressed by $\exp(-M/T)$, hence negligible. This worked very well due to the IR scale being the thermal scale $\pi T$. However, here the the situation is different: the thermal scale is the UV scale and the IR scale is the mass scale of the $\Phi$ field.

%\jh{Where to find the heavy DoFs?}

With the scale $M$, there was clearly a heavy field, $\Chi$, and a light field, $\Phi$, Eq.~\eqref{eq:exlagI}. Here, we only see one field, $\Phi$, in the Lagrangian, Eq.~\eqref{eq:piTLagrep}. Finding the light field is now a little more subtle. Computationally however, everything will be basically the same as above.

Let us look at the Euclidean action describing the thermal scale,
\begin{equation}
    S_{\pi T}[\Phi]=\int_0^\beta\dd \tau\int_{\bf{x}}\mathscr{L}_{\pi T}\,.
\end{equation}
The thermal effects come from the Euclidean time, $\tau$, being finite with the extent of the inverse temperature, $\beta=T^{-1}$. (See Appendix~\ref{app:imtimform}.) The bosonic field, $\Phi$, is periodic in the Euclidean time:
\begin{equation}
    \Phi(\tau=0)=\Phi(\tau=\beta)\,.
\end{equation}
(Fermionic fields would be anti-periodic.)

In thermal situation, we can find the separation into the heavy and light fields by Fourier decomposing the field:
%In general, the IR field degrees of freedom, compared to the thermal scale, can be found by decomposing the fields into their Matsubara modes:
\begin{equation}\label{eq:matsubmodes}
    \Phi(\tau,\,\mathbf{x}) = \sum_{n} \Phi_n(\mathbf{x}) \exp(i\omega_n\tau), \quad\omega_n=2\pi Tn,
\end{equation}
%which is the Fourier decomposition in the Euclidean time dimension, $\tau$ (see Appendix~\ref{app:imtimform}).
The components, $\Phi_n$, are called the Matsubara modes and the frequencies, $\omega_n$, Matsubara frequencies. The zero Matsubara mode, $\Phi_0$, will be the light degree of freedom, and the non-zero Matsubara modes, $\Phi_{n\neq0}$, will be heavy.

%\jh{Three dimensionality!}

Importantly, the Matsubara modes are three dimensional, $\Phi_n(\mathbf{x})$. Let us look at the momentum-space propagator from this point of view:
\begin{align}
    \frac{1}{P^2+m_{\pi T}^2}&=\frac{1}{\spat{p}^2+\underbrace{\omega_n^2+m_{\pi T}^2}_{\equiv M^2_n}}, \qquad P=(\omega_n, \spat{p}) \nonumber\\
    &=\frac{1}{\spat{p}^2+M^2_n}\,.
\end{align}
In three dimensions, the only momentum is the three-dimensional momentum, $\spat{p}$. Thus, the Matsubara frequencies are absorbed into the masses, $M_n^2$.

Due to the thermal scale hierarchy, $\pi T\gg m_{\pi T}$, the Matsubara modes with $n\neq0$, $\Phi_{n\neq0}$, are much heavier than the zero mode, $\Phi_0$,
\begin{equation}
    M^2_{n\neq0}=\omega_{n\neq0}^2+m_{\pi T}^2 \gg m_{\pi T}^2\,.
\end{equation}
Consequently, we have
\begin{align*}
    \text{Heavy:}&\quad\Phi_{n\neq0}\,,\\
    \text{Light:}&\quad\Phi_0\,,
\end{align*}
and for the scales
\begin{align*}
    \text{UV:}&\quad\Phi_{n\neq0}\,,\;\Phi_0^{\text{UV}}\,,\\
    \text{IR:}&\quad\Phi_0^{\text{IR}}\,.
\end{align*}
(\textit{Cf}.\ Fig.~\ref{fig:scalesI}.)

%This can be observed from the bare momentum space propagator,\jh{Explicit three D stuff!}
%\begin{equation}
%    \frac{1}{\mathbf{p}^2+\underbrace{(\omega_n)^2+m_{\pi T}^2}}\,,
%\end{equation}
%where $\omega_n^2$ can be viewed as an addition to the mass of the non-zero Matsubara modes.

%Thus, we have now identified our IR scale field to be $\Phi_0$, the zero Matsubara mode (\textit{cf.} Fig.~\ref{fig:scalesI}).

%The zero Matsubara mode, $\Phi_0(\bf{x})$, is a three-dimensional field. Thus, the effective description is three dimensional. This does not require any drastic measures. Some cosmetic changes for the description are done at the end of this part for the $\Phi_0$ effective Lagrangian.

The IR degrees of freedom, compared to the thermal scale, are in general bosonic zero Matsubara modes, e.g.\ $\Phi_0(\bf{x})$. Consequently, the three dimensionality is a generic feature of the effective descriptions, not a one-time phenomenon. In the high-temperature literature, this is called \textit{high-temperature dimensional reduction}. It was first discussed in Refs.~\cite{Ginsparg:1980ef,Appelquist:1981vg,Nadkarni:1982kb,Landsman:1989be}, and then refined in Refs.~\cite{Kajantie:1995dw,Braaten:1995cm,Farakos:1994kx,Braaten:1995jr}.

Computationally, we can apply Sec.~\ref{sec:method} straightforwardly: We expand the diagrams in the light mass, $m^2_{\pi T}$, and in light external momenta, $\spat{k}$. The external momenta are three-dimensional due to coming from $\Phi_0^{\text{IR}}(\spat{x})$.

As discussed around Eq.~\eqref{eq:freenergyfromeft}, the free-energy density obtains direct contributions from vacuum-bubble diagrams. %the computation of free energy contributions from the scale $\pi T$ can be understood as finding the unit operator for the effective Lagrangian of the scale $m_{\pi T}$. Thus we use the same recipe from Section~\ref{sec:method} to compute the relevant vacuum bubble contributions: only 1$\Phi_0$I contributions, with the light mass $m_{\pi T}$ treated perturbatively, and the renormalization scale run to the factorization scale between $m_{\pi T}$ and $\pi T$. Due to no running in the contributions computed here, the latter does not manifestly show at all. 
To the order $x^{3/2} T^4$, we need the one-loop and two-loop diagrams:
\begin{equation}
    \irqe{\feynalignscaled{0.6}{\oneloopbubble{}}}\,, \quad\irqe{\feynalignscaled{0.6}{\twoloopbubble{}}}\,.
\end{equation}
More specifically, we need the two leading terms from the one-loop IR-quantity expansion:
\begin{align}
    \irqe{\feynalignscaled{0.6}{\oneloopbubble{}}}&=\frac{1}{2}\sumint{P}\ln(P^2)+\frac{m_{\pi T}^2}{2}\sumint{P}\frac{1}{P^2}+\dots\nonumber\\
    &\equiv\massless{\feynalignscaled{0.6}{\oneloopbubble{}}}\;+\;\massless{\feynalignscaled{0.6}{\cammassinsert{}}}+\dots
\end{align}
On the last line, the crossed dot represents a mass insertion and the propagators are massless.

%The IR mass is treated with strict \jh{IRq} perturbation theory, which basically means handling the mass term in the effective Lagrangian in Eq.~\eqref{eq:piTLag} as a two-point interaction. This leads to two relevant one-loop diagrams,
%\begin{equation}
%    \feynalignscaled{0.75}{\oneloopbubble{}}\;, \quad\feynalignscaled{0.75}{\cammassinsert{}},
%\end{equation}
%where the crossed dot represents a mass insertion. Also, the leading two-loop vacuum bubble diagram, 
%\begin{equation}
%    \feynalignscaled{0.75}{\twoloopbubble{}}\,,
%\end{equation}
%contributes at $\order{xT^4}$.

%The leading free energy is given by
The diagrams can be computed to give
\begin{align}
    \massless{\feynalignscaled{0.6}{\oneloopbubble{}}}&=\frac{1}{2}\sumint{P}\ln(P^2)=-\frac{\pi^2 T^4}{90}+\order{\epsilon}\,,\label{eq:oneloopthermcomp}\\
    \massless{\feynalignscaled{0.6}{\cammassinsert{}}}&=\frac{m_{\pi T}^2}{2}\sumint{P}\frac{1}{P^2} =\frac{m_{\pi T}^2T^2}{24}+\order{\epsilon}\,,\label{eq:hidingsunset}\\
    \massless{\feynalignscaled{0.6}{\twoloopbubble{}}}&=\frac{\lambda}{8}\left(\sumint{P}\frac{1}{P^2}\right)^2=\frac{\lambda T^4}{1152}+\order{\epsilon}\,.\label{eq:twoloopthermcomp}
\end{align}
Thus, the free energy density contribution from the thermal scale is
\begin{equation}
    f_{\pi T}=-\frac{\pi^2 T^4}{90}+\frac{m_{\pi T}^2T^2}{24}+\frac{\lambda T^4}{1152}\,.
\end{equation}

Note, that the contribution from the scale $m$ is not present in $f_{\pi T}$. It is indeed removed by the IR-quantity expansion. For example in Eq.~\eqref{eq:oneloopthermcomp},
\begin{equation}
    \frac{1}{2}\sumint{P}\ln(P^2)\xrightarrow{n=0}\frac{T}{2}\int_{\spat{p}}\ln(\spat{p}^2)=0\,.
\end{equation}
These contributions are taken into account with the scale $m$ in Sec.~\ref{sec:scalemI} below.
%we did not even try to include the contributions from the zero mode. It was just projected out by the strict \jh{IRq} perturbation theory in the light mass, as discussed in Section~\ref{sec:method}. The zero mode contributions require further resummations, and they are then taken into account within the scale $m$ EFT in Eq.~\eqref{eq:lightfree}.

The only relevant $\Phi_0^{\text{IR}}$ dependent contribution from the scale $\pi T$ is the mass correction,
\begin{align}\label{eq:thermalmass}
    \irqe{\feynalignscaled{0.6}{\selfenergyfour{dotted,thick}{}}}&=\frac{\lambda\,(\Phi_0^{\text{IR}})^2}{4}\sumint{P}\frac{1}{P^2}+\order{x}\nonumber\\
    &=\frac{\lambda T^2}{48}(\Phi_0^{\text{IR}})^2+\order{\epsilon,x}\,,
\end{align}
where the dotted external lines refer to the IR field, $\Phi_0^{\text{IR}}$.

%Similarly to the mass contribution in Eq.~\eqref{eq:theonlylagtermuvI}, the term here is a leading contribution to the 1PI potential for $\Phi_0$.

%Alike with the unit terms above, the $\Phi_0$ contributions were projected out. They would be taken into account by the scale $m$ EFT in a two-loop computation.

This nearly concludes the construction of the effective action
\begin{equation}\label{eq:mLag}
    \mathscr{L}_{\rmi{eff}}=f_{\pi T}+
    \frac{1}{2}(\partial_\mu\Phi_0^{\text{IR}})^2+\frac{m^2_{\rmi{eff}}}{2}(\Phi_0^{\text{IR}})^2+\frac{\lambda}{4!}(\Phi_0^{\text{IR}})^4\,,
\end{equation}
where the effective mass is
\begin{equation}\label{eq:meffnosoexplicit}
    m^2_{\rmi{eff}}=m^2_{\pi T}+\frac{\lambda T^2}{24}\sim x T^2\,.
\end{equation}

The finalizing touch comes from the fact that the IR field, $\Phi_0^{\text{IR}}$, does not depend on the Euclidean time $\tau$
%but only on the spatial dimensions $\mathbf{x}$
(\textit{cf.} Eq.~\eqref{eq:matsubmodes}).
%It essentially lives in the three physical spatial dimensions.
One can integrate over the Euclidean time in the effective action,
\begin{equation}\label{eq:noticing3d}
    S_{\rmi{eff}}[\Phi_0^{\text{IR}}]=\int_0^\beta\dd \tau\int_{\bf{x}}\mathscr{L}_{\rmi{eff}}=\beta\int_{\bf{x}}\mathscr{L}_{\rmi{eff}}\,,
\end{equation}
yielding an awkward factor of $\beta$ in front of a three dimensional action.
%describing a three dimensional field $\Phi_0^{\text{IR}}$.
This can be handled neatly by scaling the field and the Lagrangian:
\begin{align}
    \phi&=\beta^{1/2}\,\Phi_0^{\text{IR}}\,,\\[3pt]
    \mathscr{L}_{\rmi{3d}}&=\beta\mathscr{L}_{\rmi{eff}}\,.\label{eq:lagscaling}
\end{align}
The result is
\begin{align}
    S_{\rmi{eff}}[\phi]&=\int_{\bf{x}}\mathscr{L}_{\rmi{3d}}\,,\\
    \mathscr{L}_{\rmi{3d}}&=\beta f_{\pi T}+
    \frac{1}{2}(\partial_i\phi)^2+\frac{m^2_{\rmi{eff}}}{2}\phi^2+\frac{\lambda T}{4!}\phi^4\,,\label{eq:3dlag}
\end{align}
where the kinetic term has the canonical normalization.
%Also, the term $\tfrac{1}{2}(\partial_0\phi)^2$ was dropped due to being zero.

\subsection{Scale $m$}\label{sec:scalemI}

This is the lowest scale of the system, and all that remains is to compute the one-loop contribution to the free energy using the effective Lagrangian in Eq.~\eqref{eq:3dlag}.

We have already made all the needed mass resummations while constructing the effective field theory for the scale $m$,
\begin{equation}\label{eq:explicitmeff}
    m^2_{\rmi{eff}}=m^2(\mu)-\frac{g^2}{(4\pi)^2}\left(1-\ln(\frac{M^2}{\mu^2})\right)+\frac{\lambda T^2}{24}\,.
\end{equation}
These mass corrections came from Eqs.~\eqref{eq:theonlylagtermuvI} and \eqref{eq:thermalmass}. Thus, we do not need to worry about resummations at all.

The one-loop diagram is given by
\begin{equation}\label{eq:lightfree}
    \feynalignscaled{0.6}{\oneloopbubble{dotted,thick}}=\frac{1}{2}\int_{\bf{p}}\ln({\bf{p}}^2+m_{\rmi{eff}}^2)=-\frac{m_{\rmii{eff}}^3}{12\pi}+\order{\epsilon}\,.
\end{equation}
It contributes to the unit term, which is $\beta f_{\pi T}$. Thus, it has to be multiplied by $T$ in order to obtain the free-energy-density contribution.

The full free energy density is thus
\begin{equation}\label{eq:freeenergy}
    f=-\frac{\pi^2 T^4}{90}+\frac{m_{\pi T}^2T^2}{24}+\frac{\lambda T^4}{1152}-\frac{m_{\rmii{eff}}^3T}{12\pi}+\order{x^2T^4}\,.
\end{equation}
Note, that the mass $m_{\pi T}^2$ in Eq.~\eqref{eq:effpiTmass} has only corrections from the scale $M$ whereas the mass $m_{\rmii{eff}}^2$ in Eq.~\eqref{eq:explicitmeff} has also contributions from the thermal scale.

\subsection{Discussion}

We have now demonstrated the construction of effective descriptions using the method presented in Section~\ref{sec:method} for a model with three scales in Fig.~\ref{fig:scalesI}.

To construct these effective descriptions, we only needed to compute the contributions in Eqs.~\eqref{eq:effpiTmass} and \eqref{eq:thermalmass}, and add them to the tree-level Lagrangian. Along the way, we picked up the thermal contributions, Eqs.~\eqref{eq:oneloopthermcomp}, \eqref{eq:twoloopthermcomp} and \eqref{eq:lightfree}, to the free energy in Eq.~\eqref{eq:freeenergy}.

%\jh{Drop everything else than the sunset diagram!}

Here, we will exemplify the usefulness of leveraging the scales of a system in a computation
%Then, we will first discuss resummations and at the same time a bit about different temperature ranges of the model.
%One way to illustrate, how much the use of scales eased our calculation, is to 
by looking at the sunset diagram:
\begin{equation}\label{eq:horrible}
    \feynalignscaled{0.75}{\sunsetintuit}\,.
\end{equation}
%where the solid propagators are for the full $\Phi$ field.
It contributes thermally to the free energy at our order $x^{3/2}T^4$. However, we didn't seem to need to tackle it at all.
Let us see how this diagram was naturally embedded into our computation.

We want to first note that even though the actual computation handled the sunset diagram fluently, the deconstruction of the handling may seem laborious. This is actually further proof of the usefulness of EFTs: A direct computation a single diagram, such as the sunset in Eq.~\eqref{eq:horrible}, can be very difficult because the diagram would need to be \textit{deconstructed} into calculable parts. On the other hand, the EFT computation is \textit{constructed} out of easily calculable pieces, which the EFT computation then naturally assembles into more complicated diagrams. 

On the scale $M$, Sec.~\ref{sec:scaleMI}, we didn't need to consider vacuum bubble diagrams due to their thermal corrections being exponentially suppressed. Hence, we didn't need to compute the pure scale $M$ contribution of the sunset, with the $\Phi$ propagator momenta being order $M$.

On the thermal scale, the contributions of the sunset diagram are encoded into the one-loop diagram with a mass insertion, Eq.~\eqref{eq:hidingsunset}. Diagrammatically, this can be understood in the following way:
\begin{align}
    \irqe{\feynalignscaled{0.75}{\intuitionmass}}\quad &\xrightarrow{\substack{\text{Zooming}\\ \text{out}}}\quad\,\feynalignscaled{0.75}{\selfenergytree{plain}{dot,minimum size=2mm}}\,,\label{eq:zooommmmm}\\
    \feynalignscaled{0.75}{\sunsetintuit}\quad\;\; &\xrightarrow{\substack{\text{Zooming}\\ \text{out}}}\quad\feynalignscaled{0.6}{\cammassinsertB{dot,minimum size=2.66mm}}\,.
\end{align}
Equation~\eqref{eq:zooommmmm} represents constructing the EFT for the scale $\pi T$, where we accounted for the mass contribution in Eq.~\eqref{eq:theonlylagtermuvI}. The arrow represents zooming out from the scale $M$ to the scale $\pi T$, which makes the contributing diagram local. (See Sec.~\ref{sec:diagunderstanding}.) Thus, the dot on the right-hand side is a contribution to the effective mass, $m_{\pi T}^2$. On the next line, the same localization happens for either of the loops due to scale $M$ momentum. The crux is that the bottom-right, one-loop diagram is within the aforementioned one-loop diagram, Eq.~\eqref{eq:hidingsunset}. This happens due to the fact that the mass insertion, $m_{\pi T}^2$, contains the mass correction from the top-left diagram.
%in Eq.~\eqref{eq:theonlylagtermuvI}.

The diagram is also involved in the thermal corrections from the scale $m$ in Sec.~\ref{sec:scalemI},
\begin{equation}
    \feynalignscaled{0.75}{\sunsetintuitB}\quad\longrightarrow\quad\feynalignscaled{0.6}{\cammassinsertC{dot,minimum size=2.66mm}{dotted,thick}}\,,
\end{equation}
where we have a dotted, scale-$m$ propagator.
%is only a part of the full $\Phi$ propagator of Eq.~\eqref{eq:horrible}.
The mass correction on the right needed to be resummed into the mass, $m_{\rmii{eff}}$. This has been done in the EFT construction, Eq.~\eqref{eq:explicitmeff}. Consequently, the contribution is taken into account by the scale-$m$ one-loop diagram in Eq.~\eqref{eq:lightfree}.

We can thus see that the sunset diagram, Eq.~\eqref{eq:horrible}, takes part into the second and fourth term in the final free energy in Eq.~\eqref{eq:freeenergy}.

We can even play the game further. At the order $x^2T^4$, there is another relevant contribution from the thermal scale of the diagram in Eq.~\eqref{eq:horrible}:
\begin{equation}
    \feynalignscaled{0.75}{\sunsetintuit}\quad\longrightarrow\quad\feynalignscaled{0.6}{\twoloopbubbleB{dot,minimum size=2.66mm}}\,.
\end{equation}
Here, only the heavy $\Chi$ propagator is on the scale $M$ and the $\Phi$ propagators are on the thermal scale. This contribution would be taken into account by including the four-point correction in Eq.~\eqref{eq:fourpointeff} and computing the thermal-scale two-loop contribution, Eq.~\eqref{eq:twoloopthermcomp}, with the modified self-coupling of Eq.~\eqref{eq:modifiedselfcoupling}.

The above discussion gives understanding how the EFT construction splits a horribly complicated diagram into smaller pieces that are easily calculable. As mentioned above, the power of an EFT computation is that one doesn't need to start from a complicated diagram, which needs deconstruction. One can start from the simple pieces to build up the calculation. This is demonstrated by the initial computation of the free energy, which didn't need to explicitly compute the sunset diagram of Eq.~\eqref{eq:horrible}, nor do any of the above dissection of the diagram.

\section{Example 2:\\first-order symmetry-breaking\\phase transition}\label{sec:exampleII}

%\jh{State the goals precisely! Computational ones and then the physical one.}

There are two main goals for this example. Computationally, we want to show, how to handle derivative contributions to the effective action and resummations of an external field. Physically, we want to show an application of the EFT construction method onto a phase transition.

The discussion of the aforementioned computational methods takes place in Sec.~\ref{sec:scaleMII}. The physical discussion is a bit more broad: In Sec.~\ref{sec:scaleMII}, we will observe a dynamically produced mass hierarchy at the phase transition. In Sec.~\ref{sec:scalemII}, we compute the ingredients to a gravitational-wave spectrum as far as it is possible with pen and paper, which means the duration and the strength of the transition~\cite{Hindmarsh:2017gnf}.

%The main goal with this example is to show the appearance of the derivative expansion in the EFT construction and also show how to handle resummations in external fields. The example model below was used in Ref.~\cite{Gould:2021ccf} to study nucleation rates in first-order phase transitions. Here, we will follow the power countings of the reference but also compute free energy difference between the phases. \jh{Gravitational wave spectrum!}

%\jh{Why do we cite the thermal scale? An example has already been given above, done correctly in the Refs. and we want to get going!}

The same example has been used in Ref.~\cite{Gould:2021ccf} to illustrate nucleation rates. We will use the same power countings, initially motivated by Ref.~\cite{Arnold:1992rz}. Our results do not disagree with Ref.~\cite{Gould:2021ccf}. Thus, there is no good reason to explicitly reproduce the full computation. Also, citing results will enable us to highlight the above motivations more.

%The computations have been done in Appendix A of Ref.~\cite{Hirvonen:2020jud} to excruciating detail. Here, we will be using simpler computational methods to streamline the calculations and make them practical.

The example model is given by
\begin{align}\label{eq:exLagII}
    \mathscr{L}_{\rmi{3d}}=&\beta f_{\pi T}+\frac{1}{2}(\partial_i\phi)^2+\frac{m_3^2+\cou{m_3^2}}{2}\phi^2 +\frac{\lambda_3}{4!}\phi^4\nonumber\\
    +&\frac{1}{2}(\partial_i\chi)^2+\frac{M_3^2}{2}\chi^2+\frac{f_3}{4!}\chi^4+\frac{g_3^2}{4}\phi^2\chi^2\,.
\end{align}
In this model, the $\phi$ field will undergo a symmetry-breaking phase transition. The $\chi$ field will also play an important role: It renders the phase transition to be first order.

Note, that the theory is already three dimensional. This comes from the fact that the thermal scale has already been integrated out (\textit{cf.} Sec.~\ref{sec:scalepitI} and the Lagrangian in Eq.~\eqref{eq:3dlag}).
%The unit term is the free energy contribution from the thermal scale similar to the one in Eq.~\eqref{eq:3dlag}.
Due to the fact that the thermal scale has been integrated out, the Lagrangian coefficients depend on the temperature (\textit{cf}.\ Eqs.~\eqref{eq:meffnosoexplicit}, \eqref{eq:3dlag}).

\begin{figure}
    \centering
    \includegraphics[width=0.8\columnwidth]{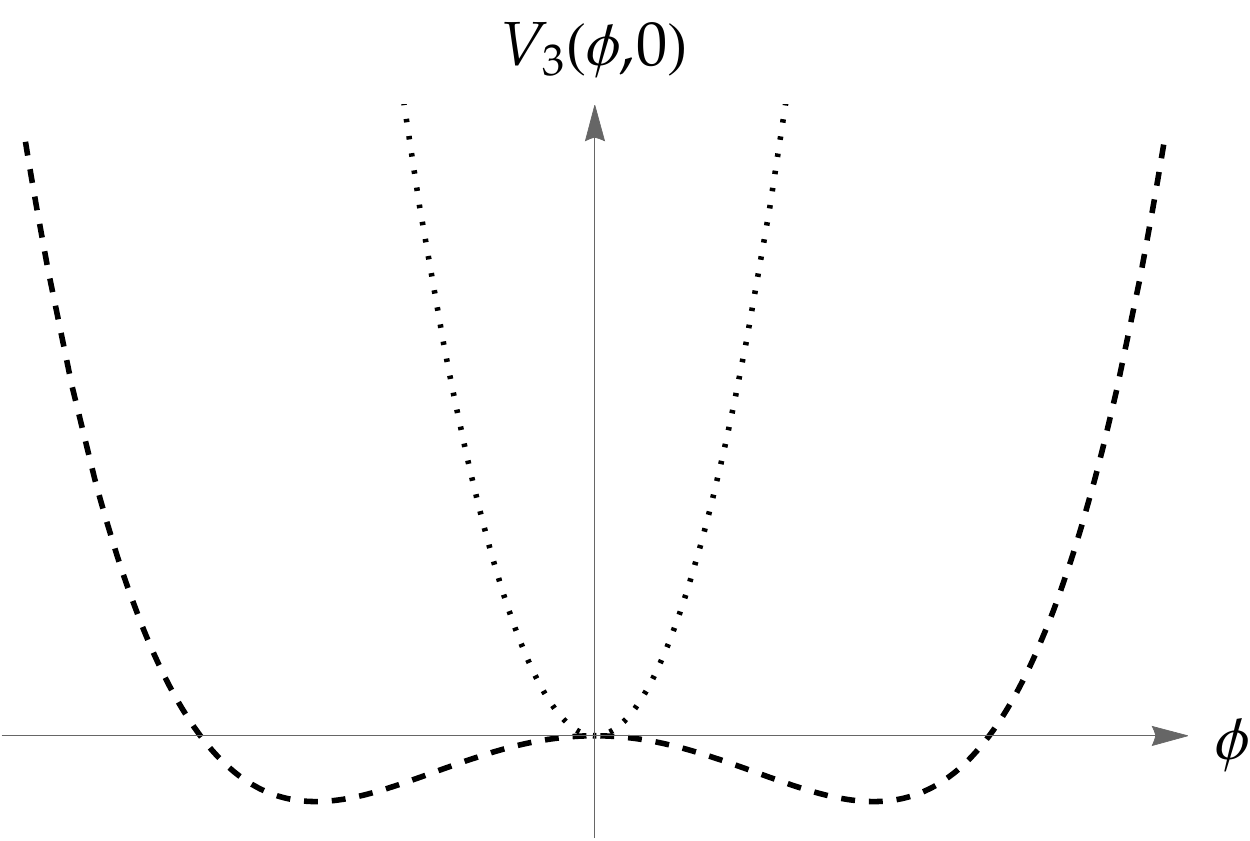}
    \caption{Here, we have plotted the potential of the Lagrangian in Eq.~\eqref{eq:exLagII} at different temperatures. The potentials plotted are on different sides of the phase transition, as the mass parameter, $m_3^2$, has changed sign. Note, that radiative contributions from the $\chi$ field, which are not present in the plotted potential, play an important role in the phase transition. Thus, the nature of the transition cannot be inferred from the plot.}
    \label{fig:masscrosszero}
\end{figure}

Most importantly, the mass of the $\phi$ field, $m_3^2$, depends strongly on temperature: The $\phi$ field undergoes a symmetry-breaking phase transition. Correspondingly, the mass parameter goes through zero near the transition (Fig.~\ref{fig:masscrosszero}).

%\jh{First state the phase transition!}

%\jh{The thermal mass goes through zero!}

We now lay down a proper power counting in accordance with Ref.~\cite{Gould:2021ccf}.
%Then we will highlight the parts that will be crucial to our computation, and discuss the first-order transition.
%\jh{Power countings copied!}
For the dimensionless power counting parameter, we define
\begin{equation}\label{eq:exIIpowparam}
    x\equiv\frac{g_3^2}{M_3}\ll 1\,.
\end{equation}
In relation to $M^2_3$ and $g_3^2$,
\begin{equation}\label{eq:exIIotherpows}
    m_3^2\sim xM_3^2\,,\qquad\lambda_3\sim x g_3^2\,,\qquad f_3\sim g_3^2\,.
\end{equation}

%In this model, the $\chi$ field induces the first-order phase transition for the $\phi$ field radiatively through the relatively strong cross-coupling,
The cross-coupling, $g_3^2$, is chosen to be relatively strong to the $\phi$ self coupling, $\lambda_3$: 
\begin{equation}
    g_3^2\gg \lambda_3\,.
\end{equation}
This enables the $\chi$ field to induce the first-order nature of the phase transition radiatively.

Due to being close to a symmetry breaking phase transition, the $\phi$ field is naturally lighter than the inducing $\chi$ field,
\begin{equation}
    m_3^2\ll M_3^2\,.
\end{equation}
%as we will discuss below. This leads to a scale hierarchy that allows us to construct an EFT for the nucleating $\phi$ field.
Around Eq.~\eqref{eq:dynmasshierarchyyy} below, we will find, that the specific mass power counting of Eq.~\eqref{eq:exIIotherpows} arises naturally in the phase transition.

Due to the mass hierarchy, we can see that
\begin{align*}
    \text{Heavy:}&\quad\chi\,,\\
    \text{Light:}&\quad\phi\,.
\end{align*}
Thus, the scales are given by
\begin{align*}
    \text{UV:}&\quad\chi\,,\;\phi_{\text{UV}}\,,\\
    \text{IR:}&\quad\phi_{\text{IR}}\,.
\end{align*}

%\jh{A bit of discussion of nucleation!}

Due to the symmetry-breaking phase transition, the field value of the $\phi_{\text{IR}}$ field changes from $\phi_{\text{IR}}=0$ to $\phi_{\text{IR}}\neq0$. We will power count it as
%Another key point is that the background field of the nucleating $\phi$ field is
\begin{equation}\label{eq:phisizeFirst}
    \phi_{\text{IR}}^2\sim\frac{M_3^2}{g_3^2}= x^{-1}M_3\,.
\end{equation}
This power counting actually arises naturally at the phase transition (shown around Eq.~\eqref{eq:phisize} below).
%This can only be shown after finding the effective description for the $\phi$ field (\jh{Eq. what}).
An important consequence of this power counting is that we need to resum the effects of the background field, $\phi_{\text{IR}}$, to the mass of the $\chi$ field.
%\jh{Put later:}The $\phi$ background gives a leading order contribution to the $\chi$ mass through the interaction in the Lagrangian. Thus, the mass contribution has to be taken into account along side with the renormalized mass $M^2$.

Finally, we note that the power counting of the derivatives is important as well. We want to compute the nucleation rate. There, the background field will be a spatially-dependent critical bubble. Consequently, the derivatives on the tree-level Lagrangian are already power counted according to
\begin{align}
    &\partial_i\phi_{\text{CB}}\sim m_3\,\phi_{\text{CB}}\\[3pt]
    \Rightarrow\;&\partial_i^2\sim m_3^2\sim xM_3^2\,,\label{eq:nuclscalederiv}
\end{align}
where $\phi_{\text{CB}}$ is the critical bubble. This power counting for the derivative follows from the fact that the nucleation physically takes place on the energy scale given by $m_3$. Consequently, the size scale of the nucleating bubbles is $L_{\text{bubble}}\sim m_3^{-1}$~\cite{Gould:2021ccf}.

%the derivative can be power counted on the tree-level of the effective Lagrangian as the effective mass of the nucleating field due to the bubble background~\cite{Gould:2021ccf}\jh{maybe other refs. as well?}. If the order of the nucleating field mass is given by the mass parameter $m^2$, we have
%\begin{equation}\label{eq:derivativepower}
%    \partial_i^2\sim m^2\sim xM^2\,.
%\end{equation}
%This will lead to the leading term of the one-loop derivative expansion being important.

In terms of the free-energy density, the aim is to obtain the accuracy of
\begin{equation}
    \beta f \sim m_3^3\sim x^{3/2}M_3^3\,.
\end{equation}
%so that the one-loop of the $\phi$ field can be accounted for. \jh{Inverse duration to large terms.}
Then, the scale $m_3$ contributes as well. Even more importantly for the gravitational wave spectrum, we will notice, that we obtain all of the large terms, $\order{x^{-n}}$, to the nucleation rate.

\subsection{Scale M}\label{sec:scaleMII}

%\jh{Move the more thorough discussion to the appendix.}

%\jh{Shorten the shit out of this!}

%\jh{First show the needed diagrams!}

%\jh{Pure $\phi$ diagrams zero under the IR-quantity expansion.}

Now, we get to computing the effective action for $\phi_{\text{IR}}$ with the method of Sec.~\ref{sec:method}.

%Now, we are ready to compute the effective action for the $\phi$ field by using the recipe from Section~\ref{sec:method}. First, we compute the one-loop potential terms, where we can see the resummations in external fields. Then, we compute the first derivative correction. Finally, we list the relevant two-loop results. 

%\jh{Leading part in external momentum:}

A very important thing to note immediately is that the background field, $\phi_{\text{IR}}$, affects the mass of the heavy field at the leading order:
\begin{equation}\label{eq:backgroundfieldmass}
    M_3^2(\phi_{\text{IR}})\equiv M_3^2+\frac{g_3^2}{2}\phi_{\text{IR}}^2\,.
\end{equation}
The mass $M_3^2(\phi_{\text{IR}})$ can be found from the Lagrangian in Eq.~\eqref{eq:exLagII} by splitting the light field into the IR and UV modes:
\begin{equation}
    \phi=\phi_{\text{IR}}+\phi_{\text{UV}}\,.
\end{equation}
It is now important to keep the mass $M_3^2(\phi_{\text{IR}})$ as a full package. Otherwise, we would use a mass, $M_3^2$, that is not correct to the leading order.%
\footnote{There is another, much more laborious way, in which one only has $M_3^2$ as the mass. Then, one has to explicitly resum infinite classes of diagrams. See Ref.~\cite{Coleman:1973jx}.}

The diagrams that we need to compute are
\begin{align}\label{eq:exIIdiags}
    \irqe{\feynalignscaled{0.6}{\oneloopbubble{dashed}}}\,,\quad\irqe{\feynalignscaled{0.6}{\twoloopbubble{dashed}}}\,,\quad\irqe{\feynalignscaled{0.6}{\sunset{}{dashed}}}\,,
\end{align}
where the dashed lines refer to the $\chi$ field and the solid lines to the $\phi$ field. The two explicit $\phi_{\text{IR}}$ legs in the last diagram come from the interaction term
\begin{equation}
    \frac{g_3^2}{2}\,\phi_{\text{IR}}\,\phi_{\text{UV}}\chi^2
\end{equation}
in the Lagrangian, Eq.~\eqref{eq:exLagII}.

From each diagram, we will need the leading term in the IR-quantity expansion. From the one-loop diagram, we will also need the first term of the external momentum, $\spat{k}$, expansion. This will become a derivative correction to the effective action. We start from the potential terms of $(\spat{k}^2)^0$. After that, we will move on to the derivative correction of $(\spat{k}^2)^1$.

The one-loop potential term from the $\chi$ field is given by
\begin{align}\label{eq:fullchione}
    \kOrd{\feynalignscaled{0.6}{\oneloopbubble{dashed}}}{0}&=\frac{1}{2}\int_{\bf{p}}\ln({\bf{p}}^2+M_3^2(\phi_{\text{IR}}))\nonumber\\
    &=-\frac{1}{3(4\pi)}M_3^3(\phi_{\text{IR}})+\order{\epsilon}\,.
\end{align}
Here, we have now successfully resummed in the external $\phi_{\text{IR}}$ field the one-loop potential term.
%This is necessary, because we cannot expand the resulting radical in either of the terms due to the scalings in Eq.~\eqref{eq:phisize}.
We have computed a three-dimensional analog to the Coleman--Weinberg potential~\cite{Coleman:1973jx}.
%where one needs to resum the external legs of the one-loop diagram

The two other potential terms are
\begin{align}
    \kOrd{\feynalignscaled{0.6}{\twoloopbubble{dashed}}}{0} &=\frac{g_3^2}{8}\left(\int_{\bf p}\frac{1}{{\bf p}^2+M_3^2(\phi_{\text{IR}})}\right)^2\nonumber\\
    &=\frac{f_3}{8(4\pi)^2}M_3^2(\phi_{\text{IR}})+\order{\epsilon}\,,\label{eq:twoloopsnowII}
\end{align}
and
\begin{align}
&\kmOrd{\feynalignscaled{0.6}{\sunset{}{dashed}}}\nonumber\\
=&-\frac{g_3^4\phi_{\text{IR}}^2}{4}\int_{\bf p}\int_{\bf q}\frac{1}{({\bf p}^2+M_3^2(\phi_{\text{IR}}))({\bf q}^2+M_3^2(\phi_{\text{IR}}))({\bf p}+{\bf q})^2}\nonumber\\
=&-\frac{g_3^4\phi_{\text{IR}}^2}{8(4\pi)^2}\left(\frac{1}{2\epsilon}+\ln(\frac{\mu_3^2}{4M_3^2(\phi_{\text{IR}})})+1\right)+\order{\epsilon}\,,\label{eq:twoloopIIsun}
\end{align}
where $\mu_3$ is the $\MSbar$ renormalization scale of the three-dimensional EFT.

Similarly to the discussion in Example 1 below the Lagrangian in Eq.~\eqref{eq:piTLag}, the divergence in the sunset diagram cancels with the counter term in the Lagrangian in Eq.~\eqref{eq:exLagII}, and the running cancels the running in the mass parameter, $m_3^2(\mu_3)$~\cite{Gould:2021ccf,Hirvonen:2020jud}.

There will be no counterterms in the effective Lagrangian. Hence, we do not need any of the $\order{\epsilon}$ terms.

%\jh{Drop this diagram altogether!}

%The 1PI one-loop potential term from the $\phi$ field requires an expansion in the light mass, $m^2$. It is given by
%\begin{align}
%    \feynalignscaled{0.6}{\oneloopbubble{}}&=\frac{1}{2}\int_{\bf{p}}\left(\ln({\bf{p}}^2)+\frac{m^2}{{\bf{p}}^2}+\dots\right)=0\,,
%\end{align}
%where the solid line is a $\phi$ propagator.
%It vanishes in dimensional regularization due to being a scale free integral.
%Same applies to the following terms in the mass expansion, e.g.\ the next one:
%The $\phi$ one-loop contribution is projected out in the EFT construction and then taken consistently into account within the EFT. \jh{Derivative term discussion to the derivative term section!}
%This holds for the derivative contributions from the $\phi$ one-loop diagram as well.

%\jh{Post-pone this to the end of the section!}

Now, we will explore how the derivative series arises from the one-loop diagram of the $\chi$ field.
%It follows the same recipe as before: 1PI action contributions for the $\phi$ field with the two steps implemented.
The most straightforward way, presented in Appendix A of Ref.~\cite{Hirvonen:2020jud}, is to compute directly the action contribution in the presence of the general IR background, $\phi_{\text{IR}}$. It can easily become very cumbersome. Hence, we present two other methods in this article.

%Obtaining also the derivative corrections is somewhat more complicated than just the potential terms. There are few methods for obtaining the derivative corrections to the 1PI action. The most straightforward (and the most gruesome) one was used in Appendix A of Ref.~\cite{Hirvonen:2020jud}. Here, we will show two more streamlined methods. \jh{Only one here and the other in the appendix!}

Both of the more streamlined methods are based on a simplification. We split the IR background into a homogeneous background and IR fluctuations around the background:
\begin{equation}\label{eq:exIIflatNfluct}
    \phi_{\text{IR}}=\varphi+h_{\text{IR}}\,.
\end{equation}
The full Lagrangian term, that we are after, can now be expanded in the IR fluctuations:
\begin{align}\label{eq:backgroundhomogNfluctsZ}
    &\frac{Z_{\text{eff}}(\phi_{\text{IR}})}{2}(\partial_i\phi_{\text{IR}})^2\nonumber\\
    =\;&\frac{Z_{\text{eff}}(\varphi)}{2}(\partial_i h_{\text{IR}})^2+\frac{Z_{\text{eff}}'(\varphi)}{2}h_{\text{IR}}\,(\partial_i h_{\text{IR}})^2+\dots
\end{align}
The achieved simplification is that we can compute the two-point function of the fluctuations, $h_{\text{IR}}$, on a homogeneous background, $\varphi$. This way, we can obtain the first term on the right-hand side, and consequently the field normalization, $Z_{\text{eff}}(\varphi)$.%
\footnote{The method used by Ref.~\cite{Hirvonen:2020jud} computes $Z_{\text{eff}}(\phi_{\text{IR}})(\partial_i\phi_{\text{IR}})^2/2$ directly. Hence, it needs all $n$-point functions for $\phi_{\text{IR}}$ on the general background of $\phi_{\text{IR}}$.}

Here, we will only show one method. The other one can be found from Appendix~\ref{app:anoutherderiv}. The method shown here computes a contribution to the self-energy for the fluctuations, $h_{\text{IR}}$:
\begin{equation}\label{eq:selfenergydiscussion}
    Z_{\text{eff}}(\varphi)\spat{k}^2+m_{\text{eff}}^2\,.
\end{equation}
This method is more widely in use (see e.g.\ Refs.~\cite{WeinbergVacDec,Garny:2012cg,Hirvonen:2021zej}) as it is computationally simpler.
%on the background of $\varphi$.
The other method computes directly the action contribution (see e.g.\ Ref.~\cite{Laine:2016hma}, Sec.~6). Its benefit is that it shows the origins for the locality of the EFT much more clearly. Another benefit is that it generalizes readily to higher orders in derivatives.

%With the above result in mind, we can start computing the leading derivative term. First, we use the more practical method, in which one computes a self-energy contribution, and then a more clumsy, but possibly more transparent, method of computing an action term.

There are two distinct self-energy contributions from the one-loop level in the $\chi$ field:
One is
\begin{equation}\label{eq:chimass1}
    \irqe{\feynalignscaled{0.6}{\selfenergyfour{}{dashed}}}=\frac{g_3^2}{2}\int_{\bf p}\frac{1}{{\bf p}^2+M_3^2(\varphi)}\,,
\end{equation}
but it only leads to a mass contribution (already taken into account by the potential term in Eq.~\eqref{eq:fullchione}). The other one is
\begin{align}\label{eq:kselfenergy}
    \irqe{\feynalignscaled{0.75}{\intuitionselfenergy}}\,.
\end{align}
It depends on the external momentum, ${\bf k}$, leading to a derivative series. The cubic interactions are from the Lagrangian term
\begin{equation}
    \frac{g_3^2\varphi}{2}\,h_{\text{IR}}\chi^2
\end{equation}
from the Lagrangian in Eq.~\eqref{eq:exLagII}. The external legs of the diagram correspond to $h_{\text{IR}}$, and the constant background, $\varphi$, is absorbed into the coefficient.
%in the latter diagram are caused by the constant background $\varphi$ from the $\phi^2\chi^2$ term of the Lagrangian in Eq.~\eqref{eq:exLagII}. We have also defined a shorthand notation for the background dependent mass:
%\begin{equation}\label{eq:fullchimass}
%    M_3^2(\varphi)=M^2+\frac{g_3^2}{2}\varphi^2\,.
%\end{equation}

Without any expansions, the diagram is given by
\begin{align}\label{eq:kselfenergyreprep}
    &\feynalignscaled{0.75}{\intuitionselfenergy}\nonumber\\
    =&-\frac{g_3^4\varphi^2}{2}\int_{\bf p}\frac{1}{(({\bf p}+{\bf k})^2+M_3^2(\varphi))({\bf p}^2+M_3^2(\varphi))}\,.
\end{align}
We only need the first order in the expansion in $\spat{k}^2$. Thus, we compute:
\begin{align}
    &\kOrd{\feynalignscaled{0.75}{\intuitionselfenergy}}{1}=-\frac{g_3^4\varphi^2}{2}{\bf k}^2\times\nonumber\\
    &\times\int_{\bf p}\left(-\frac{1}{({\bf p}^2+M_3^2(\varphi))^3}+\frac{4}{d}\frac{{\bf p}^2}{({\bf p}^2+M_3^2(\varphi))^4}\right)\nonumber\\[5pt]
    &=\frac{g_3^4\varphi^2}{48(4\pi)M_3^3(\varphi)}{\bf k}^2+\order{\epsilon}\,,
\end{align}
where $d$ is the number of spatial dimensions, and we have solved the $\spat{p}$ integral for the last line.

The obtained contribution is a direct contribution to the self energy in Eq.~\eqref{eq:selfenergydiscussion}. Thus, we can read off the contribution to the field normalization:
\begin{equation}\label{eq:resforDZ}
    \Delta Z_{\text{eff}}(\phi_{\text{IR}})=\frac{g_3^4\varphi^2}{48(4\pi)M_3^3(\phi_{\text{IR}})}\,.
\end{equation}

%Slightly anti-climatically, we will look at the zeroth order term to show that it is already included to the one-loop result in Eq.~\eqref{eq:fullchione}. The zeroth order term is given by

Now, we can put together the effective Lagrangian to the order $x^{3/2}M_3^3$,
\begin{align}
	\mathscr{L}_{\rmi{eff}}=&\beta f_{\pi T}+\frac{Z_{\text{eff}}(\phi_{\text{IR}})}{2}\qty(\partial_i\phi_{\text{IR}})^2 + V_{\rmi{eff}}(\phi_{\text{IR}})\,, \label{eq:exampleTwoNuclAction}\\
	V_{\rmi{eff}}(\phi_{\text{IR}})&=\frac{\tilde{m}^2(\phi_{\text{IR}})}{2}\phi_{\text{IR}}^2 -\frac{M_3^3(\phi_{\text{IR}})}{3(4\pi)}  +\frac{\lambda_3}{4!}\phi_{\text{IR}}^4\,,\label{eq:defVeff}
\end{align}
where the parameters are
\begin{align}
	Z_{\text{eff}}(\phi_{\text{IR}})=&1+\frac{1}{48(4\pi)}\frac{g_3^4\phi_{\text{IR}}^2}{M_3^3(\phi_{\text{IR}})} \,,\label{eq:WaveFuncRenorm}\\
	\tilde{m}^2(\phi_{\text{IR}})=&m_3^2(\mu)+\frac{f_3g_3^2}{8(4\pi)^2}\nonumber\\
	&-\frac{g_3^4}{4(4\pi)^2}\qty[1+\ln(\frac{\mu^2}{4M_3^2(\phi_{\text{IR}})})] \,, \label{eq:exampleTwoNuclMass}
\end{align}
with $M_3^2(\phi_{\text{IR}})$ from Eq.~\eqref{eq:backgroundfieldmass}.

%\jh{Later with the real power counting we can discuss this.}
Finally, we can ground the power counting for the background field, Eq.~\eqref{eq:phisizeFirst}, and the mass $m_3^2$, Eq.~\eqref{eq:exIIotherpows}. Importantly, we will see below that the mass hierarchy,
\begin{equation}
    m_3^2\ll M_3^2\,,
\end{equation}
is a result of the symmetry breaking phase transition.
%As a side note regarding the first-order phase transition of the model, 
%The one-loop $\chi$ term is responsible for another minimum forming to the effective potential of $\phi$.

\begin{figure}
    \centering
    \includegraphics[width=0.8\columnwidth]{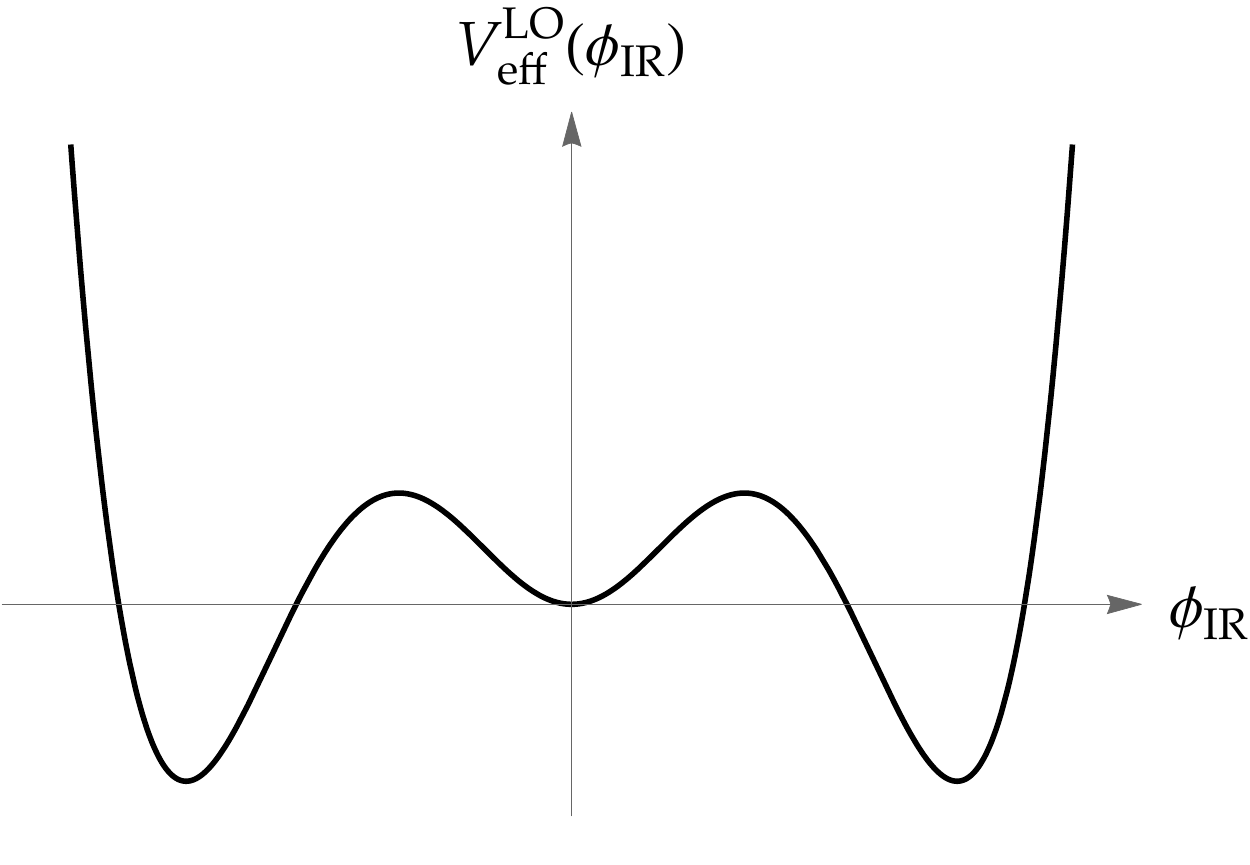}
    \caption{Here, we have plotted the leading order potential in Eq.~\eqref{eq:LOpot} slightly below the critical temperature. The symmetric phase is still metastable, but the $\phi_{\text{IR}}\neq0$ minima are favoured.}
    \label{fig:nuclpot}
\end{figure}

Let us look at the leading order potential,
\begin{equation}\label{eq:LOpot}
    V_{\rmi{eff}}^{\rmii{LO}}(\phi_{\text{IR}})=\frac{m_3^2}{2}\phi_{\text{IR}}^2-\frac{1}{3(4\pi)}\left(M_3^2+\frac{g_3^2}{2}\phi_{\text{IR}}^2\right)^{3/2}+\frac{\lambda_3}{4!}\phi_{\text{IR}}^4\,,
\end{equation}
shown in Fig.~\ref{fig:nuclpot}.

Note that the middle term from to the $\chi$-one-loop diagram may induce a $\phi_{\text{IR}}\neq0$ minimum, even if the $\phi_{\text{IR}}$ mass is positive at $\phi_{\text{IR}}=0$. This corresponds to a first-order symmetry-breaking phase transition.

With the power counting established in the beginning of this section in Eqs.~\eqref{eq:exIIpowparam}, \eqref{eq:exIIotherpows}, we can see that all of the potential terms in Eq.~\eqref{eq:LOpot} have the same order $\phi_{\text{IR}}$ derivatives at
\begin{equation}\label{eq:phisize}
    \phi_{\text{IR}}^2\sim x^{-1}M_3\,.
\end{equation}
This shows that the minimum induced by the $\chi$ field is there, which leads to the power counting for $\phi_{\text{IR}}$ in Eq.~\eqref{eq:phisizeFirst}.

Let us then turn to the mass hierarchy.

Note, that if the mass $m_3^2$ was $m_3^2\gg xM^2_3$, there could not be another minimum. The derivative of the potential near $\phi_{\text{IR}}=0$ would be dominated by the mass term, and away from $\phi_{\text{IR}}=0$, it would be dominated by the quartic term.

The mass has to be
\begin{equation}
    m_3^2\lesssim xM_3^2
\end{equation}
for there to be another minimum. Thus, the $\phi_{\text{IR}}$ mass has to obey the above relation at the phase transition. This leads to the mass hierarchy.

The upper limit of the above relation,
\begin{equation}\label{eq:dynmasshierarchyyy}
    m_3^2\sim xM_3^2\,,
\end{equation}
holds near the critical temperature, below which the $\phi_{\text{IR}}\neq0$ minimum becomes favoured.

%The scale hierarchy between the masses is produced by the occurrence of the phase transition. At other temperatures, the scale hierarchy does not hold.
%\jh{Highlight the dynamical nature!}
%\jh{Dymanical mass hierarchy!}

\subsection{Scale $m$}\label{sec:scalemII}

Here, we will apply the obtained EFT to study the first-order phase transition present in the model. We can use it to compute both the strength of the transition, $\alpha$, and the inverse duration of the transition, $\beta$ (which is not to be confused with the inverse temperature).
%In the scalar model, there are no subtleties related to gauge invariance. We will however make remarks at the end on retaining it when $\chi$ is replaced by a gauge field.

%\jh{Mod!}
%Treating nucleation rates via EFTs
%The nucleation rate computation via EFTs was
%has been shown to be consistent in Ref.~\cite{Gould:2021ccf}.
The EFT approach to bubble nucleation was put forward in Ref.~\cite{Gould:2021ccf}, and was shown to resolve a number of inconsistencies in non-EFT approaches.
%A consistent way to handle nucleation rates with EFTs has been shown in Ref.~\cite{Gould:2021ccf}.
An explicit computation showing, how to retain the gauge invariance in an EFT computation of a nucleation rate, has been done in Ref.~\cite{Hirvonen:2021zej}. The question of gauge invariance would be relevant if the $\chi$ field was replaced by a gauge field.
%in this model was discussed in Ref.~\cite{Gould:2021ccf}. We will not be able to go deep into the nuances of nucleation as was done in the reference, nor discuss the solved gauge independence of the nucleation rate~\cite{Hirvonen:2021zej,Lofgren:2021xxx}, which would be relevant is $\chi$ was replaced by a gauge field. % omputation contains certain nuances for which we refer the reader to Ref.~\cite{Gould:2021ccf}.\jh{Mixed with gauge shit}

The strength of a transition is given by~\cite{Hindmarsh:2017gnf}
\begin{equation}
    \alpha=\frac{4\Delta\theta}{3w_+}\,.
\end{equation}
Here, the enthalpy, $w$, is evaluated in the metastable phase marked with $+$ and $\Delta\theta$ is the difference of the energy-momentum trace between the stable and metastable phase,
\begin{equation}
    \theta=(e-3p)/4\,.
\end{equation}
with $e$ as the energy density and $p$ as the pressure.

All of the thermodynamic quantities above can be found with the free energy density. The strength of a transition in terms of the free energy density, $f$, and temperature, $T$, is
\begin{equation}
    \alpha=\frac{T\partial_T\Delta f-4\Delta f}{3T\partial_T f_+}\,.
\end{equation}
Thus, we only need to find the free energy density of our system.

%For the strength of the transition, we need to obtain the free energy of the system. Its computation takes place on a flat background, $\varphi$. Thus, the derivative correction does not affect our computation at order $x^{3/2}M_3^3$ in contrast to nucleation. The background can either be the $\mathbb{Z}_2$-symmetric phase, $\phi=0$, or then the broken phase created radiatively by the corrections from the $\chi$ field. Here, we will keep it general to apply for both minima.

Let us start from $\Delta f$ by computing the scale $m_3$ contributions to the free-energy density.
We begin by expanding the effective Lagrangian to second order in fluctuations, $h_{\text{IR}}$, around a background, $\varphi$:
%, to obtain the one-loop contribution from the scale $m$:
\begin{align}
	\mathscr{L}_{\rmi{eff}}=&V_{\rmi{eff}}(\varphi) + \frac{1}{2}\qty(\partial_i h_{\text{IR}})^2+\frac{1}{2}V_{\rmi{eff}}''(\varphi)h_{\text{IR}}^2\,.
\end{align}
The linear term vanishes since we assume the background to be a minimum of the potential. The minima correspond to the different phases.
% due to being at the minimum.

The one-loop diagram is given by
\begin{align}
    \feynalignscaled{0.6}{\oneloopbubble{}}&=\frac{1}{2}\int_{\bf p}\ln({\bf p}^2+V_{\rmi{eff}}''(\varphi))\nonumber\\
    &=-\frac{V''(\varphi)^{3/2}}{12\pi}+\order{\epsilon}\,.
\end{align}
Thus, the free energy difference is given by
\begin{align}\label{eq:freeendiff}
    \Delta f=T\Big(&V_{\rmi{eff}}(\varphi_-)-V_{\rmi{eff}}(0)\nonumber\\
    &\left.-\frac{V_{\rmi{eff}}''(\varphi_-)^{3/2}}{12\pi}+\frac{V_{\rmi{eff}}''(0)^{3/2}}{12\pi}\right)\times\left(1+\order{x^2}\right)\,,
\end{align}
%where $V_{\rmi{eff}}$ is defined in Eq.~\eqref{eq:defVeff},
$\varphi_-\neq0$ is the field value at the broken minimum of the potential, $V_{\text{eff}}$. The factor of temperature appears from the scaling in Eq.~\eqref{eq:lagscaling}. Note importantly, that the thermal scale free-energy contributions, $f_{\pi T}$, cancel.

%\jh{trace anomaly GWs}

The $f_+$ part of the strength of the transition, $\alpha$, on the other hand is dominated by the thermal scale contributions (\textit{cf.} the free energy of Example 1 in Eq.~\eqref{eq:freeenergy}):
\begin{equation}
    f_+=f_{\pi T}+\order{T M_3^3}\sim T^4\,.
\end{equation}
The parameter $f_{\pi T}$, found from the initial effective Lagrangian in Eq.~\eqref{eq:exLagII}, cannot be computed within the effective description.

Here, we will cut some corners. We simply state that the theory describing the thermal scale only contains the four-dimensional counterparts to $\phi$ and $\chi$: $\Phi$, $\Chi$. (The description is the same as in Refs.~\cite{Gould:2021ccf,Hirvonen:2020jud}.) The parameter $f_{\pi T}$ can be computed to be
\begin{align}
    f_{\pi T}&=\thermalscale{\feynalignscaled{0.6}{\oneloopbubble{}}}+\thermalscale{\feynalignscaled{0.6}{\oneloopbubble{dashed}}}+\order{x^2T^4}\nonumber\\
    &=-\frac{\pi^2 T^4}{45}+\order{\epsilon,x^2T^4}\,.
\end{align}

%Thus, its computation requires knowledge on the parent theory of our example theory in Eq.~\eqref{eq:exLagII}. It is possible that there are fields that do not explicitly show in the effective description, heavy bosons ($\gtrsim\pi T$) or fermions. Here, we assume the parent theory to be the minimal one, only containing the counter parts to $\phi$ and $\chi$, as in Refs.~\cite{Gould:2021ccf,Hirvonen:2020jud}.

%The result, up to our accuracy of $\Delta f$, is just the one-loop result,
%\begin{equation}\label{eq:oneloopthermcomp}
%    \feynalignscaled{0.6}{\oneloopbubble{}}\;+\;\feynalignscaled{0.6}{\oneloopbubble{dashed}}=-\frac{\pi^2 T^4}{45}+\order{\epsilon}\,,
%\end{equation}
%from an equivalent computation to the one in Eq.~\eqref{eq:oneloopthermcomp}. This leads to
%\begin{equation}
%    f_+=-\frac{\pi^2 T^4}{45}\times\left(1+\order{x^2}\right)\,.
%\end{equation}
%The order $x^2$ error comes from the power countings of the parent theory being
%\begin{align}
%    g^2\sim x^2\,\quad f&\sim x^2\,,\quad\lambda\sim x^3\,,\\
%    M_3^2\sim x^2\,&\qquad m^2\sim x^3\,.
%\end{align}
%Thus, the equivalents of Eqs.~\eqref{eq:hidingsunset} and \eqref{eq:twoloopthermcomp} would be $\order{x^2T^4}$.

We obtain for the strength of the transition
\begin{equation}\label{eq:ptstreng}
    \alpha=\frac{15}{4\pi^2 T^4}\left(4\Delta f-T\partial_T\Delta f\right)\times\left(1+\order{x^2}\right)\,,
\end{equation}
where $\Delta f$ is given in Eq.~\eqref{eq:freeendiff}.

Now, we turn to the inverse duration of the phase transition, $\beta$, given by
\begin{equation}
    \beta=H_*T\dv{T}\ln\Gamma\,,
\end{equation}
where $\Gamma$ is the nucleation rate and $H_*$ is the Hubble parameter at the time of the transition~\cite{Enqvist:1991xw}.

The nucleation rate is then given by
\begin{align}\label{eq:nuclrate}
    \Gamma=A\,e^{-S_{\rmii{eff}}[\phi_{\text{CB}}]+S_{\rmii{eff}}[\varphi_+]}\,,
\end{align}
where the exponential contributions to the rate are given by the effective action from the Lagrangian of Eq.~\eqref{eq:exampleTwoNuclAction},
\begin{equation}\label{eq:fulleffact}
    S_{\rmi{eff}}[\phi_{\text{IR}}]=\int_{\bf x}\mathscr{L}_{\rmi{eff}}\,,
\end{equation}
evaluated on the critical bubble, $\phi_{\text{CB}}$, and the metastable minimum, $\varphi_+$~\cite{Gould:2021ccf}. Note, that the unit term from the thermal scale cancels in the exponent as well.

The critical bubble is found as an extremum of $S_{\text{eff}}$:
\begin{align}
    &\frac{\delta S_{\rmi{eff}}}{\delta\phi}[\phi_{\text{CB}}]=0\\
    \Leftrightarrow\;& Z_{\text{eff}}(\phi_{\text{CB}})\pdv[2]{\phi_{\text{CB}}}{r}\nonumber\\
    &+\frac{2}{r}Z_{\text{eff}}(\phi_{\text{CB}})\pdv{\phi_{\text{CB}}}{r}+\frac{1}{2}Z'(\phi_{\text{CB}})\qty(\pdv{\phi_{\text{CB}}}{r})^2\nonumber\\ 
    &= V_{\rmi{eff}}'(\phi_{\text{CB}})\,,\label{eq:bubbleeom}
\end{align}
where $r$ is the radial coordinate for the spherically symmetric critical bubble, $\phi_{\text{CB}}(r)$.

A major numerical simplification can be obtained by treating the correction to the field normalization, $\Delta Z_{\text{eff}}$ in Eq.~\eqref{eq:resforDZ}, perturbatively~\cite{Gould:2021ccf}. The action term, corresponding to $\Delta Z_{\text{eff}}$,
\begin{align}
    \int_{\bf{x}}\frac{\Delta Z_{\text{eff}}(\phi_{\text{CB}})}{2}(\partial_i\phi_{\text{CB}})^2\,,
\end{align}
can simply be evaluated on the critical bubble found from
\begin{align}
    \pdv[2]{\phi_{\text{CB}}}{r}+\frac{2}{r}\pdv{\phi_{\text{CB}}}{r}= V_{\rmi{eff}}'(\phi_{\text{CB}})\label{eq:bubbleeomsimpl}
\end{align}
to our accuracy.

Using the nucleation rate formula in Eq.~\eqref{eq:nuclrate}, we obtain for the inverse duration:
\begin{equation}\label{eq:invdur}
    \frac{\beta}{H_*}=-T\dv{T}\big(S_{\rmi{eff}}(\phi_{\text{CB}})-S_{\rmi{eff}}(0)\big)+\order{\ln x}\,,
\end{equation}
where the uncertainty of $\ln x$ arises from the prefactor $A$.

Let us finally show, that for the order of $x^{-1/2}$, we need the two-loop order in Eqs.~\eqref{eq:twoloopsnowII}, \eqref{eq:twoloopIIsun} and the derivative correction in Eq.~\eqref{eq:resforDZ}. The power counting for the action can be done as
\begin{equation}
    \text{bubble volume}\ \times\ \text{Lagrangian term}\,,
\end{equation}
and the volume is given by~\cite{Gould:2021ccf}
\begin{equation}
    \text{bubble volume}\sim L_{\text{bubble}}^3 \sim m^{-3}_3\,.
\end{equation}
With these formulas and the power counting of parameters set in Eqs.~\eqref{eq:exIIpowparam}, \eqref{eq:exIIotherpows}, \eqref{eq:phisize}, \eqref{eq:nuclscalederiv}, we indeed obtain that the aforementioned contributions come at order $x^{-1/2}$. The uncertainty in the inverse duration, $\beta$, is $\order{\ln x}$, shown in Eq.~\eqref{eq:invdur}.

%%%%%%%%%%%%%%%%%%%%%%%%%%%%%%%%%%%%%
%%%%%%%%%%%% SECTION II %%%%%%%%%%%%% 
%%%%%%%%%%%%%%%%%%%%%%%%%%%%%%%%%%%%%

\section{Integrating out a UV scale:\\a more thorough discussion} \label{sec:methodlong}

Here, we will study EFTs again from the point of view of purely integrating out the UV scale. The discussion repeats Sec.~\ref{sec:method} with more details.

Purely integrating out the UV scale will lead us to a method of constructing EFTs that is extremely similar to computing a 1PI effective action for the longer-scale (lighter) fields, i.e.\ the 1LPI action. As we will see, there are only two differences regarding the 1LPI action. Both of the differences originate from the fact that the IR scale has not been integrated out.

%\av{References to the motivation?}\jh{Prob no need}

%In perturbative high-temperature QFTs, there are often at least least two length scales in the theory: a UV scale corresponding to the thermal fluctuations and an IR scale, which is possibly an emergent, longer length scale. There can also additionally be scale hierarchies between the masses of fields.

%The behavior of an IR scale is not usually manifest in the action describing the total system. For example, there can be strong Debye screening from a UV scale. To obtain the long-scale IR behavior of the theory, one must first integrate out the short-scale UV contributions. This leads to an effective description for the long length scale.

In Sec.~\ref{sec:findingeffact}, we find the abstract form for the effective action for the EFT, and discuss the relation to just integrating out the heavier fields. In Sec.~\ref{sec:diagrammaticexp}, we discover the computational similarity to the 1LPI action, and in Secs.~\ref{sec:separatingIRUV} and \ref{sec:subsecepsilon} we find the two differences regarding the 1LPI action. In between, we discuss a diagrammatic way of understanding the novel method in Sec.~\ref{sec:diagunderstanding}. This both aids seeing the full picture of the procedure already summarized in Sec.~\ref{sec:method} and eases the discussion of the second difference in Sec.~\ref{sec:subsecepsilon}.

%As we will discover, there are only two differences between the computation of the 1PI action for the lighter fields and the construction of a consistent EFT for the fields: an expansion in IR quantities of the 1PI diagrams shown in Box~\ref{diff:one} in Sec.~\ref{sec:separatingIRUV}, and the possibility of $\order{\epsilon}$ terms contributing.\jh{Where do we find this?}
%There will only be two small modifications to the 1PI action calculation that need to be implemented for a consistent EFT calculation, which we will discover here.
%\jh{More explanation on what is in this section.}

\subsection{Effective action for an IR scale}\label{sec:findingeffact}

Let us start with a theory, which contains a short UV scale and a longer IR scale. The thermodynamics of the theory are described by a partition function,
\begin{equation}\label{eq:statdesc}
    Z=\int\mathcal{D}\Phi\, e^{-\Seu[\Phi]}\,,
\end{equation}
where $\Phi$ represents the full field content of the theory and $\Seu$ is its Euclidean action (See. Appendix~\ref{app:imtimform}).
%\jh{How to handle ITF?}\jh{How about the applicability to vacuum computations?}
%The partition function is sufficient to study the model in thermal equilibrium (\textit{cf.} Appendix~\ref{app:imtimform}).

Borrowing from Sec.~\ref{sec:method}, Eq.~\eqref{eq:fluctdivreprep}--\eqref{eq:effactreprep}, we can rewrite the partition function using the scales:
\begin{align}
    Z&=\int\mathcal{D}\ir\, e^{-S_{\text{eff}}[\ir]}\,,\label{eq:effstatdesc}\\
    S_{\text{eff}}[\ir] &= -\ln \int\mathcal{D}\uv\,e^{-S_{\mathrm{E}}[\ir+\uv]}\,,\label{eq:effact}\\
    \mathcal{D}\Phi
    &=\mathcal{D}\ir \times \mathcal{D}\uv\,.\label{eq:fluctdiv}
\end{align}
Here, $\ir$ are the IR-scale field modes and $\uv$ are the UV-scale field modes.

\begin{comment}

The idea for constructing an effective description for the long IR scale is straightforward: We separate the short scale fluctuations, $\uv$, from the long-scale fluctuations, $\ir$,
\begin{equation}\label{eq:fluctdiv}
    \mathcal{D}\Phi
    =\mathcal{D}\ir \times \mathcal{D}\uv\,,
\end{equation}
We then integrate over the shorter scale fluctuations in the partition function, $Z$, to produce an EFT for the long length scale:
\begin{equation}\label{eq:effstatdesc}
    Z=\int\mathcal{D}\ir\, e^{-S_{\text{eff}}[\ir]}\,.
\end{equation}
The effective action, $\Sef$, is given by
\begin{equation}\label{eq:effact}
    S_{\text{eff}}[\ir] = -\ln \int\mathcal{D}\uv\,e^{-S_{\mathrm{E}}[\ir+\uv]}\,.
\end{equation}
The long-scale fluctuations, $\ir$, appear as a background for the short scale fluctuations in the action:\linebreak $S_{\mathrm{E}}[\ir+\uv]$. Here, we have split the full field content in the action with $\Phi=\ir+\uv$.
%\og{$\ir$ is a background field.}

Importantly, the effective action, $\Sef$ is local (given that the original action, $\Seu$, is local). Physically, this is due to the effective action containing contributions only from the shorter UV-scale fluctuations. These fluctuations appear point-like from the perspective of the longer-length IR scale. Correspondingly, the contributions from the UV-scale fluctuations lead to local terms in the effective action.
%for the IR-scale fluctuations.
%\jh{Local action terms from the UV scale.}

\end{comment}

We will now first focus on how to divide the field content of the theory into the IR and UV parts as in Eq.~\eqref{eq:fluctdiv} and then show how to compute the effective action in Eq.~\eqref{eq:effact} as a slightly modified 1LPI action.

Let us state for concreteness that the scale hierarchy originates from a mass hierarchy, $M^2\gg m^2$, with a field $\heavy$ being much heavier than the other fields $\light$.
If one of the scales was given by the temperature, everything would be computationally exactly analogous. (This is shown in Example 1, Sec.~\ref{sec:exampleI}.)
%The example in Sec.~\ref{sec:exampleI} covers both possibilities, temperature giving the UV scale and the IR scale.
%\footnote{In the high-temperature case, $\heavy$ contains all of the non-zero Matsubara modes (see Appendix~\ref{app:imtimform}), and $\light$ contains all the light zero Matsubara modes. An example is given in Section~\ref{sec:exampleI}.}

It would be compelling to identify the heavier field as the UV fluctuations and the light fields as the IR fluctuations: $\uv=\heavy$ and $\ir=\light$. This is because the correlation length of the heavy field, $\heavy$, is constrained to be on the shorter scale due to its large mass.
%, whereas this does not hold for the light fields $\light$ due to their light masses.

More rigorously, the heavy field decouples regarding the IR scale~\cite{Appelquist:1974tg}. This means that integrating over $\heavy$ with a background of IR-scale fields, $\ir$, leads to only local action terms for the IR-scale. Therefore, we can indeed integrate over the heavy field in Eq.~\eqref{eq:effact}.

The identification of $\uv=\heavy$ and $\ir=\light$ would simply correspond to integrating out the heavier field. This is briefly discussed below.

The identification above is however not sufficient for constructing a local effective description.
The light fields can have a \textit{momentum} that is on the UV scale, $p\sim M$. These high-momentum fluctuations live on the shorter UV scale. Hence, they cannot be included to $\ir$.

Another way of understanding this is that the high-momentum fluctuations probe the distances of the $\heavy$ correlation length, $M^{-1}$. As a consequence, integrating out only the heavy field, $\heavy$, would make the description non-local for the high-momentum modes
%of the light fields non-local
(see e.g.\ Ref.~\cite{collins_1984}). Therefore, the high-momentum modes must be integrated out as well for a consistent local effective description for the longer IR scale.

%However, the identification is not sufficient for constructing a local effective description. The light fields can have a momentum that is on the UV scale, $p\sim M$. These high-momentum fluctuations live on the shorter UV scale, probing the distances of the $\heavy$ correlation length. As a consequence, integrating out only the heavy field, $\heavy$, would make the description for the high-momentum modes of the light fields non-local (e.g.\ Ref.~\cite{collins_1984}). Therefore, the high-momentum modes cannot be a part of $\ir$. They must be integrated out as well for a consistent local effective description for the longer IR scale.

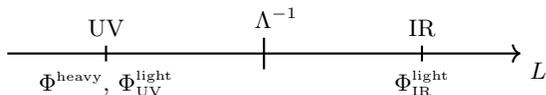
\begin{figure}
    \centering
    \begin{tikzpicture}[scale=1.05]
        \draw[->,thick] (-0.25,0) -- (6.25,0) node[anchor=north west] {$L$};
        \draw[thick] (5,-0.1) node[anchor= north] {$\lightlow$} -- (5,0.1) node[anchor=south] {IR};
        \draw[thick] (3,0.2) node[anchor= south] {$\;\;\;\,\Lambda^{-1}$} -- (3,-0.2) ;
        \draw[thick] (1,-0.1) node[anchor=north] {$\heavy,\,\lighthigh$} -- (1,0.1) node[anchor=south] {UV};
    \end{tikzpicture}
    \caption{The UV and IR length scales, $L$, presented. The high-momentum modes $\lighthigh$ of the light fields $\light$ are separated from the low-momentum modes $\lightlow$ by a factorization scale $\Lambda$. All of the fluctuations of the heavy field $\heavy$ reside in the UV scale.}
    \label{fig:IRUV}
\end{figure}

The division of the fluctuations into scales is done according to
\begin{equation}\label{eq:refinedscaleseparation}
    \mathcal{D}\ir = \mathcal{D}\lightlow\,,\qquad\mathcal{D}\uv
    =\mathcal{D}\lighthigh\, \mathcal{D}\heavy\,,
\end{equation}
where $\lightlow$ and $\lighthigh$ are are the IR-momentum and UV-momentum parts of the light field $\light$ respectively. The momentum modes are separated by a factorization scale, $\Lambda$, which is between the scales, $m\ll\Lambda\ll M$, as shown in Fig.~\ref{fig:IRUV}.

%\jh{Doesn't anymore discuss the factorization scale. Should it?}
%The factorization scale eventually becomes the ultraviolet regulator for the IR scale effective description. This is due to everything above the factorization scale been integrated out.

%\jh{Note, that integrating over the $\lighthigh$ modes is not as pure as integrating over the heavy fields. There will be a corresponding change in the counterterms as well.}\jh{It is definitely not a subtlety at all!}

%In this article, we will use cut-off regularization as a source for intuitive understanding of integrating out a scale. The reason is that the separation of $\lighthigh$ and $\lightlow$ modes can be defined very concretely using the factorization scale, $\Lambda$, as above. The actual procedure of integrating out a scale is however given in a more commonly used dimensional regularization. It makes the computations of the EFT construction much more fluent (see e.g.\ Ref~\cite{Manohar:2018aog}). The (comparatively very low) price to pay is that the same separation of $\lighthigh$ and $\lightlow$ becomes more abstract. This is discussed in Sec.~\ref{sec:separatingIRUV}.

%Before making the connection between the effective action in Eq.~\eqref{eq:effact} an the 1PI action, and showing how to easily separate the UV momentum fluctuations, $\lighthigh$, in a computation, we want to pause to discuss the similarities with integrating out a heavy field $\heavy$.

Below, we will make the connection between the effective action in Eq.~\eqref{eq:effact} and the 1LPI action, and show how to separate easily the UV-momentum fluctuations, $\lighthigh$, in a calculation. But before that, we want to pause to discuss the similarities with integrating out the heavy field, $\heavy$.

Had we chosen the tempting identifications of $\uv=\heavy$ and $\ir=\light$, the steps in Eqs.~\eqref{eq:statdesc}--\eqref{eq:effact} would reduce to merely integrating out the heavy field, $\heavy$. We can thus note that the method presented here is just integrating out the heavy field and 
%along with
the UV fluctuations of the light fields as well.

The problems of integrating out a heavy field, which are cured by integrating of the UV scale entirely, are illustrated by the first example of Ref.~\cite{WeinbergVacDec}. In this excellent article on radiative vacuum decay, only a heavy field is initially integrated out. This would leave the calculation inconsistent due to the non-localities for the high-momentum modes of a light field. These show up as uncancelled divergences in the computation.%
\footnote{Note, that these uncancelled divergences are not stated explicitly in the article but are still present. They are the counter-terms cancelling the divergences of the sunset diagram in their Fig.~2(b).}

After-the-fact however, another important contribution to the calculation was found, repairing the computation. The crucial contribution follows from integrating out the UV fluctuations of the light field, $\lighthigh$. The computation now abides the separation in Eq.~\eqref{eq:refinedscaleseparation}. Thus, a consistent local EFT was constructed.

\subsection{Diagrammatic expansion}\label{sec:diagrammaticexp}

Let us now press on with the effective action in Eq.~\eqref{eq:effact} to find its diagrammatic representation. This way, we will also discover the connection to the 1LPI action.

We can directly see that in a perturbative computation only the connected diagrams contribute to the effective action. This is due to the logarithm in Eq.~\eqref{eq:effact} cancelling the disconnected contributions.

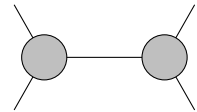
\begin{figure}
    \centering
    \begin{tikzpicture}
        \node[] at (0,0) {$\feynalignscaled{1}{\PRdiag}$};
    \end{tikzpicture}
    \caption{A 1$\light$-reducible diagram. Solid line represents a $\light$ field, the external legs are from the IR background of $\lightlow$ (\textit{cf.} Eqs.~\eqref{eq:effact}, \eqref{eq:refinedscaleseparation}) and the grey blobs are 1$\light$-irreducible contributions. The reducible propagator carries an IR momentum from the external legs.}
    \label{fig:reducible}
\end{figure}

Besides the disconnected diagrams, there is another class of diagrams that do not contribute:
%Another class of diagrams that do not contribute are
the 1$\light$-reducible diagrams. This means the diagrams that are reducible by cutting one light-field propagator. An example is in Fig.~\ref{fig:reducible}. Diagrams that are reducible in heavy field propagators do enter the diagrammatic expansion.

The reason for 1$\light$-reducible diagrams not contributing is that the reducible propagators would be on the IR scale: The reducible $\light$ propagators can only carry a momentum from the external legs, and the external legs are from the IR-scale background. (\textit{cf.} Fig.~\ref{fig:reducible} and Eqs.~\eqref{eq:effact}, \eqref{eq:refinedscaleseparation}). Thus, the momentum in the reducible light-field propagator is on the IR scale.

IR-scale fluctuations of the light fields, $\lightlow$, are not integrated over in the effective action of Eq.~\eqref{eq:effact}.  Consequently, the reducible $\light$ propagators do not appear in the computation of the effective action. Therefore, there are no 1$\light$-reducible diagrams.
%are actually $\lightlow$ propagators, which do not enter the effective action. This is due to the IR modes of the light fields not being integrated over in Eq.~\eqref{eq:effact}.

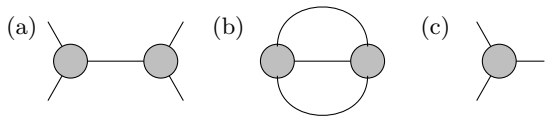
\begin{figure}
    \centering
    \begin{tikzpicture}
        \node[] at (-1.25,0.5) {(a)};
        \node[] at (0,0) {$\feynalignscaled{0.75}{\PRdiag}$};
        \node[] at (1.5,0.5) {(b)};
        \node[] at (2.75,0) {$\feynalignscaled{0.75}{\basketball}$};
        \node[] at (4.25,0.5) {(c)};
        \node[] at (5.25,0) {$\feynalignscaled{0.75}{\PIdiag}$};
    \end{tikzpicture}
    \caption{The 1$\light$-reducible diagram in (a) would produce the contribution in (b) when integrating over the low momentum modes $\lightlow$. However, the contribution already follows from the 1$\light$-irreducible contribution (c) in the effective action.}
    \label{fig:accounting}
\end{figure}

Note, that all the contributions are still consistently accounted after integrating over the IR modes of the light fields, $\lightlow$. This is illustrated in
%As exemplified
Fig.~\ref{fig:accounting}.
%All the IR contributions are correctly obtained from the 1$\light$-irreducible diagrams in the effective action.
Everything is accounted for once and only once.
%A contribution (b), that would result from a 1$\light$-reducible diagram (a), is already obtained from the effective action only containing 1$\light$-irreducible diagrams (c). Thus, every contribution is counted once and only once, hence no double counting either.

%\og{Define which kind of a 1PI action is in question.}
We can now see the correspondence between the effective action and the 1LPI action clearly: They both have the same diagrammatic representation containing only the 1LPI diagrams.
%\footnote{By 1PI action, we mean it as defined in terms of a perturbative expansion containing only 1PI diagrams~\cite{Fukuda:1974ey} and not as a Legendre transformation~\cite{Jackiw:1974cv}. Note however that the effective action does not inherit the non-convexity problem of 1PI action as it arises from the IR-scale contributions.\jh{ref.}}
Therefore, $\Sef$ is almost the same as the 1LPI action.

The only difference is that the $\lightlow$ fluctuations are not yet being integrated over. This affects the meaning of the diagrams: The internal propagators are $\lighthigh$ propagators, and the external legs are $\lightlow$ legs.

\subsection{Separating the IR and UV scales}\label{sec:separatingIRUV}

We will now move to the final piece of the puzzle, which is how to separate the modes $\lightlow$ and $\lighthigh$ in a perturbative calculation.

In cut-off regularization, one could separate the IR and UV fluctuations of the light fields, $\lightlow$ and $\lighthigh$, with an infrared cut-off at the factorization scale, $\Lambda$, for the $\light$ propagators~\cite{PhysRevB.4.3174,WILSON197475} (\textit{cf}.\ Fig.~\ref{fig:IRUV}),
\begin{equation}\label{eq:ircutoff}
    \int_{\Lambda}^{\Lambda_{\text{UV}}}\dd^D p\,.
\end{equation}
Here, $p$ is a loop momentum and $\Lambda_{\rmii{UV}}$ is the cut-off for the full theory. This restricts the light field propagators to contain solely the high-momentum modes, $\lighthigh$.

For the more practical dimensional regularization, using a cut-off is not feasible. This is because the dimensionful regulator is a renormalization scale, $\mu$, rather than a cut off.%
\footnote{Here, we would like to note a very recent article on the physical equivalence of dimensional regularization and Wilsonian renormalization group, Ref.~\cite{Branchina:2022jqc}.}
One has to be a bit more clever to only take into account the $\lighthigh$ fluctuations in the effective action in Eq.~\eqref{eq:effact}.
This then pays off with simple computations that differ quite minimally from the 1LPI-action construction.

%Here, we will first provide the resolution and the related computational intuition.\jh{Watch out for this!} Then, we will work up to the resolution leaning on the clear, concrete cut-off version.

The method to only keep the UV-scale contributions of $\heavy$ and $\lighthigh$ fluctuations in the 1LPI diagrams is the following:
\begin{align}\tag*{1}
    \fbox{
        \begin{minipage}{7.5cm}
            \vspace{0.1cm}
            Expand the diagrams in IR quantities, i.e.\ the light masses, $m^2$, and the low external momenta, $k$,
            \vspace{-0.2cm}
            $$\frac{1}{(p+k)^2+m^2} \to\frac{1}{p^2}-\frac{2p\cdot k+k^2+m^2}{p^4}+\dots\,,$$
            \vspace{-0.5cm}
            $$\frac{1}{(p+k)^2+M^2} \to\frac{1}{p^2+M^2}-\frac{2p\cdot k+k^2}{(p^2+M^2)^2} +\dots\,,$$
            where $p$ is a loop momentum and $M^2$ is the heavy mass.
            \vspace{0.1cm}
        \end{minipage}
    }
\end{align}

To understand the above IR-quantity expansion, we need to discuss three central facts:
\begin{itemize}
    \item The UV-scale contributions remain intact.
    \item The IR scale is now absent from the loop integrals.
    \item Instead of the IR scale, there are infrared divergences in the loop-momentum region of $p^2\ll M^2$.
\end{itemize}

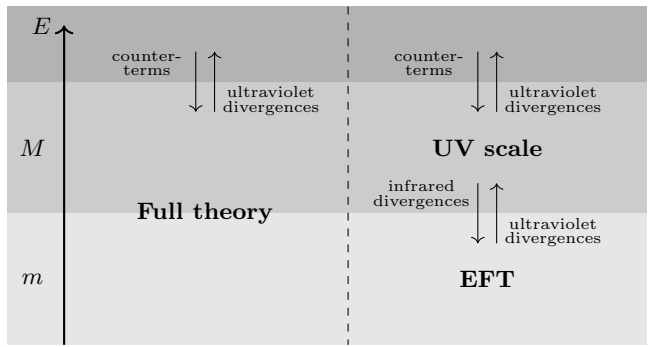
\begin{figure}
    \centering
    \begin{tikzpicture}[scale=1.0]
        \filldraw[draw=none, fill=black!30, very thick](0,4.5) rectangle (8.5,3.5);
        \filldraw[draw=none, fill=black!20, very thick](0,3.5) rectangle (8.5,1.75);
        \filldraw[draw=none, fill=black!10, very thick](0,1.75) rectangle (8.5,0);
        \draw[dashed] (4.525,4.5) -- (4.525,0);
        \node[] at (0.33,2.625) {$M$};
        \node[] at (0.33,0.875) {$m$};
        \draw[->,thick] (0.75,0) -- (0.75,4.25);
        \node[] at (0.45,4.25) {$E$};
        %\node[] at (0.33,1.75) {$\Lambda$};
        \node[] at (2.625,1.75) {\textbf{Full theory}};
        \draw[->] (2.5,3.9) -- (2.5,3.1);
        \node[] at (1.85,3.75) {$\substack{\text{counter-}\\ \text{terms}}$};
        \draw[->] (2.75,3.1) -- (2.75,3.9);
        \node[] at (3.5,3.25) {$\substack{\text{ultraviolet}\\ \text{divergences}}$};
        \node[] at (6.375,2.625) {\textbf{UV scale}};
        \draw[->] (6.25,3.9) -- (6.25,3.1);
        \node[] at (5.6,3.75) {$\substack{\text{counter-}\\ \text{terms}}$};
        \draw[->] (6.5,3.1) -- (6.5,3.9);
        \node[] at (7.25,3.25) {$\substack{\text{ultraviolet}\\ \text{divergences}}$};
        \draw[->] (6.25,2.15) -- (6.25,1.35);
        \node[] at (5.5,2) {$\substack{\text{infrared}\\ \text{divergences}}$};
        \draw[->] (6.5,1.35) -- (6.5,2.15);
        \node[] at (7.25,1.5) {$\substack{\text{ultraviolet}\\ \text{divergences}}$};
        \node[] at (6.375,0.875) {\textbf{EFT}};
    \end{tikzpicture}
    \caption{The EFT organization of a computation is presented schematically on the right. The ultraviolet divergences of the UV scale cancel against the counterterms of the full theory. The infrared divergences of the UV scale cancel against the ultraviolet divergences of the EFT.%\jh{Show explicitly the cancellations here.}
    }
    \label{fig:regulation}
\end{figure}

One may feel especially wary about the last point. It does not seem like a good idea to purposefully induce divergences to one's calculation. Here however, the infrared divergences turn out to be central: The resulting EFT requires ultraviolet regularization, i.e.\ counterterms. The infrared divergences become the correct counterterms for the EFT (Fig.~\ref{fig:regulation})!

After discussing the UV-scale contributions of the $\heavy$ and $\lighthigh$ fluctuations, we note that the EFT contains the final contributions from the $\lightlow$ fluctuations. Thus, it marks a successful implementation of the division of the fluctuations given in Eq.~\eqref{eq:refinedscaleseparation}.

Here, we want to also note the achieved computational simplification. One does not need to perform any explicit mass resummations in the diagrams. In the construction of the effective action of Eq.~\eqref{eq:effact}, the IR propagators are rather expanded in the masses, or ``anti-resummed'', by the above expansion. Then, the UV-scale contributions are resummed by construction into the effective action. Thus, they are already resummed for the IR-scale fluctuations, which are computed within the EFT.

Let us start from the $\lighthigh$ contributions, i.e.\ the UV modes of the light fields. Physically, the light fields are ultrarelativistic on the UV scale, $M$. Thus, they are massless at the leading order, obtaining higher-order corrections from their masses:
\begin{equation}\label{eq:lightmassExp}
    \frac{m^2}{M^2}\ll 1\,.
\end{equation}
Computationally, this is reflected as expanding in the light masses, $m^2$, against the UV-scale loop momentum, $p^2\sim M^2$, in the IR-quantity expansion.

Similarly to the light masses, one can expand in the IR-scale external momenta, $k^2\sim m^2$, against the UV scale,
\begin{equation}\label{eq:derivativeExp}
    \frac{k^2}{M^2}\ll1\,.
\end{equation}
In the light-field propagators, the UV scale is in the loop momentum, $p^2\sim M^2$, and in the heavy-field propagator the UV-scale already follows from the heavy mass, $M^2$.

The expansion in the external momenta eventually results in a derivative expansion of the effective action,
\begin{equation}\label{eq:toDerivExp}
    k^2\to -\partial^2
\end{equation}
(see the example in Sec.~\ref{sec:exampleII}). Physically, the derivative expansion contains the effects of a non-constant IR-scale background field on the UV-scale contributions.

We have now discussed, how the UV-scale contributions of $\heavy$ and $\lighthigh$ remain intact in the IR-quantity expansion: Basically, one expands smaller IR quantities against the UV scale.

In a practical computation, this expansion is truncated to some desired accuracy,
\begin{equation}
    \order{\qty(m^2/M^2)^n,\,\qty(k^2/M^2)^m}\,,
\end{equation}
which is then reflected in the accuracy of the obtained physical quantities.

Let us move on to the $p^2\ll M^2$ region of the light-field propagators. It is clearly altered by the IR-quantity expansion.

Firstly, it is important to note that the IR-quantity expansion removes every reference to the IR scale from the integrals, i.e.\ the light masses, $m^2$, and the external momenta, $k$. For example,
\begin{align}\label{eq:theRemovalofIR}
    &\irqe{\int_p\frac{1}{p^2+m^2}}\nonumber\\[5pt]
    =&\int_p\frac{1}{p^2} -m^2\int_p\frac{1}{p^4}+\frac{m^4}{2}\int_p\frac{1}{p^6}+\dots
\end{align}
Thus, the momentum integrals do not see the IR scale at all, and consequently the $\lightlow$ contributions are not present in the integrals. 

Since we do not have an infrared cut off in our integrals, we still have the loop-momentum region of $p^2\ll M^2$ present. Due to the absence of the IR scale, the region of $p^2\ll M^2$ only yields infrared divergences.

As already mentioned above, these infrared divergences are a crucial part of the EFT construction. They are the counterterms for the EFT. Therefore, it may be better to directly think of them as the infrared divergences of the UV-scale -- simply an important part of the UV-scale contributions.

In slightly more detail, there are two types of divergences in the UV-scale contributions: ultraviolet and infrared divergences. The former divergences are unaltered by the IR-quantity expansion. Thus, they cancel against the counterterms of the full theory. The infrared divergences are left over, and they go on to instate the correct counterterms for the EFT.

Figure~\ref{fig:regulation} illustrates the fact that the infrared divergences of the UV scale are the counterterms of the EFT.
%is illustrated in Fig.~\ref{fig:regulation}.
An example on the level of diagrams is given in Sec.~\ref{sec:diagunderstanding} and the general proof (along the lines of Ref.~\cite{Manohar:2018aog}) is shown in Appendix~\ref{app:IRdivs}.

To summarize, the UV-scale contributions are unaffected by the IR-quantity expansion. The expansion removes the IR-scale from the integrals completely and replaces it with the infrared divergences. These infrared divergences of the UV scale are a crucial part of the computation, because they are the correct counterterms of the EFT. Thus, we get rid of the unwanted $\lightlow$ contributions and instead get the correct counterterms!

We have now discussed the first part of accounting for the $\light$ contributions. The $\lighthigh$ contributions are taken into account by the EFT construction. Let us briefly explain, how the EFT then only takes into account the final missing contributions from the $\lightlow$ fluctuations.

A very similar story applies here: There is no heavy mass scale $M$ present in the EFT. Thus, the momentum integrals of the EFT only produce ultraviolet divergences in the region of $p^2\gg m^2$. The ultraviolet divergences of the EFT then cancel against the infrared divergences of the UV scale (Fig.~\ref{fig:regulation}).

There is a poetic way of thinking about these cancelling intermediate divergences: We tear the contributions of a single theory into two more manageable pieces by leveraging the scale hierarchy (Fig.~\ref{fig:regulation}). Neither of these pieces can see the scale of the other piece. Quite like jigsaw-puzzle tabs and indentations, the intermediate divergences interlock the contributions back together in the physical results.

Here, we still want to be more specific about the non-existence of the mass scale $M$ in the EFT. There are two key components to this: One is clearly integrating over the heavy field, $\heavy$. The other one is a bit more subtle. It is the external-momentum expansion against the heavy mass, $M$, in the heavy-field propagator, Box~\ref{diff:one}. If the expansion was not conducted, the EFT loop integrals (in which one integrates over the momentum $k$) would still see the mass scale $M$ at $k^2\gg m^2$ through dependence alike
\begin{equation}\label{eq:scaleHiding}
    \frac{1}{k^2+M^2}\,.
\end{equation}
With the external-momentum expansion, the scale is no longer present in the EFT loops:
\begin{equation}
    \irqe{\frac{1}{k^2+M^2}}=\frac{1}{M^2}-\frac{k^2}{M^4}+\dots
\end{equation}
This is the reason, why the full IR-quantity expansion in Box~\ref{diff:one} is required for separating the scales, and for the ultraviolet divergences of the EFT to cancel with the infrared divergences of the UV scale.

%Here, we have seen that we can indeed separate the UV scale the IR scale from each other. contributions using dimensional regularization

We have now seen that the UV-scale contributions can indeed be separated from the IR contributions in dimensional regularization, even though we do not have the factorization-scale cut off, $\Lambda$, at our disposal. In more technical terms, the UV-scale contributions are analytic in the light masses, $m^2$, and the IR-scale contributions non-analytic \textit{by definition} in dimensional regularization. (See e.g.\ Ref.~\cite{Laine:2016hma}.) The non-analytic, IR-scale contributions vanish due to the expansion in the light masses in Box~\ref{diff:one}
(e.g.\ Refs.~\cite{Manohar:2018aog,Schicho:2020xaf}), and we are just left with the UV-scale contributions.

%Finally, we note that the separation of the high- and low-momentum modes, $\lighthigh$ and $\lightlow$, does not require any reference to the factorization scale, $\Lambda$ (Fig.~\ref{fig:IRUV}). %This is a rather curious fact, as the factorization scale was the starting point for our intuition, similarly to Ref.~\cite{Braaten:1995cm}.

%Finally, we note that the EFT construction method does not make any reference to the factorization-scale cut off, $\Lambda$.
%By not having any reference to the factorization scale,
%This way, dimensional regularization eases practical computations: one does not need to juggle around with $\Lambda$, which cancels from the physical quantities anyway. In a sense, dimensional regularization automatizes the handling of $\Lambda$~\cite{Manohar:2018aog}, as it does not need to be introduced to the computation in the first place. (Note, that the renormalization scale, $\mu$, is of course still present. It plays a similar role to the factorization scale and is also denoted by $\Lambda$ sometimes.)

\subsection{Diagrammatic understanding}\label{sec:diagunderstanding}

%\av{This is general diagrammatic discussion. Specific, actual computations in examples!}

%In the previous sections, we have, very compactly, presented the method to construct an EFT by directly integrating out a scale. \jh{The following no longer simply true. (Neither is the previous one...)} Before we summarize the method in Sec.~\ref{sec:summary}, we want to discuss diagrammatic view on integrating out a scale. This is due to the fact that diagrams provide a very powerful way of understanding EFT construction.

In the previous sections, we have presented the central computational ideas of integrating out a scale. Here, we want to step back to look at the bigger picture using diagrams, which provide a very powerful way of understanding EFT construction.

This section is nearly completely auxiliary regarding practical computations. Thus, it is not required in any way for performing computations. However, it hopefully makes the understanding of the EFT construction more robust. Only the very end of {\bf \emph{Reorganization of the computation and local action terms}} discusses a practical shortcut for the EFT construction.

We will first discuss the reorganization of the perturbation theory and the 1LPI diagrams becoming local action terms. We will see also the power of the EFT arrangement of a computation: complicated diagrams being effortlessly constructed out of simple pieces.

Then, we move on to discussing the infrared divergences of the UV scale and the ultraviolet divergences of the EFT via an example. This sheds light on the cancellation between these intermediate divergences (which again means that the infrared divergences of the UV scale are the correct counterterms of the EFT). The example will also illuminate the scale separation in a practical context.
%\jh{Does it?}

%as the counterterms of the EFT. \jh{Also, we see the divregences giving us the split.} \jh{Not anymore here:} Finally, we note the final computational difference between the 1PI action for the light fields and the effective action mentioned at the beginning of Sec.~\ref{sec:method}.

%\jh{Computationally auxiliary.}

We will be using the following notational convention: The heavy-field propagator is a dashed line and the light-field propagator is a solid line,
\begin{align*}
    \text{Heavy:}&\quad\feynalignscaled{0.95}{\propagator{dashed}}\;,\\
    \text{Light:}&\quad\feynalignscaled{0.95}{\propagator{}}\;.
\end{align*}

\parttitle{Reorganization of the computation and local action terms}

%\jh{What will be found here?}

Here, we will first find a diagrammatic way for representing the EFTs, which has been used before e.g.\ in Ref.~\cite{Laine:2016hma}. Then, we continue with the representation, making it match the EFT construction more precisely. The representation can be used to diagrammatically understand, how the UV-scale contributions are local action terms in the effective action, and how the EFT organization rearranges the computation naturally according to scales.

%When one integrates out a UV scale, the contributions from the UV scale to the IR-scale behavior become local terms to the effective action in Eq.~\eqref{eq:effact}. This is due to the fact that the shorter UV-scale fluctuations look point-like from the perspective of the longer IR scale.
The act of constructing an EFT from a full theory can be regarded in the following way: First, we ``look'' at the plasma at the shorter, UV-scale distances, Eq.~\eqref{eq:statdesc}: 
\begin{equation}\label{eq:statdescrep}
    Z=\int\mathcal{D}\Phi\, e^{-\Seu[\Phi]}\,.
\end{equation}
Then, we ``zoom'' out to look at the plasma on the longer, IR-scale distances, Eq.~\eqref{eq:effstatdesc}:
\begin{equation}\label{eq:effstatdescrep}
    Z=\int\mathcal{D}\ir\, e^{-S_{\text{eff}}[\ir]}\,.
\end{equation}
The UV fluctuations are point-like from the IR-scale perspective. These are contained in the local effective action, Eq.~\eqref{eq:effact}:
\begin{equation}\label{eq:effactrep}
    S_{\text{eff}}[\ir] = -\ln \int\mathcal{D}\uv\,e^{-S_{\mathrm{E}}[\ir+\uv]}\,.
\end{equation}

%\jh{Full theory diagram. Then how the EFT sees the diagram.}

The above can be translated into a diagrammatic language. We can take a diagram from the full-theory partition function in Eq.~\eqref{eq:statdescrep}. Then, we zoom out to the longer EFT length scale to see how the original diagram looks on this scale. 

As the UV fluctuations are point-like from the IR-scale perspective, the zooming can be represented diagrammatically as the UV parts of a diagram shrinking to points. We will refer to these shrunken UV parts as \textit{UV vertices}. These UV parts can either be a heavy-field propagator,
\begin{equation}\label{eq:zoomingEFTinteraction}
    \feynalignscaled{0.75}{\intuitionfourpoint}\quad \xrightarrow{\substack{\text{Zooming}\\ \text{out}}}\quad \feynalignscaled{0.75}{\fourpointtree{dot,minimum size=1.77mm}} \,,
\end{equation}
or a loop containing a UV-scale loop momentum,
\begin{align}\label{eq:zoomingEFTmass}
    \feynalignscaled{0.75}{\intuitionmass}\quad \xrightarrow{\substack{\text{Zooming}\\ \text{out}}}\quad
    \underbrace{\feynalignscaled{0.75}{\selfenergytree{plain}{dot,minimum size=2mm}}}_{\text{UV mom.}}\;
    +\;\underbrace{\feynalignscaled{0.75}{\selfenergyfourdot{dot,minimum size=2mm}{}}}_{\text{IR mom.}}\,.
\end{align}
(\textit{Cf}.\ the division into UV and IR in Eq.~\eqref{eq:refinedscaleseparation}.) In Eq.~\eqref{eq:zoomingEFTmass}, the second diagram follows from the heavy propagator shrinking to a point.
%The same diagrammatic representation has been used, for example, in Ref.~\cite{Laine:2016hma}.

On the left-hand sides of the arrows, we have a diagram from the full theory and on the right-hand sides we see how the EFT organization perceives the initial diagram. The UV contributions become local UV vertices on the EFT side. This is the origin of the EFT organization of the computation: We first compute the UV vertices in the EFT construction. Then, we can compute the EFT diagrams with the UV vertices taken into account.

%Now, we can see diagrammatically, what is the role of the UV contributions in the EFT: The UV dots in the above diagrams are masses and interactions from the point-of-view of the EFT.
%Correspondingly, they are contributions to the local terms in the effective action of Eq.~\eqref{eq:effact}.

%\jh{Note the universality.}

Before pressing on with the diagrammatic understanding of the EFT organization, we want to make an important remark: There is clearly some universality to the UV vertices. The UV vertex in Eq.~\eqref{eq:zoomingEFTinteraction} was formed by shrinking a heavy-field propagator into a point. Exactly the same thing happens in the last diagram in Eq.~\eqref{eq:zoomingEFTmass}.
%Below, we will see that one only needs to compute the former vertex to obtain the effects of the latter.
This is foundational for the usefulness of the EFT organization, as we will note below around Eq.~\eqref{eq:dissection}.
%In the EFT construction, we can compute the UV vertex of 

The diagrammatic representation of Eqs.~\eqref{eq:zoomingEFTinteraction} and \eqref{eq:zoomingEFTmass} already helps us to understand some aspects of EFT construction. Mainly, it helps to compare the EFT organization to the ``direct'' computation of the full theory: In the EFT construction, one first computes the UV contributions. These then become UV vertices for the EFT, which then deals with the IR scale.

However, the above diagrammatic representation does not yet represent the natural course of an actual EFT computation. It rather shows the EFT organization glued on top of the ``direct'' computation. This makes the EFT seem more clumsy than the ``direct'' computation, which is not true. Let us go a little further to find a diagrammatic representation to the actual computational process.

%In the EFT organization, one first computes the UV-scale contributions.  and then the IR-scale (i.e.\ the EFT) contributions.

%As a small disclaimer, we want to note that the above diagrams do not yet represent the natural progression of an EFT computation. \jh{Not a small disclaimer but a crucial way to get to the next step!}

As discussed in the sections above, we take solely the UV-scale contributions into account in the EFT construction.
This is achieved
%can be encapsulated as the 1-light-field-irreducible diagrams, in which the internal propagators are from the UV-scale and the external legs are from the IR scale. (The internal propagators were filtered to be only from the UV scale by the IR-quantity expansion of the above section.)
by the IR-quantity expansion of Box~\ref{diff:one}. Thus, these entire diagrams shrink to points when zooming out to the IR scale:
\begin{align}
    \irqe{\feynalignscaled{0.75}{\intuitionfourpoint}}\quad &\xrightarrow{\substack{\text{Zooming}\\ \text{out}}}\quad \feynalignscaled{0.75}{\fourpointtree{dot,minimum size=1.77mm}} \,,\label{eq:onlyuv1}\\
    \irqe{\feynalignscaled{0.75}{\intuitionmass}}\quad &\xrightarrow{\substack{\text{Zooming}\\ \text{out}}}\quad
    \feynalignscaled{0.75}{\selfenergytree{plain}{dot,minimum size=2mm}}\,.\label{eq:onlyuv2}
\end{align}
Obtaining these contributions to the effective action corresponds to computing the UV-scale contributions, i.e.\ to the EFT construction.
%This corresponds to obtaining local action terms to the effective action in Eq.~\eqref{eq:effactrep}.
%Note, that there is a derivative expansion on the right-hand sides of the zooming and also a light-mass expansion on the bottom right.

%We can finally see the EFT reorganization in action. We have first computed the UV-scale contributions, such as Eqs.~\eqref{eq:onlyuv1}, \eqref{eq:onlyuv2}, and added them to the effective action, $\Sef$.

%\jh{Not separate UV vertices}

After obtaining the UV-scale contributions to the effective action, we can simply compute the EFT contributions. These are the diagrams coming out from the effective action, for example
\begin{equation}\label{eq:mass1EFTEFR}
    \feynalignscaled{0.75}{\selfenergyfourdot{}{}}.
\end{equation}

Note, that the UV contributions do not appear as separate UV vertices (\textit{cf}.\ Eq.~\eqref{eq:zoomingEFTmass}) but as parts of the EFT masses and interactions:
The mass of the effective action contains the mass contribution from Eq.~\eqref{eq:onlyuv2}. This has automatically implemented the mass resummations from the UV scale. Also, the EFT vertex contains the UV contribution from Eq.~\eqref{eq:onlyuv1}.

%At the leading order of the derivative expansion, the UV contribution of Eq.~\eqref{eq:onlyuv2} contributes to the effective mass, $\meff$. Therefore, the mass resummations from the UV scale are already done. We can simply compute, for example, the diagram\jh{Eq.~\eqref{eq:mass1EFTEFR}}in the EFT.

%\jh{Blatantly state that the ``missing'' piece is in the above diagram.}

Now, we have an adequate diagrammatic representation for the EFT computation. First, we compute the UV contributions, Eqs.~\eqref{eq:onlyuv1}, \eqref{eq:onlyuv2}. Those are local action terms and are added to the effective action. Then, we compute the EFT contributions, Eq.~\eqref{eq:mass1EFTEFR}.

One may be wondering, where the latter contribution in Eq.~\eqref{eq:zoomingEFTmass} was taken into account. This actually happens in the EFT diagram of Eq.~\eqref{eq:mass1EFTEFR} due to the aforementioned universality of the UV vertices.
%, the contribution is in the EFT diagram of Eq.~\eqref{eq:mass1EFTEFR}!

%One may wonder, where the IR-momentum contribution in Eq.~\eqref{eq:zoomingEFTmass} is, as it was not computed as a part of the UV-scale contribution in Eq.~\eqref{eq:onlyuv2}.
%The contribution is clearly part of EFT contributions due to the IR-scale loop momentum.
%The vertex comes from the UV scale, as it is a heavy propagator shrunken to a point (\textit{cf}.\ Eq.~\eqref{eq:zoomingEFTinteraction}). Conversely, the loop has been left for the EFT. Therefore, the contribution has to be a part of the EFT one-loop diagram in Eq.~\eqref{eq:mass1EFTEFR}, where the loop is computed.

%The contribution is clearly part of EFT contributions due to the IR-scale loop momentum. By diagrammatic inspection, it could be expected that the IR-momentum contribution is contained within the EFT diagram in Eq.~\eqref{eq:mass1EFTEFR}.

%Let us remember that also the vertex in the EFT received a UV correction of Eq.~\eqref{eq:onlyuv1}. Also, note that the vertex of the IR contribution of Eq.~\eqref{eq:zoomingEFTmass} is a shrunken heavy-field propagator.

We will now ``zoom'' back into the vertex of the EFT diagram, $\lambda_{\text{eff}}$. Due to the EFT construction, it is the sum of the UV contribution in Eq.~\eqref{eq:onlyuv1} and the original vertex, $\lambda$, in the initial action, $\Seu$:
\begin{equation}\label{eq:dissection}
    \underbrace{\feynalignscaled{0.6}{\selfenergyfour{}{}}}_{\lambda_{\rmi{eff}}} \quad\xrightarrow[\text{vertex}]{\substack{\text{Zooming}\\\text{into}}}\quad 
    \underbrace{\feynalignscaled{0.6}{\selfenergyfour{}{}}}_{\lambda}\;
    +\;
    \underbrace{\feynalignscaled{0.6}{\intuitzoomone}\;
    +\;
    \feynalignscaled{0.6}{\intuitionzoomtwo}}_{\text{UV contribution}}\,.
\end{equation}
The last two diagrams come from the UV contribution of Eq.~\eqref{eq:onlyuv1} to the effective coupling, $\lambda_{\text{eff}}$. The middle term on the right-hand side is the sought-after contribution in Eq.~\eqref{eq:zoomingEFTmass}.

The last contribution on the right-hand side corresponds to the following diagram of the full theory:
\begin{equation}\label{eq:zoomingagain}
    \feynalignscaled{0.6}{\intuitionlast}\quad \xrightarrow{\substack{\text{Zooming}\\ \text{out}}}\quad
    \feynalignscaled{0.75}{\selfenergytree{plain}{dot,minimum size=2mm}}\;
    +\;\feynalignscaled{0.75}{\selfenergyfourdot{dot,minimum size=2mm}{}}\,.
\end{equation}
%Notice the similarities to the previously mentioned contribution in Eq.~\eqref{eq:zoomingEFTmass} -- especially the fact that the UV vertex in the last diagram comes from a shrunken heavy-field propagator. This is why it was also found in Eq.~\eqref{eq:dissection}.
The latter contribution was found in Eq.~\eqref{eq:dissection} due to the fact that it comes from a heavy-field propagator shrinking to a UV vertex as well.

%Thus, one obtains the two IR-scale contributions from Eqs.~\eqref{eq:zoomingEFTmass}, \eqref{eq:zoomingagain} by computing the contributions in Eqs.~\eqref{eq:onlyuv1} and \eqref{eq:mass1EFTEFR}.

Dissecting diagrams to find certain contributions may seem cumbersome. However, EFTs are just the tool to bypass all of the complications of accounting each contribution. We only need to compute the diagrams in Eqs.~\eqref{eq:onlyuv1}, \eqref{eq:onlyuv2} and \eqref{eq:mass1EFTEFR} without worrying about anything more.

The punchline is that \textit{EFTs construct complicated diagrams out of simple pieces -- automatically}.

Finally, we note that there is also a reorganization happening on the level of the diagrammatic expansion, not only on the level of energy scales.

All of the diagrams containing only light fields are zero under the IR-quantity expansion. For example,
\begin{align}
    \irqe{\feynalignscaled{0.6}{\oneloopbubble{}}} =&\frac{1}{2}\int_p\ln(p^2) +\frac{m^2}{2}\int_p\frac{1}{p^2}
    -\frac{m^4}{4}\int_p\frac{1}{p^4}
    +\dots\nonumber\\[2pt]
    &=0\,,\label{eq:lightoneloop}
\end{align}
where all of the terms vanish as scale-free integrals.

As a consequence, one only needs to compute the diagrams containing heavy-field propagators for the UV-scale contributions.
Diagrams containing solely light fields need to be computed only when evaluating the IR-scale contributions, where there are no diagrams containing heavy-field propagators. Thus, no diagrams need to be computed twice.

The above goes even a bit further giving diagrammatic shortcuts with respect to the 1LPI action: Some diagrams need not to be computed at all. If a diagram contains even a single loop integral containing solely light fields, it is zero due to the loop integral being scale free. For example,
\begin{align}\label{eq:firstzerofound}
    \irqe{\feynalignscaled{0.6}{\twoloopbubbleC}} \propto&\int_p\frac{1}{p^2+M^2} \left(\int_q\frac{1}{q^2}-m^2\int_q\frac{1}{q^4}+\dots\right)\nonumber\\[2pt]
    &=0\,.
\end{align}
Hence, it is zero on the UV scale, and does not appear on the IR scale due to the heavy-field propagator.

The information of the diagram is, of course, still taken into account. First, one computes the mass contribution from the heavy field in
\begin{equation}\label{eq:EFTonloopExfourdUVpart}
    \irqe{\feynalignscaled{0.75}{\selfenergyfourdot{}{dashed}}}\quad \xrightarrow{\substack{\text{Zooming}\\ \text{out}}}\quad \feynalignscaled{0.75}{\selfenergytree{plain}{dot,minimum size=2mm}} \;.
\end{equation}
After obtaining this mass resummation, the information is contained within the one-loop diagram of the EFT:
\begin{equation}\label{eq:EFTonloopExfourd}
    \feynalignscaled{0.6}{\oneloopbubble{}}\;.
\end{equation}
This can be regarded as first computing the UV loop and then the IR loop with resummations.

%Similarly to Eq.~\eqref{eq:firstzerofound}, the UV part in Eq.~\eqref{eq:zoomingagain} is also zero:
%\begin{equation}
%    \irqe{\feynalignscaled{0.6}{\intuitionlast}}=0\,.
%\end{equation}

\parttitle{Cancellation of the intermediate divergences and the scale separation}

%\jh{Is this anything more than cancellations?}

%\jh{Change the nature of this part: It also shows how the EFT does not see the UV contributions. Note, that this analysis differs from an actual calculation.}

We will be looking at an example diagram of the full theory:
\begin{align}\label{eq:irdivsexdiag}
    \feynalignscaled{0.75}{\sunsetIR{}{dashed}}\;.
\end{align}
The EFT organization divides it into two pieces, a purely UV piece and an EFT loop diagram:
\begin{align}\label{eq:weirdmonstrocity}
    \feynalignscaled{0.6}{\sunsetIR{}{dashed}}
    \quad \xrightarrow{\substack{\text{Zooming}\\ \text{out}}}\quad
    \feynalignscaled{0.75}{\selfenergytree{plain}{dot,minimum size=2mm}}\;
    +\;\feynalignscaled{0.75}{\selfenergyfourIRdot{dot,minimum size=2mm}{}}\;.
\end{align}
Note importantly that there is an ultraviolet divergence in the EFT loop diagram and an infrared divergence in the UV piece. Our goal here is to find these intermediate divergences of the two pieces and to show that they cancel.%
\footnote{The reason, why one should expect that the intermediate divergences should cancel on the right-hand side of Eq.~\eqref{eq:weirdmonstrocity}, is that there are none before the ``zooming''. This turns out to be the case for every full-theory diagram~\cite{Manohar:2018aog}. We only show the cancellation of divergences on the level of the whole EFT construction in Appendix~\ref{app:IRdivs} and here for this particular diagram.}
Thus, the UV-scale piece counters the divergence from the EFT piece.

Let us start from the infrared divergence of the UV-scale contribution. The UV-scale piece can be found with the IR-quantity expansion of the full diagram (\textit{cf}.\ Eqs.~\eqref{eq:onlyuv1}, \eqref{eq:onlyuv2}):
\begin{align}\label{eq:irdivuvexp}
    \irqe{\feynalignscaled{0.6}{\sunsetIR{}{dashed}}}\;.
\end{align}
Due to the IR-quantity expansion, the light-field propagators are massless. Thus, they may cause divergences in the infrared regime, $p^2\ll M^2$, of the diagram.

Note, that the light-field propagators are on a longer length scale than the heavy-field loop ($p^2\ll M^2$) in the infrared-divergent part of the UV contribution.
%in Eq.~\eqref{eq:irdivuvexp}.
Correspondingly, the heavy-field loop looks like a local UV vertex to the light-field propagators:
\begin{equation}\label{eq:exofirdiv}
    \left[\irqe{\feynalignscaled{0.6}{\sunsetIR{}{dashed}}} \right]_{\substack{\text{ir}\\\text{divs.}}}
    =
    \left[\irqe{\feynalignscaled{0.75}{\selfenergyfourIRdot{dot,minimum size=2mm}{}}} \right]_{\substack{\text{ir}\\\text{divs.}}}.
\end{equation}
(This holds, of course, only for the infrared divergent part of the diagram.)

Let us be more precise about the UV vertex in Eq.~\eqref{eq:exofirdiv}. We will notice two central points: Firstly, the UV vertex is exactly the same as the one in the EFT part in Eq.~\eqref{eq:weirdmonstrocity}. Secondly, the infrared divergence is ``blind'' to both scales, the UV and the IR. This blindness will hold for the ultraviolet divergence of the EFT as well.

%{
%Let us look at the infrared divergence more closely. It comes from the momentum region, where the loop momentum through the light-field propagators is $p^2\ll M^2$. Thus, the propagators are on a longer scale than the heavy-field loop. Diagrammatically for the infrared divergence, we have \jh{Eq.~\eqref{eq:exofirdiv}}. The diagram, whose infrared divergence we are after, is the same as the EFT diagram in Eq.~\eqref{eq:weirdmonstrocity}. It does not contain the IR scale due to the IR-quantity expansion, but the UV dot is equivalent.
%}

The UV vertex in Eq.~\eqref{eq:exofirdiv} originates from 
\begin{align}\label{eq:anotherexamplechop}
    \feynalignscaled{0.6}{\irdivsex}
\end{align}
on the left-hand side. The external legs connect to the light-field propagators. As a consequence, the external momentum of Eq.~\eqref{eq:anotherexamplechop} is much below $M^2$ in the infrared-divergent part. Thus, we can perform the IR-quantity expansion (which corresponds to expanding in the light external momenta):
\begin{align}\label{eq:anotherexample}
    \irqe{\feynalignscaled{0.6}{\irdivsex}}&\propto
    \irqe{\int_q\frac{1}{((q+p)^2+M^2)(q^2+M^2)}}\nonumber\\[5pt]
    &=a_0+a_2\,\frac{p^2}{M^2}+a_4\,\frac{p^4}{M^4}+\dots\,,
\end{align}
where $p^2\ll M^2$ is the external momentum coming from the light-field propagators. This is now exactly the same UV vertex as in the EFT diagram.

Now, we can also see, why the infrared-divergent part is blind to both of the scales: In the integrals corresponding to the part,
\begin{equation}\label{eq:irdivpartjoo}
    \left[\irqe{\feynalignscaled{0.75}{\selfenergyfourIRdot{dot,minimum size=2mm}{}}} \right]_{\substack{\text{ir}\\\text{divs.}}},
\end{equation}
the IR-quantity expansion removes the IR-scale quantities (\textit{cf}.\ Eq.~\eqref{eq:theRemovalofIR}). Similarly, the IR-quantity expansion in the UV vertex, Eq.~\eqref{eq:anotherexample}, expands the heavy mass, $M^2$, out from the integrals of Eq.~\eqref{eq:irdivpartjoo}. Thus, we only have scale free integrals in the infrared divergence:
\begin{align}\label{eq:itsscalefree}
    &\left[\irqe{\feynalignscaled{0.75}{\selfenergyfourIRdot{dot,minimum size=2mm}{}}} \right]_{\substack{\text{ir}\\\text{divs.}}}\nonumber\\[5pt]
    =\;\,&\dots + b_1\left[\int_p\frac{1}{p^2} \right]_{\substack{\text{ir}\\\text{divs.}}}+b_2\left[\int_p\frac{1}{p^4} \right]_{\substack{\text{ir}\\\text{divs.}}}+\dots
\end{align}
Hence, the integrals are blind to both of the scales. %In general, the coefficients, $b_i$, can depend on the heavy mass, $M^2$, the light masses, $m^2$, and the external momenta, $k^2$.

Let us now move on to the case of the ultraviolet divergences of the EFT part. They come from the loop-momentum region of $p^2\gg m^2$. Consequently, we can perform the IR-quantity expansion against this loop momentum:
\begin{equation}\label{eq:uvdivpartjoo}
    \left[\feynalignscaled{0.75}{\selfenergyfourIRdot{dot,minimum size=2mm}{}} \right]^{\substack{\text{uv}\\\text{divs.}}} =\left[\irqe{\feynalignscaled{0.75}{\selfenergyfourIRdot{dot,minimum size=2mm}{}}} \right]^{\substack{\text{uv}\\\text{divs.}}}.
\end{equation}
(The IR-quantity expansion can obviously be performed only for the ultraviolet-divergent part and not for the whole diagram.)

%{
%In the diverging integral of Eq.~\eqref{eq:EFTdivEx}, we left the mass resummations to the propagator undone. They do not affect the ultraviolet divergences. If there was an external momentum, we could have expanded in it as well. Hence, we could have applied the full IR-quantity expansion.
%}

The ultraviolet divergences are now in exactly the same form as the infrared divergences. Hence, they fit together into a single diagram:
\begin{equation}\label{eq:puzzlepiecesfit}
    \left[\irqe{\feynalignscaled{0.75}{\selfenergyfourIRdot{dot,minimum size=2mm}{}}} \right]_{\substack{\text{ir}\\\text{divs.}}}^{\substack{\text{uv}\\\text{divs.}}}.
\end{equation}
The tabs and indentations of the pieces of Eq.~\eqref{eq:weirdmonstrocity} have now been fit together.

All that remains is to show that the combination indeed adds to zero:
\begin{equation}\label{eq:divpartei}
    \left[\irqe{\feynalignscaled{0.75}{\selfenergyfourIRdot{dot,minimum size=2mm}{}}} \right]_{\substack{\text{ir}\\\text{divs.}}}^{\substack{\text{uv}\\\text{divs.}}}
    %=\irqe{\feynalignscaled{0.75}{\selfenergyfourIRdot{dot,minimum size=2mm}{}}}
    =0\,.
\end{equation}
%We have used in both equalities the fact that the corresponding integrals are scale free (\textit{cf}.\ Eq.~\eqref{eq:itsscalefree}). Firstly, the scale-free integrals in the middle are simply the sum of their divergences on the left. Secondly, the scale-free integrals in the middle vanish, resulting to the zero on the right.
This follows directly from the fact that the integral in the brackets is scale free: The ultraviolet divergences and the infrared divergences of a scale-free diagram add to zero.

The intermediate divergences cancel! Thus, the UV-piece provides the correct counterterm for the EFT contribution. This is shown generally on the level of the full EFT construction in Appendix~\ref{app:IRdivs}.

Finally, we want to draw our attention to the scale-free diagram in Eq.~\eqref{eq:divpartei}:
\begin{equation}\label{eq:dividerdiag}
    \irqe{\feynalignscaled{0.75}{\selfenergyfourIRdot{dot,minimum size=2mm}{}}}\,.
\end{equation}
It is, in some sense, in between the UV and IR pieces of Eq.~\eqref{eq:weirdmonstrocity}: The heavy-field loop, that is manifest in the UV piece of the diagram in Eq.~\eqref{eq:irdivuvexp}, turns into the UV vertex, when the momenta in the light-field propagators is lowered to $p^2\ll M^2$. Similarly, raising the loop-momentum to $p^2\gg m^2$ is the IR-scale piece, the light-field propagators can be expanded in the light masses (the IR-quantity expansion). Thus, the scale-free diagram appears at the factorization scale, $m^2\ll p^2\sim\Lambda\ll M^2$, in both the UV and IR piece.

One can think that the scale-free diagram acts as the joint at the factorization scale, $\Lambda$, between the UV and IR pieces. Momenta below $\Lambda$ in the UV piece and momenta above $\Lambda$ in the IR piece lead to divergences, which cancel,
\begin{equation}
    \left[\irqe{\feynalignscaled{0.75}{\selfenergyfourIRdot{dot,minimum size=2mm}{}}} \right]_{\substack{\text{ir}\\\text{divs.}}}^{\substack{\text{uv}\\\text{divs.}}}=0\,.
\end{equation}

%The diagram works as the point, at which the two scales are joined together, with the IR scale below and the UV scale above. In this sense it is the same as the factorization scale, $\Lambda$ (Fig.~\ref{fig:IRUV}).

%However, the EFT cannot know anything about the physics of the UV scale (which is similar to the Standard Model not knowing the physics of e.g.\ the Planck scale). Trying to go beyond the dividing diagram in Eq.~\eqref{eq:dividerdiag}, the EFT results in its ultraviolet divergences (Eq.~\eqref{eq:uvdivpartjoo}):
%\begin{equation}
%    \left[\irqe{\feynalignscaled{0.75}{\selfenergyfourIRdot{dot,m%inimum size=2mm}{}}} \right]^{\substack{\text{uv}\\\text{divs.}}}.
%\end{equation}
%Therefore, the EFT requires regularization in the form of counterterms.

%The counterterms come from the UV-scale integrals, which do not know about the IR scale. Consequently, they go below the dividing diagram (Eq.~\eqref{eq:puzzlepiecesfit}):
%\begin{equation}
%    \left[\irqe{\feynalignscaled{0.75}{\selfenergyfourIRdot{dot,minimum size=2mm}{}}} \right]^{\substack{\text{uv}\\\text{divs.}}}_{\substack{\text{ir}\\\text{divs.}}}=0\,.
%\end{equation}

\subsection{The $\order{\epsilon}$ terms of the UV scale}\label{sec:subsecepsilon}

We are now ready to make a remark, which will be the second -- and final difference between the 1PI effective action for the light fields, i.e. the 1LPI action, and the effective action of Eq.~\eqref{eq:effact}. In a 1PI effective action, all of the fluctuations have been integrated over. As a consequence, there are no divergences left, and one can take the physical limit in the dimensional modifier: $\epsilon\to0$. In the effective action, only the UV scale has been integrated over. Hence, there may still be $\epsilon$ poles present, and the limit cannot yet be taken.

In the effective action, the above leads to the fact that one cannot simply drop the UV-scale contributions that are suppressed by $\epsilon$. Those contributions can still multiply with the divergences of the EFT to produce contributions to physical quantities that are of the physical order $\order{\epsilon^0}$.
\begin{align}\tag*{2}
    \fbox{
        \begin{minipage}{7.5cm}
            \vspace{0.1cm}
            The $\order{\epsilon}$ terms of the UV-scale contributions must be kept if they can multiply with the $\epsilon$ poles of the EFT diagrams to become at least of the physical order of $\epsilon^0$.
            \vspace{0.1cm}
        \end{minipage}
    }
\end{align}
To clarify, we use $\order{\epsilon}$ term here to include all powers of $\epsilon$ that are $\epsilon^n$, $n\geq 1$.
%The only reason these terms are special in comparison to $\epsilon^n$, $n< 1$ is that they vanish in the physical limit of $\epsilon\to0$. Thus, they need an $\epsilon$ pole to render their effects relevant regarding physical quantities.

Before examining the matter further, we want to note that this brings closure to our quest of finding the effective action of Eq.~\eqref{eq:effact}. We know, which diagrams to evaluate for the effective action: the 1LPI diagrams. These are expanded in the IR-quantities,
\begin{equation}
    \frac{m^2}{M^2}\ll1\,,\quad\frac{k^2}{M^2}\ll1\,,
\end{equation}
to remove the IR-scale contributions. Due to the choice of dimensional regularization, they are also expanded in the dimensional modifier, $\epsilon$. All of these expansions are then truncated to get a desired accuracy for physical quantities.

A nice example for the need of $\order{\epsilon}$ terms can be found from Eqs.~\eqref{eq:EFTonloopExfourdUVpart}, \eqref{eq:EFTonloopExfourd}. The EFT diagram in Eq.~\eqref{eq:EFTonloopExfourd},
\begin{equation}\label{eq:diagonelooprepexforeps}
    \feynalignscaled{0.6}{\oneloopbubble{}}\;,
\end{equation}
contains a divergence of the form
\begin{equation}
    \frac{\meff^4}{\epsilon}
\end{equation}
in four dimensions. In this example, the mass in the EFT receives a UV correction from the diagram in Eq.~\eqref{eq:EFTonloopExfourdUVpart}:
\begin{equation}
    \irqe{\feynalignscaled{0.75}{\selfenergyfourdot{}{dashed}}} = a+b\,\epsilon+\order{\epsilon^2}\,.
\end{equation}
The mass must be known to the order of $\epsilon$,
\begin{equation}
    \meff^2 = m^2+a+b\,\epsilon+\order{\epsilon^2}\,,
\end{equation}
so that the diagram in Eq.~\eqref{eq:diagonelooprepexforeps} can be known to the physical order $\order{\epsilon^0}$:
\begin{align}
    \frac{\meff^4}{\epsilon}=\frac{\qty(m^2+a)^2}{\epsilon}+2\qty(m^2+a)b+\order{\epsilon}\,.
\end{align}

This may make the need for $\order{\epsilon}$ terms seem abundant. However, it is not in high-temperature computations. None of the thermal examples in this article need them, nor a state-of-the-art computation on a first-order phase transition in Ref.~\cite{Gould:2021dzl}. In high-temperature calculation, the terms were needed for the first time in the computation of the QCD pressure to $\order{g^6\ln g\,T^4}$~\cite{Kajantie:2002wa}, which used the matching procedure.

This somewhat uglier side of EFTs is suppressed in high-temperature EFTs due to the scarcity of $\epsilon$ poles. Dimensional regularization is very merciful in three dimensions, and the high-temperature EFTs are three dimensional (\textit{cf}. Example 1 Sec.~\ref{sec:scalepitI}). On the one-loop order in three dimensions, there are no $\epsilon$ poles -- hence no need for the $\order{\epsilon}$ terms. On higher order, it is more subtle that one does not immediately need these terms.
%This will be discussed in Sec.~\ref{sec:epshight} and also in the second example in Sec.~\ref{sec:exampleII}.

One may be a bit baffled that there are $\order{\epsilon}$ terms in the effective action. However, these are just simply terms that appear when one integrates out a scale (or does the matching via Green's functions), and that need to be kept along in a perturbative computation.

A greater question may also arise: Should we have $\order{\epsilon}$ terms also in the action for SMEFT, for example? The answer here is no! All of the physical effects of the $\order{\epsilon}$ terms can be reproduced by an equivalent effective action that does not contain any $\order{\epsilon}$ terms. The general algorithm to finding this action is shown in Sec.~\ref{sec:findingproper}. However, this is basically never necessary, but merely a curiosity.

%In Sec.~\ref{sec:ordereps}, there is more discussion on the $\oeps$ terms.

Finally, we note that the running of a mass parameter in a simple theory is computed in Appendix~\ref{app:rengrouprunwithprotoefts} using the idea of action containing $\order{\epsilon}$ terms. This may help placing the matter into a wider context. Also, there is more discussion on the $\oeps$ terms in Sec.~\ref{sec:ordereps}.

%%%%%%%%%%%%%%%%%%%%%%%%%%%%%%%%%%%%%
%%%%%%%%%%%% SECTION III %%%%%%%%%%%% 
%%%%%%%%%%%%%%%%%%%%%%%%%%%%%%%%%%%%%

\section{More on $\order{\epsilon}$ terms}\label{sec:ordereps}

In this section, we will continue with a more detailed discussion of the $\oeps$ terms needed in the EFT. We begin by giving a very transparent example, where one needs to take into account these terms, in Sec.~\ref{sec:exatzerotempeps}. Then, we move on to a more analytical discussion for the need and reproduce the rule in Box~\ref{diff:two} in Sec.~\ref{sec:analundereps}. Finally, we note that there is a general method for removing the $\oeps$ terms from the action by modifying the coefficients of the action, Sec.~\ref{sec:findingproper}. In Sec.~\ref{sec:anothExEFTprop}, we compute another example for this.

Note that in these last two sections we will use non-standard terminology. There, we will refer to an EFT containing $\oeps$ terms as a proto-EFT, and to the corresponding EFT without the $\oeps$ terms as the EFT proper.

\subsection{An example at zero temperature}\label{sec:exatzerotempeps}

%\jh{A simple example to build upon}

Here, we want to give a simple and transparent example on the need for the $\oeps$ terms.
%The aim here is to understand how the need for $\order{\epsilon}$ terms arise.
Therefore, we will make an example computation in a case, where both direct loop expansion and EFT are valid. We can easily gain firm ground by the loop expansion to study the EFT construction.

We also want to bring the need of $\order{\epsilon}$ terms to low orders in perturbation theory. This keeps computations as clean as possible. For this end, we will perform a computation for the cosmological constant at zero temperature to two-loop order.

The model is given by a Lagrangian,
\begin{align}
    \mathscr{L}=&\frac{1}{2}(\partial_\mu\Phi)^2+\frac{m^2+\cou{m^2}}{2}\Phi^2\nonumber\\
    &+\frac{1}{2}(\partial_\mu\Chi)^2+\frac{M^2+\cou{M^2}}{2}\Chi^2\nonumber\\
    &+\frac{\lambda}{4}\Phi^2\Chi^2+\Omega+\cou{\Omega}\,,
\end{align}
containing two fields $\Phi$ and $\Chi$.
The scalings of the constants are given by
\begin{equation}\label{eq:vacexscale}
    \lambda M^2\ll m^2 \ll M^2
\end{equation}
to make both the EFT for the $\Phi$ field and the direct loop expansion valid. The former comes from the mass hierarchy ($m^2 \ll M^2$) and the latter follows from no need to resum mass contributions from the scale $M$ ($\lambda M^2\ll m^2$).

From finite temperature perspective, the cosmological constant, $\Lambda$, can be understood as the zero-temperature limit of the free energy density,
\begin{equation}
    \Lambda=\lim_{T\to0}-\frac{T}{V}\ln Z\,,
\end{equation}
where $Z$ is the Euclidean partition function of the theory.
In the loop expansion, this is given by
\begin{align}\label{eq:loopcosmconst}
    \Lambda=&\Omega+\cou{\Omega}
    +\;\feynalignscaled{0.5}{\oneloopbubble{dashed}}\; +\;\feynalignscaled{0.5}{\oneloopbubble{}}\nonumber\\
    &+\;\feynalignscaled{0.5}{\twoloopbubbleC}\;+\;  \feynalignscaled{0.5}{\cammassinsertG{dot,minimum size=2.66mm}} \;+\;\feynalignscaled{0.5}{\cammassinsertB{dot,minimum size=2.66mm}}\,,
\end{align}
where solid lines represent $\Phi$ propagators, dashed lines $\Chi$ propagators, and the dots represent mass counterterms.

We will be especially interested how the EFT calculation handles the first term on the second line.

One could do the computation with explicit results for the integrals. This would make the notation more cluttered than it needs to be. Hence, we will use
\begin{align}
    \int_p\ln(p^2+m^2)&=\frac{a_1}{\epsilon}+b_1+\order{\epsilon}\,,\label{eq:onelepsint}\\
    \int_p\frac{1}{p^2+m^2}&=\frac{a_2}{\epsilon}+b_2+c_2\,\epsilon+\order{\epsilon^2}\,. \label{eq:epsint}
\end{align}
The integrals for the heavier field are denoted by swapping the parameters: ($m^2\to M^2$) $a\to A$, $b\to B$ and $c\to C$. The formulas above are enough to yield the cosmological constant in Eq.~\eqref{eq:loopcosmconst} to the physical accuracy of $\order{\epsilon^0}$.

The counterterms can be given in the $\MSbar$ scheme as
\begin{align}
    \cou{m^2}=-\lambda\frac{A_2}{2\epsilon}\,, \quad\cou{M^2}=-\lambda\frac{a_2}{2\epsilon}\,,\label{eq:vacmasscounts}\\
    \cou{\Omega}=-\frac{A_1}{2\epsilon}-\frac{a_1}{2\epsilon} -\lambda\frac{a_2A_2}{4\epsilon^2}\;\;\;\;\label{eq:vacvaccount}
\end{align}
in terms of the integrals above.

For the cosmological constant, Eq.~\eqref{eq:loopcosmconst}, we obtain
\begin{equation}\label{eq:cosmconstres}
    \Lambda=\Omega+\frac{B_1+b_1}{2}-\lambda\frac{b_2B_2}{4}\,.
\end{equation}

Let us now turn to integrating out the scale $M$.

Notable for integrating out a scale from above is that
%the $\order{\epsilon}$ term from the loop integral in Eq.~\eqref{eq:epsint} has an $\order{\epsilon^0}$ effect at intermediate states in the computation. 
all of the diagrams on the second line of Eq.~\eqref{eq:loopcosmconst} carry dependence on either $C_2$ or $c_2$ at the order of $\epsilon^0$. 
This is due to the $\epsilon$ poles.
%of the one-loop integrals.
%It has to cancel from the final result, shown in Eq.~\eqref{eq:cosmconstres}.
%This kind of divergences were not present in the thermal computation of Sec.~\ref{sec:exampleI} (\textit{cf.} Eqs.~\eqref{eq:oneloopthermcomp}, \eqref{eq:lightfree}).
The dependence on $C_2$ and $c_2$ cancels eventually from the cosmological constant, Eq.~\eqref{eq:cosmconstres}, but this is not know by the procedure of integrating out a scale.
%Hence, one has to be careful in the computation.

%For the unit term, we would in principle have all of the diagrams in Eq.~\eqref{eq:loopcosmconst} to compute.
%All of the diagrams in Eq.~\eqref{eq:loopcosmconst} are 1PI diagrams. In principle, we would have to compute all of them for the unit term of the effective action.
%Due to the IR-quantity expansion in Box~\ref{diff:one}, only
In the EFT construction we only need to compute
\begin{align}\label{eq:vacUV}
    \;\irqe{\feynalignscaled{0.6}{\oneloopbubble{dashed}}}\;
    +\;  \irqe{\feynalignscaled{0.6}{\cammassinsertG{dot,minimum size=2.66mm}}}
\end{align}
%remain non-zero.
for the cosmological constant.

The only other contribution from the scale $M$ at this order is a mass term:
\begin{equation}\label{eq:vacmassUV}
    \irqe{\feynalignscaled{0.6}{\selfenergyfour{}{dashed}}} =-\frac{\lambda}{4}\qty(\frac{A_2}{\epsilon}+B_2+C_2\,\epsilon)\Phi^2+\order{\epsilon^2}\,.
\end{equation}
Here, we have kept the $\order{\epsilon}$ term. The reason for this is that the mass parameter for the scale $m$ will mix together with the divergence of the one-loop diagram.
% in Eq.~\eqref{eq:onelepsint}.

%There is also another way of understanding this, which leans directly on the loop expansion: As discussed in \jh{diagrammatic understanding section}, the computation of the above mass correction can be understood as computing the heavy loop of the $\feynalignscaled{0.3}{\twoloopbubbleC}$ diagram in Eq.~\eqref{eq:loopcosmconst} before computing the full diagram within the EFT. In loop expansion, one needs to account for the $\order{\epsilon}$ term in the integral corresponding to the integral in Eq.~\eqref{eq:vacmassUV}. Correspondingly, one needs the $\order{\epsilon}$ term here as well.

%Here, we want to note that this was not necessary at our high temperature examples. Neither of the one-loop diagrams in Eq.~\eqref{eq:oneloopthermcomp} contained an $\epsilon$ pole, nor did the three-dimensional one-loop diagram in Eq.~\eqref{eq:lightfree}. Thus, an $\order{\epsilon}$ term wouldn't affect anything. Below in Sec.~\ref{sec:epshight}, we will see that this feature is much more ubiquitous in the three-dimensional effective descriptions.

We have now our effective parameters:
\begin{align}
    \Omega_{\text{eff}}&=\Omega+\frac{B_1}{2}+\lambda\frac{a_2C_2}{4}+\order{\epsilon}\,,\label{eq:cosmconeff}\\
    \cou{\Omega_{\text{eff}}}&=-\frac{a_1}{2\epsilon}+\lambda\frac{a_2B_2}{4\epsilon}\,,\\
    m^2_{\text{eff}}&=m^2-\frac{\lambda}{2}\qty(B_2+C_2\,\epsilon) +\order{\epsilon^2}\,,\label{eq:vacresmas}
\end{align}
where the first two lines line follow from Eqs.~\eqref{eq:vacvaccount} and \eqref{eq:vacUV}, and the last from Eq.~\eqref{eq:vacmassUV}. Note, that there is no mass counterterm.

The resulting effective Lagrangian is
\begin{equation}
    \mathscr{L}_{\text{eff}}=\frac{1}{2}(\partial_\mu\Phi_{\text{IR}})^2+\frac{m^2_{\text{eff}}}{2}\Phi_{\text{IR}}^2 +\Omega_{\text{eff}}+\cou{\Omega_{\text{eff}}}\,.
\end{equation}

Note, that there is now an $\oeps$ term in the effective Lagrangian.
%The effective description is not an EFT proper but a proto-EFT. The $\order{\epsilon}$ term in the effective mass is unusual dependence on the regulator. This is not accepted in an EFT proper. The key point is that
It is unusual, but it is not a problem: We have merely integrated over the length-scale $M^{-1}$ which resulted in the above effective Lagrangian. It can be used to compute the physics of the length scale $m^{-1}$.
%physical correlation functions on the length-scale of $m^{-1}$ and thermodynamic quantities.

%Below, we will see how one can get rid of the $\order{\epsilon}$ term if one wants to switch the regularization scheme.

Now, we can compute the same cosmological constant within the EFT:
\begin{align}\label{eq:effcosmconst}
    \Lambda=\Omega_{\text{eff}}+\cou{\Omega_{\text{eff}}} +\;\feynalignscaled{0.5}{\oneloopbubble{}}\,.
\end{align}
%Usually, the point of using EFTs is that it resums higher scale contributions. The scale $M$ mass contributions are resummed into the effective mass in Eq.~\eqref{eq:vacresmas}. Here, we have however chosen the scalings (Eq.~\eqref{eq:vacexscale}) so that the resummations are unnecessary. Due to the scalings, we can expand the one-loop diagram:
To obtain manifestly the same result as from the direct loop expansion, we expand the effective mass, $m_{\text{eff}}^2$:
\begin{align}\label{eq:whataline}
    &\feynalignscaled{0.6}{\oneloopbubble{}} =\frac{1}{2}\int_p\ln(p^2+m^2_{\text{eff}})\nonumber\\
    =&\frac{1}{2}\int_p\ln(p^2+m^2) -\frac{\lambda\qty(B_2+C_2\,\epsilon)}{4}\int_p\frac{1}{p^2+m^2}+\order{\lambda^2} \nonumber\\
    =&\frac{a_1}{2\epsilon}+\frac{b_1}{2}-\lambda\frac{a_2B_2}{4\epsilon}-\lambda\frac{b_2B_2+a_2C_2}{4}+\order{\epsilon}\,.
\end{align}

Finally, we can put together the cosmological constant:
\begin{equation}
    \Lambda=\Omega+\frac{B_1+b_1}{2}-\lambda\frac{b_2B_2}{4}\,.
\end{equation}
It is exactly the same as above, as it should be. The $\epsilon$ adventure has reached its goal safely.

%\jh{Note the second line!}

We want to highlight one particular fact from this example: The second term on the second line of Eq.~\eqref{eq:whataline},
\begin{equation}
    -\frac{\lambda\qty(B_2+C_2\,\epsilon)}{4}\int_p\frac{1}{p^2+m^2}\,,
\end{equation}
corresponds to the two-loop diagram in Eq.~\eqref{eq:loopcosmconst},
\begin{equation}
    \feynalignscaled{0.6}{\twoloopbubbleC}\,,
\end{equation}
(without the heavy-loop divergence). In the EFT construction, we computed the heavy-field loop, Eq.~\eqref{eq:vacmassUV}, and only after the construction the light-field loop, Eq.~\eqref{eq:whataline}. Consequently, we need to retain the $\oeps$ term of Eq.~\eqref{eq:vacmassUV} in the EFT construction.

%\subsection{High-temperature setting}\label{sec:epshight}

%\jh{Position changed!!!}

%\begin{itemize}
%    \item Dimensional regularization merciful
%    \item Thermal scale leading comes from two loop high
%    \item Note about SMEFT holding without $\order{\epsilon}$ if new physics UV for EWPT
%    \item Below thermal scale EFT superrenormalizable
%    \item Sunset avoids mass $\epsilon$
%\end{itemize}

\subsection{Analytical understanding on the $\order{\epsilon}$ terms}\label{sec:analundereps}

In Secs.~\ref{sec:method} and \ref{sec:methodlong}, we have stated the following: There can be $\epsilon$ poles in the EFT due  to ultraviolet divergences. As a consequence, one has to keep the $\order{\epsilon}$ terms from the UV-scale contributions, that affect the physical quantities to a desired order. (See Box~\ref{diff:two}.) Here, we will
%be more analytical about
study this statement in more detail, and show that it is indeed true.

The above statement is in some sense framed in terms of the EFT: We need to have the $\oeps$ terms in the EFT action, so that they give contributions to physical quantities by multiplying the ultraviolet divergences of the EFT. However, we need to look at the computation as a whole to actually prove the statement.

The origin for the need for the $\oeps$ terms comes from the intermediate divergences of the EFT calculation, Fig.~\ref{fig:regulation}, as we shall soon see. The intermediate divergences cancel among themselves. (Shown in Appendix~\ref{app:IRdivs}.) However, this is not enough. The divergences are not initially present, but induced by the EFT calculation. Consequently, \textit{all} of their artefacts must cancel from physical quantities, not just the divergences themselves.

Let us look at an illustrative example to get a better handle. We will use the diagrammatic illustration discussed in Sec.~\ref{sec:diagunderstanding}. However, we try to keep this section self-contained so that it is not obligatory to read Sec.~\ref{sec:diagunderstanding}.
%A computational example is then given in Sec.~\ref{sec:exatzerotempeps}.

We will use a diagram, whose intermediate divergences were already discussed in Sec.~\ref{sec:diagunderstanding}, Eq.~\eqref{eq:irdivsexdiag}:
\begin{align}
    \feynalignscaled{0.75}{\sunsetIR{}{dashed}}\;,
\end{align}
where the dashed line corresponds to a heavy field and the solid line to a light field.

The diagram can be separated in scales as
\begin{align}\label{eq:lolScaleSep}
    \feynalignscaled{0.6}{\sunsetIR{}{dashed}}\;
    =\;
    \feynalignscaled{0.75}{\selfenergytree{plain}{dot,minimum size=2mm}}\;
    +\;\feynalignscaled{0.75}{\selfenergyfourIRdot{dot,minimum size=2mm}{}}\;.
\end{align}
Here, the dots represent UV-scale loops that are shrunk to points. The remaining explicit loop on the right-hand side is an IR loop with momentum $p^2\sim m^2$.

In the EFT computation, we first compute the dots: The first one is given by
\begin{align}\label{eq:lolUV}
    \feynalignscaled{0.75}{\selfenergytree{plain}{dot,minimum size=2mm}}\; =\;\irqe{\feynalignscaled{0.6}{\sunsetIR{}{dashed}}}\,,
\end{align}
and the second by
\begin{align}\label{eq:lolUV2}
    \feynalignscaled{0.75}{\selfenergytree{plain}{dot,minimum size=2mm}}\; =\;\irqe{\feynalignscaled{0.6}{\irdivsex}}\,.
\end{align}
Then, we compute the remaining IR loop in the EFT:
\begin{equation}\label{eq:lolEFT}
    \EFT{\feynalignscaled{0.75}{\selfenergyfourdot{}{}}}\,.
\end{equation}

In the following, we will show, that one must keep the order $\epsilon$ term of the diagram in Eq.~\eqref{eq:lolUV2}. This ensures that the scale separation in Eq.~\eqref{eq:lolScaleSep} is done correctly.

Note importantly, that the EFT computation induces divergences: an infrared divergence to the UV contribution in Eq.~\eqref{eq:lolUV}, and an ultraviolet divergence to the EFT contribution in Eq.~\eqref{eq:lolEFT}.
%This can be depicted as
%Eq.~\eqref{eq:lolScaleSep}.

The induced intermediate divergences cancel, as shown in Sec.~\ref{sec:diagunderstanding}. This is a must because the divergences were not present on the left-hand side of Eqs.~\eqref{eq:lolScaleSep}.

Are there any other possible artefacts, that we should be mindful about, besides the divergences? Yes, there are. Let us now find these.

The infrared divergence in the UV contribution of Eq.~\eqref{eq:lolUV} comes from the momentum region of $p^2\ll M^2$ in the light-field propagators.
The infrared divergence can be represented diagrammatically as
\begin{equation}~\label{eq:exofirdivrep}
    \left[\irqe{\feynalignscaled{0.6}{\sunsetIR{}{dashed}}} \right]_{\substack{\text{ir}\\\text{divs.}}}
    =
    \left[\irqe{\feynalignscaled{0.75}{\selfenergyfourIRdot{dot,minimum size=2mm}{}}} \right]_{\substack{\text{ir}\\\text{divs.}}}.
\end{equation}
Here, the dot is exactly given by Eq.~\eqref{eq:lolUV2}. This is due to the scale separation between the infrared-diverging light-field propagators with momenta $p^2\ll M^2$ and the UV-scale heavy-field loop.

Now, we finally get to the $\oeps$ terms. As we can see from Eq.~\eqref{eq:exofirdivrep}, the infrared divergence, $\epsilon^{-1}$, \textit{promotes} the order $\epsilon$ term in the dot, Eq.~\eqref{eq:lolUV2}, to become of order $\epsilon^0$. Thus, the promoted contribution is computed in the EFT construction, Eq.~\eqref{eq:lolUV}, along with the order $\epsilon^0$. This is clearly an artefact from the infrared divergence, that is induced by the EFT computation. Hence, it must cancel from physical quantities.

The cancellation mechanism is actually already in place. We simply need to take the dot in Eq.~\eqref{eq:lolUV2} seriously, and keep its $\oeps$ terms.

In the EFT diagram, there is an equal but opposite $\epsilon$ pole from the ultraviolet divergence, $-\epsilon^{-1}$. This cancels against the infrared divergence. It also promotes the order $\epsilon$ term from the dot of Eq.~\eqref{eq:lolUV2} to order $\epsilon^0$. This cancels with the aforementioned artefact, the promotion by the infrared divergence.

As already mentioned in Secs.~\ref{sec:method} and \ref{sec:separatingIRUV}, for each infrared divergence from the UV scale, there is an equal but opposite ultraviolet divergence from the EFT. (This is shown in Appendix~\ref{app:IRdivs} along the lines of Ref.~\cite{Manohar:2018aog}.) The divergences cancel identically. Correspondingly, for each promotion done by the infrared divergences, there is an equal but opposite promotion done by the ultraviolet divergences -- as long as we retain the $\oeps$ terms in the EFT construction.

Finally, we note that not every $\oeps$ term needs to be retained in a practical computation, only the ones affecting physical quantities to a desired order.

%\color{black}

%\vspace{3cm}

\subsection{Finding an EFT proper}\label{sec:findingproper}

Here, we will discover a general method for finding an effective action, which is physically equivalent to the effective action obtained by integrating out a scale but does not contain $\oeps$ terms. We begin by examining the example in Sec.~\ref{sec:exatzerotempeps} and find the equivalent action. Then, we discuss, why it is always possible to find this equivalent action. Finally, we show the general method using the example for illustration.

Note that we will refer to the effective description with $\oeps$ terms as a proto-EFT and to the equivalent effective description without the $\oeps$ terms as an EFT proper.

%\jh{State the context...}
%\jh{State the goal!}

%\jh{First we will find the EFT proper by brute force. Then, we show that this can be done generally, Finally, we will discover a general method.}

%\jh{Continuing with the example:}

Let us start by analyzing the example in Sec.~\ref{sec:exatzerotempeps} further. We begin by introducing a split of the mass parameter in the orders of $\epsilon$,
\begin{align}
    m^2_{\text{eff}}&=\tilde{m}^2+\Delta m^2\,\epsilon\,,\label{eq:masssplit}%\\
    %\tilde{m}^2&=m^2-\frac{\lambda B_2}{2}\,,\\
    %\Delta m^2&=-\frac{\lambda C_2}{2}\,.
\end{align}
to render the notation more amenable for our purposes.

The cosmological constant contribution due to the $\oeps$ mass term came from the proto-EFT one-loop diagram, Eq.~\eqref{eq:whataline}.
This one-loop diagram can be expanded in $\Delta m^2\,\epsilon$:
\begin{align}
    &\feynalignscaled{0.6}{\oneloopbubble{}} =\frac{1}{2}\int_p\ln(p^2+m^2_{\text{eff}})\nonumber\\
    =&\frac{1}{2}\int_p\ln(p^2+\tilde{m}^2) +\frac{\Delta m^2\,\epsilon}{2}\int_p\frac{1}{p^2+\tilde{m}^2}+\order{\epsilon}\,.\label{eq:UVUVUVUV}
\end{align}
%The last term leads to the cosmological constant contribution from the $\oeps$ term, $\Delta m^2\,\epsilon$.
The cosmological constant contribution from the $\oeps$ term, $\Delta m^2\,\epsilon$ is now the last term.
Let us compute it:
\begin{align}\label{eq:firstfoundepscont}
    \frac{\Delta m^2\,\epsilon}{2}\int_p\frac{1}{p^2+\tilde{m}^2} =-\lambda\frac{a_2C_2}{4}+\order{\epsilon}\,.
\end{align}

We now make an act that may initially seem bold. We reorganize the Lagrangian parameters in the following way: We add the cosmological constant contribution in Eq.~\eqref{eq:firstfoundepscont} to the unit term and remove the $\oeps$ term from the mass parameter,
\begin{align}
    \Omega_{\text{EFF}}&=\Omega_{\text{eff}}-\lambda\frac{a_2C_2}{4}\,,\label{eq:cosmconEFT}\\
    m^2_{\text{EFF}}&=m^2_{\text{eff}}-\Delta m^2\,\epsilon\,.\label{eq:EFTcoefflast}
\end{align}
%the UV contributions
%by moving the $\order{\epsilon}$ effects from the mass parameter to the unit term:
%Thus, we obtain:
%\begin{align}
%    \Omega_{\text{EFF}}=&\Omega+\frac{B_1}{2}\,,\label{eq:cosmconEFT}\\
%    \cou{\Omega_{\text{EFF}}}=&-\frac{a_1}{2\epsilon}+\lambda\frac{a_2B_2}{4\epsilon}\,,\\
%    m^2_{\text{EFF}}&=m^2-\frac{\lambda B_2}{2}\,.\label{eq:EFTcoefflast}
%\end{align}

On a closer inspection, this is not bold at all. Both of the physical quantities, which can be computed from the effective description, i.e. the cosmological constant and the $\Phi$-pole mass, remain the same:
\begin{align}
    \Lambda&=\Omega+\frac{B_1+b_1}{2}-\lambda\frac{b_2B_2}{4}\,,\\
    m^2_{\text{pole}}&=m^2-\frac{\lambda B_2}{2}\,.
\end{align}
Thus, the physics has not changed at all. We have a physically equivalent effective description, but without the $\oeps$ term.
%It is indeed only a reshuffle of the UV contributions.

It is important to note, that the reorganization of the effective Lagrangian did not involve any of the infrared physics. The $\oeps$ term removed from the mass parameter came as a UV-scale contribution, Eq.~\eqref{eq:vacmassUV}. The contribution added to the unit term was the $\oeps$ term multiplied with an ultraviolet divergence, Eq.~\eqref{eq:firstfoundepscont}.

The above holds generally: $\order{\epsilon}$ terms from the UV scale have to be multiplied with the $\epsilon$ poles of ultraviolet divergences to make a physical $\order{\epsilon^0}$ effect. This gives a hint that we should be able to reshuffle ultraviolet contributions to obtain an equivalent effective action without $\oeps$ terms. At the moment however, this could very well be just an artefact caused by our simple example.
%Could there be infrared physics irreducibly involved, so that the construction of the equivalent effective action is impossible? The answer is no. It is always possible. 

%\begin{itemize}
%    \item This is always so: a hint
%    \item At the moment... Could there be infrared physics irreducibly involved
%\end{itemize}

%{
%Firstly, it is important to note that both the $\order{\epsilon}$ term and the $\epsilon$ pole, which together make a physical $\order{\epsilon^0}$ effect, are UV contributions. The $\order{\epsilon}$ term follows from integrating over the UV scale and the pole is an ultraviolet divergence.
%This is a hint for us: Maybe there is a way to reorganize these UV contributions so that the new Lagrangian yields the same IR physics but contains no $\order{\epsilon}$ terms.
%}

%The procedure above relies on the simplicity of the situation. For a general case, there remains two questions: Can one always do the reshuffle to find an EFT proper? Is there a simple, general procedure to execute the reshuffle? Yes and yes. At the end of this section, we will provide a simple method, which requires only the inspection of the counterterms of the proto-EFT.

%Next, we will discuss, why it is always possible to reorganize the effective action, so that there are no $\oeps$ terms. Then, we will cover the general method to do this reorganization.

A powerful way of approaching the generality of the situation is to view the EFT computation as a whole. The same approach was taken in Sec.~\ref{sec:analundereps}. In the section, we found the reason for keeping the $\oeps$ terms in a proto-EFT: The infrared divergences of the UV-scale contributions \textit{promote} $\oeps$ terms in powers of $\epsilon$. The $\oeps$ terms in the proto-EFT cancel these artificial promotions.

The cancellation takes place at the loop computation of the proto-EFT, where we obtain the ultraviolet divergences. However, we can conduct the cancellation even before the loop computation. If we remove the artificially promoted contributions already from the UV-scale contributions, we can also remove the $\oeps$ terms from the EFT. This is because their only function was to cancel the promoted contributions. 

In Lagrangian terms, this means changes in the coefficients: subtracting the promoted contributions out and dropping the $\oeps$ terms. Note, that this yields the corresponding EFT proper.

%In the example given in Sec.~\ref{sec:analundereps}, the infrared divergence in the diagram of Eq.~\eqref{eq:lolUV},
%\begin{align}\label{eq:lolUVrep}
%    \feynalignscaled{0.75}{\selfenergytree{plain}{dot,minimum size=2mm}}\; =\;\irqe{\feynalignscaled{0.6}{\sunsetIR{}{dashed}}}\,%,
%\end{align}
%promotes the $\oeps$ term in the diagram of Eq.~\eqref{eq:lolUV2},
%\begin{align}\label{eq:lolUV2rep}
%    \feynalignscaled{0.75}{\selfenergytree{plain}{dot,minimum size=2mm}}\; =\;\irqe{\feynalignscaled{0.6}{\irdivsex}}\,,
%\end{align}
%to be of order $\epsilon^0$. The promoted contribution is then cancelled by the proto-EFT $\oeps$ term, which came from the UV diagram in Eq.~\eqref{eq:lolUV2rep}.
%divergence of the EFT diagram in Eq.~\eqref{eq:lolEFT}, which contains the promoted $\oeps$ term of the diagram in Eq.~\eqref{eq:lolUV2rep}.

%If we could remove the promoted $\oeps$ term from the mass contribution given by Eq.~\eqref{eq:lolUVrep}, the EFT would not need to remember the $\oeps$ term of the mass contribution in Eq.~\eqref{eq:lolUV2rep}. This is something that we can actually carry out within proto-EFTs. A proto-EFT has the full knowledge of the infrared divergences of the UV-scale; they are the counterterms. The proto-EFT has also the full knowledge on the $\oeps$ terms that were promoted by the infrared divergences.

The removal of the artificially promoted contributions is always possible in principle. In practice however, we need to first find the promoted contributions. As we will see below, this is possible within a proto-EFT: The counterterms have the full knowledge on the infrared divergences of the UV scale. (They are the infrared divergences.) The proto-EFT also contains the $\oeps$ terms that have been promoted by the infrared divergences.

Let us get back to the example used here previously, i.e.\ that of Sec.~\ref{sec:exatzerotempeps}. The ultraviolet divergence from the one-loop,
\begin{align}
    \feynalignscaled{0.6}{\oneloopbubble{}} =\frac{1}{2}\int_p\ln(p^2+m^2_{\text{eff}})\,,
\end{align}
is cancelled by the counterterm:
\begin{align}
    \cou{\Omega_{\text{eff}}} =\frac{\tilde{m}^4}{4(4\pi)^2\epsilon}\,.
\end{align}
In the EFT computation, the counterterm is the result of infrared divergences in integrating out the scale $M$:
\begin{equation}
    \irqe{\feynalignscaled{0.6}{\oneloopbubble{}}}\;+\;\irqe{\feynalignscaled{0.6}{\twoloopbubbleC}}\;+\;\irqe{\feynalignscaled{0.4}{\threeloopbubbleC}}.\label{eq:completeinfrareddivs}
\end{equation}
%In the latter diagram, the infrared divergence promotes the $\oeps$ term,
%\begin{equation}
%    \Delta m^2\,\epsilon\,,
%\end{equation}
%to give an order $\epsilon^0$ effect on $\Omega_{\text{eff}}$.
(The last diagram is included for completeness, even though it is higher order than computed in Sec.~\ref{sec:exatzerotempeps}.)

It is crucial to note, that there is a kind of a ghostly division between the counterterm, $\cou{\Omega_{\text{eff}}}$, and the unit term, $\Omega_{\text{eff}}$. The divergence related to the parameter $\tilde{m}^2$ ends up into the counterterm, and the promoted contribution from $\Delta m^2\,\epsilon$ into the unit term. This is simply due to the contributions being different orders in $\epsilon$.

Consequently, we can recover the promoted contributions by completing the parameter $\tilde{m}^2$ to $\meff^2$ within the $\epsilon$ poles:
\begin{align}
    \cou{\Omega_{\text{eff}}}&=\frac{m_{\text{eff}}^4}{4(4\pi)^2\epsilon}\underbrace{-\frac{\tilde{m}^2\, \Delta m^2}{2(4\pi)}-\frac{(\Delta m^2)^2}{4(4\pi)}\epsilon} \,.
\end{align}
The underbraced terms, found by the completion, cancel the artificially promoted terms from the unit term, $\Omega_{\text{eff}}$.
%Here, the middle and last terms corresponding to the middle and last diagrams in Eq.~\eqref{eq:completeinfrareddivs} respectively.

Now, we have found the terms cancelling the promoted contributions. We can simply move them to the unit operator from the counterterm and ignore the $\oeps$ term in the mass:
\begin{align}
    \Omega_{\text{EFF}}&=\Omega_{\text{eff}}-\frac{\tilde{m}^2\,\Delta m^2}{2(4\pi)}-\frac{(\Delta m^2)^2}{4(4\pi)}\epsilon\,,\\
    \cou{\Omega_{\text{EFF}}} &=\frac{\tilde{m}^4}{4(4\pi)^2\epsilon}\,,\\[3pt]
    m^2_{\text{EFF}}&=\tilde{m}^2\,.
\end{align}
Neglecting the $\oeps$ term is the same as the substitution: $\meff^2\to\tilde{m}^2$.

We can neglect the last term in $\Omega_{\text{EFF}}$ for two reasons: Firstly, it is higher order than desired. Secondly, it is an $\oeps$ term in the unit term. The term only couples to gravity, which is ignored here. Thus, the $\oeps$ term cannot multiply an $\epsilon$ pole to become physical order effect.

In general, one cannot simply neglect these new $\oeps$ terms, since they are on the same footing as the initial $\oeps$ terms: It is a term that cancels an artificial promotion from the initial effective Lagrangian. The artificial promotion could multiply with an ultraviolet divergence to affect physical quantities, if the cancelling term was ignored. In a more complicated computation, this can be important.

One may worry that new $\order{\epsilon}$ terms could keep the algorithm running indefinitely, and no EFT proper would be established. However, the new terms become higher and higher order due to going through counterterms in each iteration. Eventually, they either turn into wanted Lagrangian terms (larger than $\oeps$), end up into the unit term, or become higher order than considered. And the algorithm ends.

There is still one thing to note in a more complicated situation containing multiple Lagrangian parameters with $\oeps$ terms: The above procedure can be first done with the $\oeps$ term of a single Lagrangian coefficient. Then, one can complete this for the next one, until there are no $\oeps$ terms left. 

Another example is computed in the following subsection, Sec.~\ref{sec:anothExEFTprop}.

\subsection{Another example on finding an EFT proper}\label{sec:anothExEFTprop}

In this example, we already start with a proto-EFT at zero temperature:
\begin{align}
    \mathscr{L}_{\text{eff}}=\;&
    \frac{1}{2}(\partial_\mu\Phi)^2+\frac{\meff^2+\cou{\meff^2}}{2}\Phi^2+\frac{\lambda_{\text{eff}}}{4!}\Phi^4\nonumber\\
    &+\Omega_{\text{eff}}+\cou{\Omega_{\text{eff}}}\,,\\[3pt]
    \meff^2=\;&\tilde{m}^2+\Delta m^2\,\epsilon\,,\\
    \cou{\Omega_{\text{eff}}}=\;&\frac{\tilde{m}^4}{4(4\pi)^2\epsilon} +\frac{\lambda_{\text{eff}}\,\tilde{m}^4}{8(4\pi)^4\epsilon^2}\,,\\
    \cou{m^2_{\text{eff}}}=\;&\frac{\lambda_{\text{eff}}\,\tilde{m}^2}{2(4\pi)^2\epsilon}\,.
\end{align}
The example Lagrangian is renormalized up to two loops in the cosmological constant and to one loop in the pole mass.
We assume to have
\begin{equation}
    \tilde{m}^2\sim \Delta m^2\,,
\end{equation}
which is rather natural if the leading order mass of $\tilde{m}^2$ comes in integrating out a UV scale.

We can now complete the counterterms with regarding the $\oeps$ term in the mass parameter, $\meff^2$:
\begin{align}
    \cou{\Omega_{\text{eff}}}^{(1)}&\equiv\frac{\tilde{m}^4}{4(4\pi)^2\epsilon}\\
    &=\frac{\meff^4}{4(4\pi)^2\epsilon}-\frac{\tilde{m}^2\Delta m^2}{2(4\pi)^2}+\order{\epsilon}\,,\\[3pt]
    \cou{\Omega_{\text{eff}}}^{(2)} &\equiv\frac{\lambda_{\text{eff}}\,\tilde{m}^4}{8(4\pi)^4\epsilon^2}\\ &=\frac{\lambda_{\text{eff}}\,\meff^4}{8(4\pi)^4\epsilon^2} -\frac{\lambda_{\text{eff}}\,\meff^2\Delta m^2}{4(4\pi)^4\epsilon} +\frac{\lambda_{\text{eff}}\,\qty(\Delta m^2)^2}{8(4\pi)^4}\,,\label{eq:neqcounterterm}\\
    \cou{m^2_{\text{eff}}}&=\frac{\lambda_{\text{eff}}\,\meff^2}{2(4\pi)^2\epsilon}-\frac{\lambda_{\text{eff}}\,\Delta m^2}{2(4\pi)^2}\,.
\end{align}
Here, we have split the counterterm of the unit term into two for clarity.
Note that there is a new, order $\epsilon^{-1}$ counterterm appearing in Eq.~\eqref{eq:neqcounterterm}.

Now, we can distribute the cancellations of the artificially promoted terms and change $\meff^2\to\tilde{m}^2$:
\begin{align}
    \mathscr{L}_{\text{EFF}}=\;&
    \frac{1}{2}(\partial_\mu\Phi)^2+\frac{m^2_{\text{EFF}}+\cou{m^2_{\text{EFF}}}}{2}\Phi^2+\frac{\lambda_{\text{eff}}}{4!}\Phi^4\nonumber\\
    &+\Omega_{\text{EFF}}+\cou{\Omega_{\text{EFF}}}\,,\\[3pt]
    \Omega_{\text{EFF}}=\;&\Omega_{\text{eff}}-\frac{\tilde{m}^2\Delta m^2}{2(4\pi)^2}+\frac{\lambda_{\text{eff}}\,\qty(\Delta m^2)^2}{8(4\pi)^4}\,,\\
    m^2_{\text{EFF}}=\;&\tilde{m}^2-\frac{\lambda_{\text{eff}}\,\Delta m^2}{2(4\pi)^2}\,,\\
    \cou{\Omega_{\text{EFF}}}=\;&\frac{\tilde{m}^4}{4(4\pi)^2\epsilon} -\frac{\lambda_{\text{eff}}\,\tilde{m}^2\Delta m^2}{4(4\pi)^4\epsilon} +\frac{\lambda_{\text{eff}}\,\tilde{m}^4}{8(4\pi)^4\epsilon^2}\,,\label{eq:locofnewcou}\\
    \cou{m^2_{\text{EFF}}}=\;&\frac{\lambda_{\text{eff}}\,\tilde{m}^2}{2(4\pi)^2\epsilon}\,.
\end{align}

The role of the new counterterm in Eq.~\eqref{eq:locofnewcou} is to complete the one-loop part of the full counterterm regarding the new mass parameter, $m^2_{\text{EFF}}$:
\begin{equation}
    \frac{\tilde{m}^4}{4(4\pi)^2\epsilon}-\frac{\lambda_{\text{eff}}\,\tilde{m}^2\Delta m^2}{4(4\pi)^4\epsilon} =\frac{m_{\text{EFF}}^4}{4(4\pi)^2\epsilon}+\order{\lambda_{\text{eff}}^2}\,.
\end{equation}

With a direct computation, one can indeed confirm that both effective descriptions yield
\begin{align}
    \Lambda&=\Omega_{\text{eff}}-\frac{\tilde{m}^2\Delta m^2}{2(4\pi)^2}+\frac{\lambda_{\text{eff}}\,\qty(\Delta m^2)^2}{8(4\pi)^4}\nonumber\\
    &\quad-\frac{\tilde{m}^4}{8(4\pi)}\qty(3-2\ln(\frac{\tilde{m}^2}{\mu^2}))\nonumber\\
    &\quad+\frac{\lambda_{\text{eff}}\,\tilde{m}^2\Delta m^2}{4(4\pi)^4}\qty(1-\ln(\frac{\tilde{m}^2}{\mu^2}))\nonumber\\
    &\quad-\frac{\lambda_{\text{eff}}\,\tilde{m}^4}{8(4\pi)^4}\qty(1+2\ln(\frac{\tilde{m}^2}{\mu^2})-\ln(\frac{\tilde{m}^2}{\mu^2})^2)\nonumber\\
    &\quad+\order{\lambda_{\text{eff}}^2}\,,\\[3pt]
    m_{\text{pole}}^2&=\tilde{m}^2-\frac{\lambda_{\text{eff}}\,\tilde{m}^2\Delta m^2}{2(4\pi)^2}\nonumber\\
    &\quad-\frac{\lambda_{\text{eff}}\,\tilde{m}^2}{2(4\pi)^2}\qty(\tilde{m}^2-\ln(\frac{\tilde{m}^2}{\mu^2}))
    +\order{\lambda_{\text{eff}}^2}\,.
\end{align}

\section{Mixing between light and heavy fields}\label{sec:mixingIRUV}

Here, we will push the idea of integrating out a scale
%, in some sense, to its limit:
out of the limits of usual EFT discussions:
We will discuss the possibility of integrating out a UV scale even if there is some mixing between the heavy and light fields. This underlines the generality and robustness of the method. Also, it may bring some new insight into EFTs, as the degrees of freedom of the resulting EFT are not manifestly the light-mass eigenfields.

%It might also lead to simplified computations in complicated models. There may be mixing arising due UV scale loops. One can first integrate out the UV scale without dwelling on the mixing due the UV scale. As we shall see from the example below, the effexts of the mixing become explicit in the effective action and can be removed by field redefinitions. In computations, where one is not interested in the light mass eigenstates, such as free energy computation, the mixing does not even need to be removed.

In this section, we will focus on two facts through an example:
%The resulting EFT contains all of the long-length-scale correlations and 
The mixing due to the UV scale can be undone within the EFT via field redefinitions, and the EFT contains correctly the IR scale contributions to the free energy (or the cosmological constant).
%The latter one is not strictly shown but we give general arguments for it.\jh{Maybe something}

%Here, we will show, how field redefinitions within an EFT can be used to remove mixing induced by a UV scale. This may lead to simplified computations in models more complicated than the one used below: One can first integrate out the UV scale without dwelling on mixing and then remove the mixing easily, as it is explicit in the effective action.

%This section also emphasizes the generality and robustness of integrating out a scale. The heavy field, integrated along with the UV scale, does not need to be an exact mass eigenstate for the resulting EFT to be consistent. The EFT still describes the long-length-scale behavior of the system -- even though its degrees of freedom are not manifestly the light-mass eigenstates. This hopefully provides new insight into EFTs.

We use the simplest model containing two scales and mixing:
\begin{align}\label{eq:exLagIII}
    \mathscr{L}_{\mathrm{E}}=&\;
    \frac{1}{2}(\partial_\mu\Phi)^2+\frac{m^2}{2}\Phi^2\nonumber\\
    &+\frac{1}{2}(\partial_\mu\Chi)^2+\frac{M^2}{2}\Chi^2\nonumber\\
    &+\mixm^2\Chi\Phi\,.
\end{align}
For the scale hierarchy, we require
\begin{equation}
    m^2\ll M^2\,.
\end{equation}
We also require
\begin{equation}
    \mixm^2\ll M^2\,,
\end{equation}
so that the mixing does not change the behavior of the UV scale at the leading order but can be accounted for perturbatively.

%The model is obviously appealing as an example due to its simplicity. It is so simple however that one may wonder if it can teach us anything general. The key here is the following: There is no difference for the EFT whether the mass mixing comes from UV-scale loops or the Langrangian mixing term, $\mixm^2\Chi\Phi$. \jh{Mod} Therefore, we can use the term as a proxy for mass mixing induced by UV-scale loops in a more serious model.

The example model is analytically solvable. One can perform field redefinitions in the full theory by rotating the field basis:
\begin{gather}\label{eq:rotation}
 \begin{pmatrix}
   \phiphys \\
   \chiphys 
   \end{pmatrix}
   =
   \begin{pmatrix}
   \cos \theta &
    \sin \theta\\
    -\sin \theta &
   \cos \theta 
   \end{pmatrix}
  \begin{pmatrix}
   \Phi \\
   \Chi 
   \end{pmatrix}\,,\\
   \tan{2\theta}=\frac{2\mixm^2}{M^2-m^2}\,.
\end{gather}
This way the Lagrangian of the theory becomes
\begin{align}
    \mathscr{L}_{\mathrm{E}}=&\;
    \frac{1}{2}(\partial_\mu\phiphys)^2+\frac{m_{\text{eig}}^2}{2}\phiphys^2\nonumber\\
    &+\frac{1}{2}(\partial_\mu\chiphys)^2+\frac{M_{\text{eig}}^2}{2}\chiphys^2\,,\\[5pt]
    M_{\text{eig}}^2=&\;\frac{1}{2}\qty(M^2+m^2+\sqrt{(M^2-m^2)^2+4\mixm^4})\,,\\[3pt]
    m_{\text{eig}}^2=&\;\frac{1}{2}\qty(m^2+M^2-\sqrt{(M^2-m^2)^2+4\mixm^4})\,.
\end{align}
Here, we have obtained the mass eigenstates of the theory given by the masses $m_{\text{eig}}^2$ and $M_{\text{eig}}^2$.

The long length scales of $m_{\text{eig}}^{-1}$ are described by an EFT given by
\begin{align}\label{eq:massbaslag}
    \mathscr{L}_{\text{eff}}=
    \frac{1}{2}(\partial_\mu\phiphys^{\rmii{IR}})^2+\frac{m_{\text{eig}}^2}{2}(\phiphys^{\rmii{IR}})^2\,.
\end{align}

We have now good control over the long-length-scale effective description. We can then use the analytical solution to better monitor the effects of mixing. Let us thus integrate out the UV scale without performing the field redefinitions of Eq.~\eqref{eq:rotation}.

There is a 1LPI diagram that follows from the mixing term,
\begin{align}
    &\quad\feynalignscaled{0.5}{\mixingdiag}\nonumber\\
    &=\irqe{-\frac{\mixm^4}{2}\int_{x,y}\ir(x)\ir(y)\int_k\frac{e^{i k\cdot(x-y)}}{k^2+M^2}}\nonumber\\
    &=-\frac{\mixm^4}{2M^2}\sum_{n=0}^\infty\int_{x,y}\ir(x)\ir(y)\int_k\qty(-\frac{k^2}{M^2})^ne^{i k\cdot(x-y)}\nonumber\\
    &=-\frac{\mixm^4}{2M^2}\sum_{n=0}^\infty\int_{x,y}\ir(x)\ir(y)\qty(\frac{\partial_y^2}{M^2})^n\int_k e^{i k\cdot(x-y)}\nonumber\\
    &=-\frac{\mixm^4}{2M^2}\sum_{n=0}^\infty\int_{x}\ir\qty(\frac{\partial_\mu^2}{M^2})^n\ir\,,
\end{align}
where the dashed line refers to the $\Chi$ field and the solid lines to the $\Phi$ field.
Here, we have computed directly the contribution to the effective action and not to the effective Lagrangian (\textit{cf}.\ discussion around Eq.~\eqref{eq:diagramMeaning}). The derivative expansion is done similarly as in Appendix~\ref{app:anoutherderiv}.

Note that the IR-quantity expansion must not be resummed:
\begin{equation}
    \frac{1}{M^2}\sum_{n=0}^\infty\qty(\frac{\partial_\mu^2}{M^2})^n\neq \frac{1}{\partial^2_\mu+M^2}\,.
\end{equation}
This is crucial for the mass scale $M$ to not be present in the EFT (\textit{cf}.\ discussion around Eq.~\eqref{eq:scaleHiding}).

The effective Lagrangian becomes
\begin{align}\label{eq:horribLag}
    \mathscr{L}_{\text{eff}}=\;&
    \frac{1}{2}(\partial_\mu\ir)^2+\frac{\meff^2}{2}(\ir)^2\nonumber\\
    &-\sum_{n=1}^{N_{\text{trunc}}}\frac{\mixm^4}{2M^{2(n+1)}}\ir\,\partial_\mu^{2n}\,\ir\,,\\[5pt]
    \meff^2=\;&m^2-\frac{\mixm^4}{M^2}\,,
\end{align}
where we have explicitly truncated the derivative series with $N_{\text{trunc}}$. There would be no complications in computing the unit term as well. It would follow from the one-loop diagrams with mixing term insertions. We omit it to focus on the IR scale.

As expected, the IR field $\ir$ is nearly the mass eigenfield, $\phiphys^{\rmii{IR}}$, but not identical. The effective masses coincide always to the leading order,
\begin{align}\label{eq:factorizationMass}
    m_{\text{eig}}^2
    =\meff^2\times\qty(1+\order{\frac{\mixm^4}{M^4},\frac{m^2\mixm^4}{M^6}})\,.
\end{align}
The factorization of the expansion of $m_{\text{eig}}^2$ follows from
\begin{align}
    \meff^2=0\quad\Rightarrow\quad m_{\text{eig}}^2=0\,,
\end{align}
and holds order by order when the expansion is done according to
\begin{equation}
    m^2\sim\frac{\mixm^4}{M^2}\,.
\end{equation}
The exact form for the factorization can be found by expanding $m_{\text{eig}}^2$.

The factorization plays a central role for the validity of the IR description, because it has to reproduce the physical results of the effective Lagrangian in Eq.~\eqref{eq:massbaslag}.
One fact manifest already in Eq.~\eqref{eq:factorizationMass} is that both effective descriptions become unstable at the same value of $\mixm^2$:
\begin{equation}
    \mixm^4>m^2M^2\,.
\end{equation}
To render other points as clear as possible, we will use
\begin{equation}
    \meff^2\sim m^2\sim\frac{\mixm^4}{M^2}
\end{equation}
below. 

%Here, we note that in a thermal case the free energy contributions of the IR scale can be directly computed using a Lagrangian akin to the one in Eq.~\eqref{eq:horribLag}. There, the derivative corrections on the second line would be handled perturbatively. The field redefinitions only become important when examining light mass eigenstates (or simplifying the perturbation expansion of the IR scale).

%Now that we have found the desired effective Lagrangian, Eq.~\eqref{eq:horribLag}, we would like to reiterate the central tenets of this section. The first one is that the Lagrangian describes the long-length-scale behavior of the system.\jh{Free energy}

We have not yet specified the dimensionality of the system. Let us for a moment state that the effective description is three dimensional, $d=3-2\epsilon$, akin to the one in Sec.~\ref{sec:scalemI} and compute its free energy to the order corresponding to $N_{\text{trunc}}=2$:
\begin{align}
    &\feynalignscaled{0.5}{\oneloopbubble{}}\;+\;\feynalignscaled{0.5}{\cammassinsert{}}\;+\;\feynalignscaled{0.5}{\cammassinserts{}}\nonumber\\
    =\;&\frac{1}{2}\int_{\spat{p}}\ln(\spat{p}^2+\meff^2)
    +\frac{\mixm^4}{2M^4}\int_{\spat{p}}\frac{\spat{p}^2}{\spat{p}^2+\meff^2}\nonumber\\
    &-\frac{\mixm^4}{2M^6}\int_{\spat{p}}\frac{\spat{p}^4}{\spat{p}^2+\meff^2}
    -\frac{\mixm^8}{4M^8}\int_{\spat{p}}\frac{\spat{p}^4}{(\spat{p}^2+\meff^2)^2}\nonumber\\
    =\;&-\frac{1}{12\pi}\qty(1-\frac{3\mixm^4}{2M^4}-\frac{3m^2\mixm^4}{2M^6}-\frac{27\mixm^8}{8M^8})\meff^3\nonumber\\
    =\;&-\frac{1}{12\pi}m_{\text{eig}}^3\,.\label{eq:lastOneLoop}
\end{align}
The result is the same at this order as the one obtained with the effective Lagrangian in Eq.~\eqref{eq:massbaslag}, as it should be.
The crossed dots in the diagrams represent insertions of the second line in Eq.~\eqref{eq:horribLag}.
For the last line, we used
\begin{align}
    m_{\text{eig}}^2
    =\;&\qty(1-\frac{\mixm^4}{M^4}-\frac{m^2\mixm^4}{M^6}+\frac{2\mixm^8}{M^8})\,\meff^2\,,
\end{align}
which holds to this accuracy.

%Notice the importance of the factorization in Eq.~\eqref{eq:factorizationMass} for obtaining the result. The only dimensionful parameter is $\meff^2$. Hence, the one-loop result must be proportional to $\meff^3$. From the another representation of the effective description, we know that the result must also be proportional to $m_{\text{eig}}^3$. Therefore, the only we the descriptions can be equivalent is that the masses factorize in the manner of Eq.~\eqref{eq:factorizationMass}.

We will now move onto the field redefinitions. The effective Lagrangians in Eqs.~\eqref{eq:massbaslag} and \eqref{eq:horribLag} contain the same IR-scale physics. Therefore, if we perform field redefinitions on $\ir$ in Eq.~\eqref{eq:horribLag} to obtain
\begin{align}
    \mathscr{L}_{\text{eff}}=
    \frac{1}{2}(\partial_\mu\overline{\Phi}^{\rmii{IR}})^2+\frac{\overline{m}^2}{2}(\overline{\Phi}^{\rmii{IR}})^2\,,
\end{align}
we must obtain the effective Lagrangian in the mass eigenbasis, Eq.~\eqref{eq:massbaslag}. This means that the masses are the same and the fields are equivalent,
\begin{equation}
    \overline{m}^2=m_{\text{eig}}^2\,,\qquad\overline{\Phi}^{\rmii{IR}}=\phiphys^{\rmii{IR}}\,.
\end{equation}

Let us perform this again with $N_{\text{trunc}}=2$. First, we get rid of the $\ir\partial_\mu^4\ir$ term by
\begin{align}\label{eq:removingderivs}
    \ir
    =\tilde{\Phi}^{\rmii{IR}}-\frac{\mixm^4}{2M^6} \partial_\mu^2\,\tilde{\Phi}^{\rmii{IR}}\,.
\end{align}
The truncated Lagrangian becomes
\begin{align}\label{eq:modefflag}
    \mathscr{L}_{\text{eff}}=\;&
    \frac{1}{2}(\partial_\mu\tilde{\Phi}^{\rmii{IR}})^2+\frac{\meff^2}{2}(\tilde{\Phi}^{\rmii{IR}})^2\nonumber\\
    &-\frac{1}{2}\qty(\frac{\mixm^4}{M^{4}}+\frac{\meff^2\mixm^4}{M^6})\tilde{\Phi}^{\rmii{IR}}\,\partial_\mu^{2n}\,\tilde{\Phi}^{\rmii{IR}}\,.
\end{align}
Finally, we can make the field canonically normalized by
\begin{align}\label{eq:normalizingField}
    (\tilde{\Phi}^{\rmii{IR}})^2
    =\;&\qty(1-\frac{\mixm^4}{M^4}-\frac{m^2\mixm^4}{M^6}+\frac{2\mixm^8}{M^8})\;(\overline{\Phi}^{\rmii{IR}})^2\,.
\end{align}
At the same time, we obtain
\begin{align}\label{eq:normalizingMass}
    \overline{m}^2
    =\;&\qty(1-\frac{\mixm^4}{M^4}-\frac{m^2\mixm^4}{M^6}+\frac{2\mixm^8}{M^8})\,\meff^2\nonumber\\
    =\;&m_{\text{eig}}^2\,,
\end{align}
as expected.

Let us finally discuss the importance of the non-trivial factorization in Eq.~\eqref{eq:factorizationMass} for the two results above in Eqs.~\eqref{eq:lastOneLoop}, \eqref{eq:normalizingMass}.

Notice that the terms to the free energy produced by non-mass-basis Lagrangian, Eq.~\eqref{eq:horribLag}, are $(\meff^2)^{(2n+1)/2}$ times a polynomial of $\tfrac{\mixm^4}{M^4}$ and $\tfrac{m^2}{M^2}$. Hence, the only non-analytic behavior is in $(\meff^2)^{(2n+1)/2}$. This must equal to the non-analytic behavior given by the mass-basis Lagrangian in Eq.~\eqref{eq:massbaslag}: $(m_{\text{eig}}^2)^{3/2}$. This is only possible if the non-trivial factorization holds.

%\jh{Add the free energy}
%Let us now note a non-trivial point about the factorization in Eq.~\eqref{eq:factorizationMass} related to the field redefinitions above.
For the latter result, we removed higher order derivative terms, $\ir\partial_\mu^{2n}\ir,\,n>1$, from the non-mass-basis Lagrangian, Eq.~\eqref{eq:removingderivs}.
This builds a polynomial of $\tfrac{\mixm^4}{M^4}$ and $\tfrac{m^2}{M^2}$ in front of the $\tilde{\Phi}^{\rmii{IR}}\partial_\mu^2\tilde{\Phi}^{\rmii{IR}}$ term, the second line of Eq.~\eqref{eq:modefflag}.
%Similarly, we can notice that removing higher derivative terms, $\ir\partial_\mu^{2n}\ir,\,n>1$, builds a polynomial of $\tfrac{\mixm^4}{M^4}$ and $\tfrac{m^2}{M^2}$ in front of $\tilde{\Phi}^{\rmii{IR}}\partial_\mu^2\tilde{\Phi}^{\rmii{IR}}$ (second line Eq.~\eqref{eq:modefflag}).
At the same time, the mass term remains unchanged, $\tfrac{1}{2}\meff^2(\tilde{\Phi}^{\rmii{IR}})^2$. The non-canonical normalization can be removed to a desired order by multiplying the squared field by a polynomial of $\tfrac{\mixm^4}{M^4}$ and $\tfrac{m^2}{M^2}$,
\begin{equation}
    (\tilde{\Phi}^{\rmii{IR}})^2=\qty(1+\order{\frac{\mixm^4}{M^4},\frac{m^2\mixm^4}{M^6}})\times(\overline{\Phi}^{\rmii{IR}})^2\,,
\end{equation}
Eq.~\eqref{eq:normalizingField}. The same polynomial is now a factor in the mass of the $\overline{\Phi}^{\rmii{IR}}$ field, Eq.~\eqref{eq:normalizingMass},
\begin{align}
    \overline{m}^2
    =\meff^2\times\qty(1+\order{\frac{\mixm^4}{M^4},\frac{m^2\mixm^4}{M^6}})\,.
\end{align}
The relation is of the same form as in Eq.\eqref{eq:factorizationMass}. Thus, the factorization is required for $\overline{m}^2=m^2_{\text{eig}}$ to be possible.

\section{Equivalence with matching} \label{sec:matching}

%\jh{Note the above section!}

%\jh{Field redefinitions usually at. We do them before here. Equivalence is still transparent.}

EFTs are commonly created by matching Green's functions (see e.g.\  Refs.~\cite{Braaten:1995cm,Schicho:2020xaf}). Here, we will show that integrating out a scale as presented in Sec.~\ref{sec:method} is equivalent to the matching procedure.

With matching, one first identifies the heavy fields, $\heavy$, and the light fields, $\light$, and then creates the most general effective Lagrangian for effective fields, $\effphi$ corresponding to the light fields, $\mathscr{L}_{\rmi{eff}}(\effphi)$, which respects the unbroken symmetries of the full theory. Finally, one matches the coefficients of the effective Lagrangian so that the effective description coincides with the infrared behavior of the full theory.

In perturbative theories, matching can be performed by matching the $n$-point Green's functions, $\Gamma_n$, of the effective fields to the ones of the light fields,
\begin{align}\label{eq:OGMatching}
    &\Gamma^{\rmi{EFT}}_n(k_1,\dots,k_n;\{\lambda_{\rmi{eff}}\},\{m^2_{\rmi{eff}}\})\nonumber\\
    =\;&\Gamma_n(k_1,\dots,k_n;\{\lambda\},\{m^2\};\{f\},\{M^2\})\,,
\end{align}
where $k_1,\dots,k_n$ are low external momenta, $\{\lambda_{\rmi{eff}}\},\{m^2_{\rmi{eff}}\}$ are the parameters of the effective description to be matched and $\{\lambda\},\{m^2\};\{f\},\{M^2\}$ are the light field parameters and heavy field parameters respectively in the full theory.

Here, we want to note that the effective fields, $\effphi$, do not always correspond directly to the low-momentum fluctuations of the light fields, $\lightlow$, in the matching procedure.
It is a conventional choice that there are only canonical kinetic terms present in the effective Lagrangian, $\mathscr{L}_{\rmi{eff}}(\effphi)$. Achieving the canonical kinetic terms may require field redefinitions.

The field redefinitions are commonly performed at the matching, leading to modifications to Eq.~\eqref{eq:OGMatching}. One example is to include the would-be field normalization factor, $Z$, in a matching relation so that it does not appear in the EFT:
\begin{align}
    &\Gamma^{\rmi{EFT}}_2(k_1,k_2;\{\lambda_{\rmi{eff}}\},\{m^2_{\rmi{eff}}\})\nonumber\\
    =\;&Z\,\Gamma_2(k_1,k_2;\{\lambda\},\{m^2\};\{f\},\{M^2\})\,.
\end{align}
(See for example Refs.~\cite{Kajantie:1995dw,Braaten:1995cm}.)

It however does not matter if the redefinitions are done prior to, at, or after the matching (\textit{cf}.\ Sec.~\ref{sec:mixingIRUV}). Here, we will not perform these field redefinitions at the matching so that we can demonstrate the equivalence cleanly.

There are two simplifications that can be done for the above matching equation:

The first one considers one-light-particle-reducible contributions (\textit{cf.} Fig.~\ref{fig:reduciblereprep}). These can simply be ignored when matching, because they appear identically on both sides of matching.
%\begin{equation}
%    \underbrace{{\dots}+\feynalignscaled{0.5}{\PRdiagEFT}+\dots}_{\text{EFT}}=\underbrace{{\dots}+\feynalignscaled{0.5}{\PRdiagFull}+\dots}_{\text{full theory}}\,.
%\end{equation}
%On the full theory side, one needs to correctly resum the propagator and the vertices, which has already been done in the EFT. In the end, these diagrams appear equally on both sides and can hence be dropped from the matching procedure.
Thus, we obtain
\begin{align}\label{eq:matching}
    &\Gamma^{\rmi{eff,1PI}}_n(k_1,\dots,k_n;\{\lambda_{\rmi{eff}}\},\{m^2_{\rmi{eff}}\})\nonumber\\
    =\;&\Gamma_n^{\text{1LPI}}(k_1,\dots,k_n;\{\lambda\},\{m^2\};\{f\},\{M^2\})\,.
\end{align}
The simplification corresponds to the fact that only the 1LPI diagrams contribute to the effective action in Eq.~\eqref{eq:effact}.

The second simplification is that we can expand both sides equally in the infrared quantities; we can do strict perturbation theory in $\{m^2_{\rmi{eff}}\}$ and $k_1,\dots,k_n$ on the side of the effective theory and in $k_1,\dots,k_n$, $\{m^2\}$ and other contributions to the $\light$ field masses on the full theory side.
\begin{align}\label{eq:matchingReduced}
    &\irqe{\Gamma^{\rmi{eff,1PI}}_n(k_1,\dots,k_n;\{\lambda_{\rmi{eff}}\},\{m^2_{\rmi{eff}}\})}\nonumber\\
    =\;&\irqe{\Gamma_n^{\text{1LPI}}(k_1,\dots,k_n;\{\lambda\},\{m^2\};\{f\},\{M^2\})}\,.
\end{align}
%In dimensional regularization, this leads to the regulator regulating the infrared of the light fields in the computed diagrams.
This simplification corresponds to the IR-quantity expansion in Box~\ref{diff:one}.

%Note crucially, that the expansion in the IR-quantities removes all the loop contributions to the left-hand side of the matching condition in Eq.~\eqref{eq:matching}.

%\jh{What does the conventional matching literature say. How we interpret it.}

Mathematically, the IR-quantity expansion removes the non-analytic terms in the IR quantities from both sides of the matching~\cite{Manohar:2018aog,Schicho:2020xaf}. Crucially, it removes all the loop contributions from the left-hand side of the matching condition in Eq.~\eqref{eq:matchingReduced}, and only leaves a so-called matching contribution from the loops on the right-hand side.

Here, we want to highlight the physical interpretation in accordance to Secs.~\ref{sec:method} and \ref{sec:separatingIRUV}. The loops on the left-hand side can be understood as the IR-scale contributions. As the IR-scale loops are removed by the expansion, it is clear that the expansion removes the IR-scale contributions from both sides of the equation. As a consequence, the matching contribution, which is the remaining contribution from loops on the right-hand side, can be identified as the contributions from the UV scale, which is being integrated out.

Let us now dress the matching of Green's functions into the language of actions to get to the result of Sec.~\ref{sec:method}. Equation~\eqref{eq:matching} states that all of the 1LPI $n$-point functions are the same. This can be encoded in the classic result, Refs.~\cite{Georgi:1991ch,Georgi:1992xg,Georgi:1993mps}, that the 1LPI effective actions coincide:
\begin{equation}
    \Gamma^{\text{1PI}}_{\text{eff}}[\lightlow]=\Gamma^{\text{1LPI}}[\lightlow]\,.
\end{equation}
Here, we have used the same notations as in Sec.~\ref{sec:method}, Eq.~\eqref{eq:georgi}.

We can finally obtain the result in Eq.~\eqref{eq:finalPackage},
\begin{equation}
    \Sef[\lightlow]=\Gamma^{\text{1LPI}}[\lightlow]\eval_{\substack{\text{IRq}\\\text{exp.}}}\,,
\end{equation}
by taking the IR-quantity expansion on both sides. This corresponds to the same step as for Eq.~\eqref{eq:matchingReduced}. The tree-level action of the EFT appears due to the fact that the EFT loop contributions vanish under the IR-quantity expansion. The only loop contributions are the UV-scale contributions on the right-hand side.
%(Note, that we also need to take into account the $\oeps$ terms from the right-hand side (e.g.\ Ref.~\cite{Kajantie:2002wa}).)

Thus, we have arrived to the method of integrating out a scale from the procedure of matching Green's functions.

%%%%%%%%%%%%%%%%%%%%%%%%%%%%%%%%%%%%%
%%%%%%%%%%%% DISCUSSION %%%%%%%%%%%%% 
%%%%%%%%%%%%%%%%%%%%%%%%%%%%%%%%%%%%%

\section{Conclusion}\label{sec:discussion}

In this paper, we have derived a novel method for constructing effective field theories in perturbative QFTs. It can be identified as directly integrating out the UV scale from the path integral,
\begin{align}
    S_{\text{eff}}[\ir] &= -\ln \int\mathcal{D}\uv\,e^{-S_{\mathrm{E}}[\ir+\uv]}\,,
\end{align}
leaving only the IR scale to the effective description. Therefore, this is the most direct way to construct EFTs.

We found that the formula for the effective action can be represented compactly using the 1LPI effective action of the full theory:
\begin{equation}\label{eq:conclusionsmatch}
    \Sef[\lightlow]=\Gamma^{\text{1LPI}}[\lightlow]\eval_{\substack{\text{IRq}\\\text{exp.}}}\,.
\end{equation}
On the right-hand side, the 1LPI diagrams are expanded in the IR-quantity expansion in Box~\ref{diff:one}, and the expansion in $\epsilon$ must be retained up to a desired order in physical quantities, Box~\ref{diff:two}.

The novel method streamlines computations, allowing one to reach higher perturbative orders with relative ease. One does not have to construct the Lagrangian for the EFT before matching. Equation~\eqref{eq:conclusionsmatch} dictates the effective action directly. Thus, one only needs to compute the relevant 1LPI diagrams with the IR-quantity expansion to obtain the effective action.

EFTs are widely applicable. Consequently, the novel method for their construction is as well, albeit restricted to handling perturbative UV scales. In the examples, we have demonstrated the applicability for high-temperature QFTs, and cosmological phase transitions in particular. However, the method is also straightforwardly applicable to collider physics.

%%%%%%%%%%%%%%%%%%%%%%%%%%%%%%%%%%%%%
%%%%%%%%% ACKNOWLEDGEMENTS %%%%%%%%%% 
%%%%%%%%%%%%%%%%%%%%%%%%%%%%%%%%%%%%%

\begin{acknowledgments}
We would like to thank
O.~Gould,
J.~L\"ofgren,
G.~Nardini,
M.~J.~Ramsey-Musolf,
K.~Rummukainen,
P.~Schicho,
S.~S\"appi,
T.~V.~I.~Tenkanen,
A.~Vuorinen,
and J.~\"Osterman
for enlightening discussions and
O.~Gould,
and A.~Vuorinen,
for many thoughtful and valuable comments on the manuscript.
\end{acknowledgments}

\appendix

\section{Notations and conventions}\label{app:notation}

The regularization scheme used is the $\MSbar$ scheme with dimensions $D=4-2\epsilon$. The three spatial dimensions are denoted with $d=3-2\epsilon$.

There will be three different notations for momenta:
\begin{equation}
    p\,,\quad\mathbf{p}\,,\quad P\,.
\end{equation}
The first one, $p$, is a four-dimensional Euclidean momentum corresponding to vacuum (or zero temperature). It is also used in Secs.~\ref{sec:method}, \ref{sec:methodlong} to denote a generic momentum. The second one, $\mathbf{p}$, is a three-dimensional, spatial momentum. The last one, $P$, is the notation used for a finite-temperature, four-dimensional momentum, $P=(\omega_n,\,\mathbf{p})$, where the first component, $\omega_n$, is a discrete Matsubara frequency (see Appendix~\ref{app:imtimform}). This reduces to $p$ in the limit of zero temperature, $T\to0$.

Corresponding to the three momenta, there are three different momentum integrals,
\begin{align}
    \int_p&=\qty(\frac{\mu^2e^{\gE}}{4\pi})^\epsilon\int\frac{\dd^D p}{(2\pi)^D}\,,\label{eq:vacMomInt}\\
    \int_{\mathbf{p}}&=\qty(\frac{\mu^2e^{\gE}}{4\pi})^\epsilon\int\frac{\dd^d \mathbf{p}}{(2\pi)^d}\,,\label{eq:thermSpatInt}\\
    \sumint{P}&=T\sum_{\omega_n}\int_{\mathbf{p}}\,,\label{eq:thermSumInt}
\end{align}
where $\mu$ is the renormalization scale in the $\MSbar$ scheme.
The last one contains a thermal sum over the Matsubara frequencies. The first one, the vacuum integral, is again used to denote a generic momentum integral in Secs.~\ref{sec:method}, \ref{sec:methodlong}.

The spatial integral is defined analogously:
\begin{align}
    \int_{\mathbf{x}}&=\int\dd^d \mathbf{x}\,.
\end{align}

\section{Imaginary time formalism}\label{app:imtimform}

%\jh{
%\begin{itemize}
%    \item Covering the basics (See Basics)
%    \item Path integral
%    \begin{itemize}
%        \item From partition function (No chem pot)
%        \item Analogy between the operators
%        \item Euclidean action from Minkowski
%        \item Time-extent is $\beta$
%\item Time dimension closes onto itself
%        \item Time dimension contains the information of temperature
%        \item Limits
%    \end{itemize}
%    \item Momentum integrals!
%    \item The bare propagator
%\end{itemize}
%}

We will briefly review the imaginary time formalism of thermal field theory. (For more details, see e.g.\ Ref.~\cite{Laine:2016hma}.) It is a formulation that is capable of describing equal-time correlations in plasma, and consequently contains the thermodynamic information of the plasma. It is also applicable to the leading parts of nucleation rates~\cite{Gould:2021ccf}.

First, we will look at the path integral representation of the imaginary time formalism. Then we study briefly the momentum integrals that follow from the path integral.

In the imaginary time formalism, the partition function of a theory is expanded into a path integral:
\begin{align}\label{eq:TheFundPartFunc}
    Z &=\Tr[e^{-\beta \hat{H}}] \nonumber\\
    &= \int_{\text{BCs}}\mathcal{D}\Phi\, \exp(-\int_0^\beta\dd\tau\int_{\mathbf{x}}\,\mathscr{L}_{\text{E}}) \,.
\end{align}
In the trace representation of the partition function, $\beta$ is the inverse temperature and $\hat{H}$ is the Hamiltonian describing the field theory.

In the path integral representation, $\Phi$ represents the full field content of the theory, and $\spat{x}$ labels the physical spatial dimensions. There are also two, somewhat special features in the path integral that we will discuss more closely:
\begin{itemize}
    \item the Euclidean (or imaginary) time dimension, $\tau$,
    \item the Euclidean Lagrangian, $\mathscr{L}_{\text{E}}$.
\end{itemize}

First, we will just note that the equal-time correlation functions are indeed computable using the path integral,
\begin{align}
    &\langle\hat{\Phi}(\spat{x}_1)\hat{\Phi}(\spat{x}_2)\dots\rangle=\frac{1}{Z}\Tr[\hat{\Phi}(\spat{x}_1)\hat{\Phi}(\spat{x}_2)\dots e^{-\beta \hat{H}}]\nonumber\\[5pt]
    = & \frac{1}{Z} \int_{\text{BCs}}\mathcal{D}\Phi\; \Phi(\spat{x}_1,\tau=0)\,\Phi(\spat{x}_2,\tau=0)\,\dots\, e^{-S_{\text{E}}} \,,
\end{align}
where we have labeled the exponent of Eq.~\eqref{eq:TheFundPartFunc} as the Euclidean action $\Sef$.
The correlation functions simply become equal-imaginary-time correlation functions. Here, we have chosen to set explicitly $\tau=0$.

The Euclidean time dimension can be understood to come from the analogy between the operator $e^{-\beta \hat{H}}$ and the time evolution operator, $e^{-i\hat{H} t}$. The former ``evolves'' the system by $t=-i\beta$. For this reason, the extent of the dimension is $\beta$ in the path integral.

The boundary conditions (BCs) are for the Euclidean time dimension. Due to the trace of the partition function, the dimension loops back onto itself. The bosonic fields are periodic and fermionic fields are anti-periodic:
\begin{equation}\label{eq:boundaryConditions}
    \Phi^{\text{b/f}}(\tau=0)=\pm\Phi^{\text{b/f}}(\tau=\beta)\,.
\end{equation}
%In high temperatures, the Fourier decomposition in the Euclidean time dimension becomes interesting, as we will soon see. 

%The only place, where the temperature appears in the path integral of Eq.~\eqref{eq:TheFundPartFunc}, is the extent of the Euclidean time dimension. Thus, it holds the thermal information of the system.

%At high temperatures, discussed in Sec.\ref{sec:thermalUV}, the Euclidean dimension becomes small, with respect to the length scales of interest. At low temperatures, discussed in Sec.~\ref{sec:thermalIR}, the extent becomes large. In vacuum, i.e.\ at zero temperature, the extent becomes infinite, and the Euclidean time dimension becomes similar to the spatial dimensions.

The Euclidean Lagrangian can be obtained in a straightforward manner from the regular, Minkowski counterpart by a Wick rotation:
\begin{equation}\label{eq:EuclFromMink}
    \mathscr{L}_{\text{E}}(\tau)=-\mathscr{L}_{\text{M}}(t\to- i\tau)\,.
\end{equation}
For example, the Euclidean Lagrangian of a simple scalar field theory reads:
\begin{equation}
    \mathscr{L}_{\text{E}}=\partial_\mu\phi\, \partial_\mu\phi + V(\phi)\,.
\end{equation}
A small addition to the simple rule in Eq.~\eqref{eq:EuclFromMink} is related to the temporal component of a gauge field. It is Wick rotated as well. For U(1) gauge theory, the Euclidean Lagrangian is given by:
\begin{equation}
    \mathscr{L}_{\text{E}}=\frac{1}{4}F_{\mu\nu}  F_{\mu\nu},\quad F_{\mu\nu}=\partial_\mu A_\nu-\partial_\nu A_\mu\,.
\end{equation}

Finally, we discuss the momentum integrals of the perturbative expansion produced by the partition function in Eq.~\eqref{eq:TheFundPartFunc}. The Euclidean space time contains the regular three spatial dimensions and the Euclidean time with the extent of $\beta$. Thus, it follows that the spatial momenta are continuous,
\begin{equation}
    \int_{\spat{p}}\,,
\end{equation}
from Eq.~\eqref{eq:thermSpatInt}. 

The temporal momentum component is more interesting: it is discrete. The discrete temporal momenta, called the Matsubara frequencies, differ for the bosonic and fermionic degrees of freedom due to different boundary conditions in Eq.~\eqref{eq:boundaryConditions}. The bosonic ones are periodic and the fermionic ones anti-periodic. This gives rise to frequencies of
\begin{equation}\label{eq:matsubfreqs}
    \omega^{\text{b}}_n=2\pi Tn\,,\quad\omega^{\text{f}}_n=2\pi T\qty(n+\frac{1}{2})\,,\quad n\in\mathbb{Z}\,.
\end{equation}

The full momentum integrals are given by integrating over the spatial momenta and summing over the Matsubara modes:
\begin{align}
    \sumint{P}&\equiv T\sum_{\omega^{\text{b/f}}_n}\int_{\spat{p}}
\end{align}
(\textit{cf.} Eq.~\eqref{eq:thermSumInt}).

\section{Infrared divergences of the UV scale}\label{app:IRdivs}

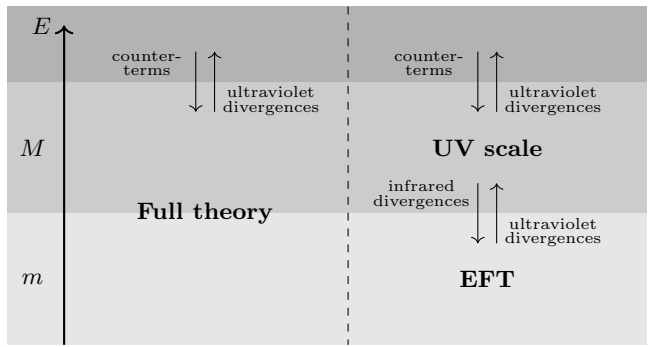
\begin{figure}
    \centering
    \begin{tikzpicture}[scale=1.0]
        \filldraw[draw=none, fill=black!30, very thick](0,4.5) rectangle (8.5,3.5);
        \filldraw[draw=none, fill=black!20, very thick](0,3.5) rectangle (8.5,1.75);
        \filldraw[draw=none, fill=black!10, very thick](0,1.75) rectangle (8.5,0);
        \draw[dashed] (4.525,4.5) -- (4.525,0);
        \node[] at (0.33,2.625) {$M$};
        \node[] at (0.33,0.875) {$m$};
        \draw[->,thick] (0.75,0) -- (0.75,4.25);
        \node[] at (0.45,4.25) {$E$};
        \node[] at (2.625,1.75) {\textbf{Full theory}};
        \draw[->] (2.5,3.9) -- (2.5,3.1);
        \node[] at (1.85,3.75) {$\substack{\text{counter-}\\ \text{terms}}$};
        \draw[->] (2.75,3.1) -- (2.75,3.9);
        \node[] at (3.5,3.25) {$\substack{\text{ultraviolet}\\ \text{divergences}}$};
        \node[] at (6.375,2.625) {\textbf{UV scale}};
        \draw[->] (6.25,3.9) -- (6.25,3.1);
        \node[] at (5.6,3.75) {$\substack{\text{counter-}\\ \text{terms}}$};
        \draw[->] (6.5,3.1) -- (6.5,3.9);
        \node[] at (7.25,3.25) {$\substack{\text{ultraviolet}\\ \text{divergences}}$};
        \draw[->] (6.25,2.15) -- (6.25,1.35);
        \node[] at (5.5,2) {$\substack{\text{infrared}\\ \text{divergences}}$};
        \draw[->] (6.5,1.35) -- (6.5,2.15);
        \node[] at (7.25,1.5) {$\substack{\text{ultraviolet}\\ \text{divergences}}$};
        \node[] at (6.375,0.875) {\textbf{EFT}};
    \end{tikzpicture}
    \caption{The EFT organization of a computation is presented schematically on the right. The ultraviolet divergences of the UV scale cancel against the counterterms of the full theory. The infrared divergences of the UV scale cancel against the ultraviolet divergences of the EFT.}
    \label{fig:regulationapp}
\end{figure}

In the UV-scale contributions that are contained in the effective action in Eq.~\eqref{eq:effactreprep}, there are infrared divergences. These divergences can be understood to follow from the IR-quantity expansion, Box~\ref{diff:one}. Here, we show that these infrared divergences of the UV scale cancel against the ultraviolet divergences of the EFT. (See Fig.~\ref{fig:regulationapp}.) Correspondingly, they are the correct counterterms to appear in the effective action describing the EFT.

Before moving onto details of the infrared divergences, it is good to note that the IR-quantity expansion does not affect the ultraviolet divergences, only the infrared divergences. Thus, the UV-scale contributions correctly cancel the counterterms of the full theory, as depicted in Fig.~\ref{fig:regulationapp}. Consequently, we only need to focus on the infrared divergences of the UV scale.

\begin{figure}
    \centering
    \begin{tikzpicture}[scale=1.0]
        \filldraw[draw=none, fill=black!20, very thick](0,3) rectangle (8.5,1.75);
        \filldraw[draw=none, fill=black!10, very thick](0,1.75) rectangle (8.5,0);
        \draw[thick] (4.05,3) -- (4.05,0);
        \draw[dashed] (6.275,3) -- (6.275,0);
        \node[] at (0.33,2.375) {$M$};
        \node[] at (0.33,0.875) {$m$};
        \draw[->,thick] (0.75,0) -- (0.75,3.25);
        \node[] at (0.45,3.25) {$E$};
        \draw[->] (2.25,2.15) -- (2.25,1.35);
        \node[] at (1.5,2) {$\substack{\text{infrared}\\ \text{divergences}}$};
        \draw[->] (2.5,1.35) -- (2.5,2.15);
        \node[] at (3.25,1.5) {$\substack{\text{ultraviolet}\\ \text{divergences}}$};
        \node[] at (2.375,2.625) {\textbf{UV scale}};
        \node[] at (2.375,0.875) {\textbf{EFT}};
        \draw[->,gray] (6.53,2.15) -- (6.53,1.35);
        \draw[->] (6.74,1.35) -- (6.74,2.15);
        \node[] at (7.633,1.65) {$\substack{\text{corresponding}\\\text{ultraviolet}\\ \text{divergences}}$};
        \node[] at (7.388,2.625) {\textbf{UV scale}};
        \node[] at (7.388,0.875) {\textbf{EFT}};
        \draw[->] (6.74,0.4) -- (6.74,-0.4);
        \node[] at (7.48,0.1) {$\substack{\text{artificial}\\\text{infrared}\\ \text{divergences}}$};
        \node[] at (5.16,1.75) {\textbf{Full theory}};
        \draw[->] (4.512,0.4) -- (4.512,-0.4);
        \node[] at (5.252,0.1) {$\substack{\text{artificial}\\\text{infrared}\\ \text{divergences}}$};
    \end{tikzpicture}
    \caption{We induce artificial infrared divergences on the right of the solid line to show the cancellation of the physical divergences on the left.}
    \label{fig:infradivsapp}
\end{figure}
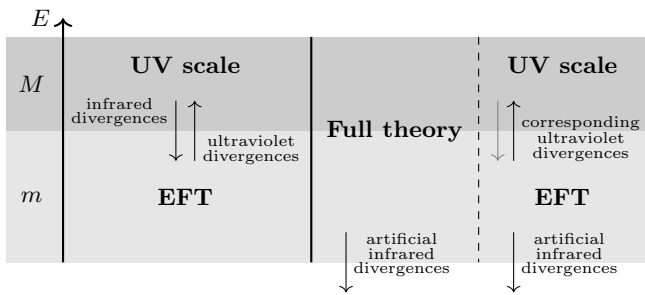

Here, we use the same strategy as in Ref.~\cite{Manohar:2018aog} to show that the infrared divergences of the UV scale indeed cancel with the ultraviolet divergences of the EFT: We imagine computing an IR-scale correlation function directly from both the full theory and the corresponding EFT. The trick will be that we artificially induce the same infrared divergences to both of the computations by conducting the IR-quantity expansion (Fig.~\ref{fig:infradivsapp}).

There will be few steps that we can then take after utilizing the above trick of inducing infrared divergences: Firstly, it will be clear that the artificial divergences of the full theory are equivalent to the infrared divergences in the UV-scale contributions. Secondly, it will also be easy to show that the artificial divergences in the loop integrals of the EFT are equal but opposite to the ultraviolet divergences of the EFT. Thirdly, the infrared divergences of the full theory and the EFT can be linked due to yielding the same IR physics. Therefore, the infrared divergences in the UV-scale contributions cancel against the ultraviolet divergences of the EFT, and they can be understood as the counterterms of the EFT.

Let us now proceed to the details.

We want to show that the UV-scale infrared divergences cancel all of the ultraviolet divergences in the correlation functions of the corresponding EFT. The ultraviolet divergences originate from loop integrals. Consequently, it is enough to focus on momentum-space $n$-point-light-field correlation functions, which are 1LPI. One can construct any correlation function for light fields out of these pieces by connecting with light-field propagators into tree graphs, i.e.\ without introducing additional loops.
%(\textit{cf}.\ Fig.~\ref{fig:accounting}).

There is also an additional reason to focus on the specific set of correlation functions: It provides a direct link to the UV-scale contributions, which consist of the 1LPI diagrams evaluated with the IR-quantity expansion. (See Secs.~\ref{sec:method}, \ref{sec:diagrammaticexp}.)

We start by showing that the artificially induced infrared divergences in the EFT cancel with its ultraviolet divergences: Due to the IR-quantity expansion, each loop integral is scale free. Thus, all of them vanish.
However, the IR-quantity expansion does not affect the ultraviolet divergences; it only induces infrared divergences. Hence, each diagram is a sum of ultraviolet and infrared divergences,
\begin{equation}\label{eq:divscancelEFT}
    \underbrace{\sum_n\frac{A_n}{\epsilon^n}}_{\text{ultraviolet}}+\underbrace{\sum_n\frac{B_n}{\epsilon^n}}_{\text{infrared}}=0\,,
\end{equation}
summing to zero.
Therefore, we find that for an ultraviolet divergence of the EFT, there is a corresponding, artificial infrared divergence with an opposing sign.

Note, that in a correct EFT computation, without the IR-quantity expansion, only the ultraviolet divergences,
\begin{equation}\label{eq:UVdivsEFT}
    \sum_n\frac{A_n}{\epsilon^n}\,,
\end{equation}
would be present (and cancelled by the corresponding counterterms).

Since the EFT reproduces the IR-scale behavior of the full theory, we can compute the same correlation functions directly using the full theory. Conducting the IR-quantity expansion for the diagrams leads to the same, artificial infrared divergences,
\begin{equation}\label{eq:IRdivsfull}
    \sum_n\frac{B_n}{\epsilon^n}\,,
\end{equation}
as in the EFT. This is also due to the corresponding IR-scale behavior.

Finally, we can link the artificial infrared divergences of the full theory to the proper infrared divergences of the UV scale: As already noted above, the diagrams computed for the particular correlation functions, the 1LPI diagrams, correspond to the diagrams evaluated for the UV-scale contributions.
Here, we also note that the UV-scale 1LPI diagrams are nothing but the full theory 1LPI diagrams evaluated with the IR-quantity expansion.
%This expansion was used to induce the infrared divergences in Eq.~\eqref{eq:IRdivsfull} to the full theory.
Therefore, the same infrared divergences are in the UV-scale contributions as well:
\begin{equation}\label{eq:IRdivsUVscale}
    \sum_n\frac{B_n}{\epsilon^n}\,.
\end{equation}

We have now showed in Eq.~\eqref{eq:divscancelEFT} that the infrared divergences of the UV scale in Eq.~\eqref{eq:IRdivsUVscale} cancel against the ultraviolet divergences of the EFT in Eq.~\eqref{eq:UVdivsEFT}. Therefore, they are the correct counterterms for the EFT. Note, that these counterterms appear automatically to the effective action in Eq.~\eqref{eq:effact} by integrating out a scale, as they should.

\section{Renormalization group running viewed through proto-EFT techniques}\label{app:rengrouprunwithprotoefts}

Here, we will study running of the Lagrangian coefficients through the lens of proto-EFTs. This adds another way of understanding the running and also hopefully makes proto-EFTs more familiar.
%and not something completely out of the blue.

The example given here is a $\lambda\Phi^4$ model in $\text{MS}$-scheme,
\begin{equation}
    V_{\text{bare}}=\frac{m^2(\mu)+\cou{m^2}}{2}\Phi^2+\mu^{2\epsilon}\frac{\lambda}{4!}\Phi^4\,.\label{eq:startingpointrunningpot}
\end{equation}
Here, we have explicitly shown the renormalization scale in the bare potential of the Lagrangian in dimensions $D=4-2\epsilon$. Our goal is to compute the running of $m^2(\mu)$ to the one-loop order.

The route taken will be to change the renormalization scale, $\mu\to\nu$ in the prefactor of the interaction term, while giving us $\order{\epsilon}$ terms. These terms can then be transformed into corrections to the coefficients using divergences as discussed in Sec.~\ref{sec:findingproper}. 

We start by
\begin{align}
    \mu^{2\epsilon}\frac{\lambda}{4!}\Phi^4 &=\nu^{2\epsilon}\qty(\frac{\mu}{\nu})^{2\epsilon}\frac{\lambda}{4!}\Phi^4\nonumber\\
    &=\nu^{2\epsilon}\qty(1+2\epsilon\ln(\frac{\mu}{\nu})+\order{\epsilon^2})\frac{\lambda}{4!}\Phi^4\,,\nonumber\\
    &=\nu^{2\epsilon}\frac{\hat{\lambda}}{4!}\Phi^4\,,\\
    \hat{\lambda}&=\qty(1+2\epsilon\ln(\frac{\mu}{\nu})+\order{\epsilon^2})\lambda\,.\label{eq:modepsselfcoupl}
\end{align}
The $\order{\epsilon^2}$ terms will be higher order than considered here, and consequently neglected below.

The $\order{\epsilon}$ term has a finite effect on physical quantities through the $\epsilon$ poles of the Feynman diagrams. These can be incorporated into the Lagrangian coefficients through the counterterms, as discussed in Sec.~\ref{sec:findingproper}.

This particular model contains only the mass counterterm. Let us complete it regarding the new four-point coupling in Eq.~\eqref{eq:modepsselfcoupl}:
\begin{align}
    \cou{m^2}=&\frac{\lambda m^2}{2(4\pi)^2\epsilon}\\
    =&\frac{\hat{\lambda} m^2}{2(4\pi)^2\epsilon}\underbrace{-\frac{\lambda m^2}{(4\pi)^2}\ln(\frac{\mu}{\nu})}\,.
\end{align}
The underbraced term is a mass contribution.

Let us move it to the mass parameter according to Sec.~\ref{sec:findingproper} and drop the $\oeps$ term, $\hat{\lambda}\to\lambda$:
\begin{align}
    V_{\text{bare}}&= \frac{m^2(\nu)+\cou{m^2}}{2} \Phi^2 +\nu^{2\epsilon}\frac{\lambda}{4!}\Phi^4\,,\\
    m^2(\nu)&=m^2(\mu)\qty(1-\frac{\lambda}{(4\pi)^2}\ln(\frac{\mu}{\nu}))\,.
\end{align}

The above bare potential is physically equivalent to the one in Eq.~\eqref{eq:startingpointrunningpot}. Thus, we have now found the running of the mass parameter using the proto-EFT techniques of Sec.~\ref{sec:findingproper}. As discussed there, these methods generalize giving an alternative method for finding running couplings.

%\section{More on derivative expansion}\label{app:derivexp}

%\subsection{Expansion on a homogeneous background}\label{app:homogback}

\section{Another method for computing derivative corrections}\label{app:anoutherderiv}

%\jh{Change the labeling!}

Here, we show another method to compute the field-normalization contribution form the one-loop diagram in Eq.~\eqref{eq:exIIdiags}:
\begin{align}
    \irqe{\feynalignscaled{0.6}{\oneloopbubble{dashed}}}\,.
\end{align}
This time we will compute the contribution directly as an action contribution.

Using the same split into a homogeneous background and fluctuations as in Eq.~\eqref{eq:exIIflatNfluct}, $\phi_{\text{IR}}=\varphi+h_{\text{IR}}$, the external momentum-dependent part of the diagram is
\begin{align}
    \feynalignscaled{0.75}{\intuitionselfenergy}=-\frac{g_3^4\varphi^2}{4}\int_{\bf{x}}\int_{\bf{y}}h_{\text{IR}}({\bf x})h_{\text{IR}}({\bf y})\times \nonumber\\
    \times\int_{\bf p}\int_{\bf q}\frac{e^{i({\bf p}+{\bf q})\cdot({\bf x}-{\bf y})}}{({\bf p}^2+M_3^2(\varphi))({\bf q}^2+M_3^2(\varphi))}\,.
\end{align}
%where the background appears both in the cubic interactions and in the $\chi$ mass, Eq.~\eqref{eq:fullchimass}.
%The fluctuations, $h$, come from the external legs, which are at different spatial points, $\bf{x}$ and $\bf{y}$.
Here, the diagram is written explicitly without the IR-quantity expansion.

Note, that the integral looks non-local, containing two spatial locations, $\spat{x}$ and $\spat{y}$. The diagram will be rendered manifestly local by the IR-quantity expansion -- more specifically by the external-momentum expansion.

%The integral structure is much more complex than the ones previously encountered. This follows from two facts: The first one is that we are keeping the external momentum in our computation, leading to $\bf p$ and $\bf q$ being different.
% If we were just computing the diagram in zero external momentum (i.e.\ just the potential term and not derivative parts), the diagram would simply be\jh{obtained later!} 
%The second reason is that the diagram is non-local, containing the two distinct spatial points.

To get to the external-momentum expansion, we need to first identify the external momentum, $\spat{k}$.
%Now, we start molding the diagram to be local. First, we extract the external momentum, 
This can be done by with ${\bf q}\to-{\bf p}-{\bf k}$:
\begin{align}
    \feynalignscaled{0.75}{\intuitionselfenergy}&=-\frac{g_3^4\varphi^2}{4}\int_{\bf{x}}\int_{\bf{y}}h_{\text{IR}}({\bf x})h_{\text{IR}}({\bf y})\int_{\bf k}e^{i{\bf k}\cdot({\bf x}-{\bf y})}\times \nonumber\\
    \times\int_{\bf p}&\frac{1}{({\bf p}^2+M_3^2(\varphi))(({\bf p}+{\bf k})^2+M_3^2(\varphi))}\,.
\end{align}
Only the external momentum $\bf k$ appears in the exponential function.

%The step (2) of the EFT construction states that we must do strict \jh{IRq} perturbation expansion in the external momentum $\bf k$. Again, we will only need to focus on the ${\bf k}^2$ term.% first and then make remarks on the ${\bf k}^0$ potential term. Note, that ${\bf k}^1$ term vanishes.

%The second order term is

Now, that we have found the external momentum, we can perform the external momentum expansion in the propagators. The first order in the expansion is given by
\begin{align}
    &\kOrd{\feynalignscaled{0.75}{\intuitionselfenergy}}{1}\nonumber\\[3pt]
    &=-\frac{g_3^4\varphi^2}{4}\int_{\bf{x}}\int_{\bf{y}}h_{\text{IR}}({\bf x})h_{\text{IR}}({\bf y})\int_{\bf k}{\bf k}^2e^{i{\bf k}\cdot({\bf x}-{\bf y})}\times \nonumber\\
    &\quad\times\int_{\bf p}\left(-\frac{1}{({\bf p}^2+M_3^2(\varphi))^3}+\frac{4}{d}\frac{{\bf p}^2}{({\bf p}^2+M_3^2(\varphi))^4}\right)\nonumber\\[5pt]
    &=\frac{g_3^4\varphi^2}{96(4\pi)M_3^3(\varphi)}\int_{\bf{x}}\int_{\bf{y}}h_{\text{IR}}({\bf x})h_{\text{IR}}({\bf y})\int_{\bf k}\spat{k}^2e^{i{\bf k}\cdot({\bf x}-{\bf y})}\,,
\end{align}
where $d$ is the number of spatial dimensions and we have performed the $\spat{p}$ integral for the last line.

%There are now two things to be done: We can compute the $\bf p$ integral according to Eq.~\eqref{eq:integralktwice}.
%We can also use
Now, we get to finally changing the external momenta into spatial derivatives:
\begin{equation}
    {\bf k}^2e^{i{\bf k}\cdot({\bf x}-{\bf y})}=\partial^{\bf x}_i\partial^{\bf y}_ie^{i{\bf k}\cdot({\bf x}-{\bf y})}\,.
\end{equation}
By integrating by parts, we obtain
\begin{align}
    &\kOrd{\feynalignscaled{0.75}{\intuitionselfenergy}}{1}\nonumber\\
    =&\frac{g_3^4\varphi^2}{96(4\pi)M_3^3(\varphi)}\int_{\bf{x}}\int_{\bf{y}}\partial^{\bf x}_ih_{\text{IR}}({\bf x})\partial^{\bf y}_ih_{\text{IR}}({\bf y})\int_{\bf k}e^{i{\bf k}\cdot({\bf x}-{\bf y})}\,.
\end{align}

Finally, the expression becomes manifestly local by integrating over $\bf k$ to obtain a delta function, $\delta({\bf x}-{\bf y})$, and then integrating over ${\bf y}$:
\begin{align}
    \kOrd{\feynalignscaled{0.75}{\intuitionselfenergy}}{1}
    =\int_{\bf{x}}\frac{g_3^4\varphi^2}{96(4\pi)M_3^3(\varphi)}(\partial_ih_{\text{IR}})^2\,.
\end{align}

We can now identify that the full action term is
%We can then identify the derivative term from above by comparing with Eq.~\eqref{eq:flatZ}. We obtain for our effective action,
\begin{align}
    \int_{\bf{x}}\frac{1}{2}\frac{g_3^4\phi_{\text{IR}}^2}{48(4\pi)M_3^3(\phi_{\text{IR}})}(\partial_i\phi_{\text{IR}})^2\,
\end{align}
%in alignment with Eq.~\eqref{eq:easyres}.
by comparing with Eq.~\eqref{eq:backgroundhomogNfluctsZ}.

\bibliography{refs}
\end{document}

%% file: feynman-diagrams.tex
\newcommand{\feynalignscaled}[2]{
    \begin{gathered}
    \begin{tikzpicture}[scale=#1, transform shape]
    \begin{feynman}
    #2
    \end{feynman}
    \end{tikzpicture}
    \end{gathered}
}

\newcommand{\selfenergyfour}[2]{
    \vertex (a) at (0,0);
    \vertex (b) at (0.5,0.5);
    \vertex (c) at (1,0);
    \diagram*{
    (a) -- [#1] (b) -- [#1] (c)
    };
    \draw[#2] (b) arc [start angle=-90, end angle=270, radius=0.6cm];
}
\newcommand{\selfenergytree}[2]{
    \vertex (a) at (0,0);
    \vertex[#2] (b) at (0.7,0.0) {};
    \vertex (c) at (1.4,0);
    \diagram*{
    (a) -- [#1] (b) -- [#1] (c)
    };
}

\newcommand{\sunset}[2]{
    \vertex (a) at (0,0);
    \vertex (b) at (0.5,0);
    \vertex (c) at (1.9,0);
    \vertex (d) at (2.4,0);
    \diagram*{
    (a) -- [#1] (b) -- [#1, half left] (c) -- [#2, half left] (b),
    (c) -- [#1] (d), (b) -- [#2] (c),
    };
}

\newcommand{\fourpointtree}[1]{
    \vertex (a) at (0,-0.7);
    \vertex (b) at (0,0.7);
    \vertex (c) at (0.7,0);
    \vertex (d) at (1.4,-0.7);
    \vertex (e) at (1.4,0.7);
    \vertex[#1] (f) at (0.7,0) {};
    \diagram*{
    (a) -- (c),
    (b) -- (c), (c) -- (d), (c) -- (e),
    };
}

\newcommand{\oneloopbubble}[1]{
    \vertex (a) at (0,0);
    \draw[#1] (a) arc [start angle=-90, end angle=270, radius=0.7cm];
}

\newcommand{\twoloopbubble}[1]{
    \vertex (a) at (0.5,0.5);
    \vertex (b) at (0.5,1.5);
    \draw[#1] (a) arc [start angle=-90, end angle=270, radius=0.5cm];
    \draw[#1] (b) arc [start angle=-90, end angle=270, radius=0.5cm];
}

\newcommand{\cammassinsert}[1]{
    \vertex (b) at (0.5,0.5);
    \vertex[crossed dot] (d) at (0.5,1.9) {};
    \diagram*{
    (b) -- [#1,half left] (d) -- [#1,half left] (b)
    };
}

\newcommand{\cammassinserts}[1]{
    \vertex[crossed dot] (b) at (0.5,0.5) {};
    \vertex[crossed dot] (d) at (0.5,1.9) {};
    \diagram*{
    (b) -- [#1,half left] (d) -- [#1,half left] (b)
    };
}

\newcommand{\cammassinsertB}[1]{
    \vertex (b) at (0.5,0.5);
    \vertex[#1] (d) at (0.5,1.9) {};
    \diagram*{
    (b) -- [half left] (d) -- [half left] (b)
    };
}
\newcommand{\cammassinsertG}[1]{
    \vertex (b) at (0.5,0.5);
    \vertex[#1] (d) at (0.5,1.9) {};
    \diagram*{
    (b) -- [half left,dashed] (d) -- [half left,dashed] (b)
    };
}
\newcommand{\cammassinsertC}[2]{
    \vertex (b) at (0.5,0.5);
    \vertex[#1] (d) at (0.5,1.9) {};
    \diagram*{
    (b) -- [#2,half left] (d) -- [#2,half left] (b)
    };
}

\newcommand{\PRdiag}{
    \fill[gray!50] (-0.3,0) circle (0.3cm);
    \draw (-0.3,0) circle (0.3cm);
    \fill[gray!50] (1.3,0) circle (0.3cm);
    \draw (1.3,0) circle (0.3cm);
    \vertex (a) at (0.0,0.0);
    \vertex (b) at (1.0,0.0);
    \vertex (c) at (1.45,0.26);
    \vertex (d) at (1.7,0.7);
    \vertex (e) at (1.7,-0.7);
    \vertex (f) at (1.45,-0.26);
    \vertex (g) at (-0.45,0.26);
    \vertex (h) at (-0.7,0.7);
    \vertex (i) at (-0.7,-0.7);
    \vertex (j) at (-0.45,-0.26);
    \diagram*{
    (a) -- (b);
    (c) -- (d);
    (e) -- (f);
    (g) -- (h);
    (i) -- (j)
    };
}

\newcommand{\basketball}{
    \fill[gray!50] (-0.3,0) circle (0.3cm);
    \draw (-0.3,0) circle (0.3cm);
    \fill[gray!50] (1.3,0) circle (0.3cm);
    \draw (1.3,0) circle (0.3cm);
    \vertex (a) at (0.0,0.0);
    \vertex (b) at (1.0,0.0);
    \vertex (c) at (1.3,0.26);
    \vertex (f) at (1.3,-0.26);
    \vertex (g) at (-0.3,0.26);
    \vertex (j) at (-0.3,-0.26);
    \diagram*{
    (a) -- (b);
    (c) --[half right] (g);
    (j) --[half right] (f)
    };
}

\newcommand{\PIdiag}{
    \fill[gray!50] (-0.3,0) circle (0.3cm);
    \draw (-0.3,0) circle (0.3cm);
    \vertex (a) at (0.0,0.0);
    \vertex (b) at (0.5,0.0);
    \vertex (g) at (-0.45,0.26);
    \vertex (h) at (-0.7,0.7);
    \vertex (i) at (-0.7,-0.7);
    \vertex (j) at (-0.45,-0.26);
    \diagram*{
    (a) -- (b);
    (g) -- (h);
    (i) -- (j)
    };
}

\newcommand{\intuitionfourpoint}{
    \vertex (a) at (-0.3,0.0);
    \vertex (b) at (1.3,0.0);
    \vertex (c) at (1.3,0.);
    \vertex (d) at (1.7,0.7);
    \vertex (e) at (1.7,-0.7);
    \vertex (f) at (1.3,-0.);
    \vertex (g) at (-0.3,0.);
    \vertex (h) at (-0.7,0.7);
    \vertex (i) at (-0.7,-0.7);
    \vertex (j) at (-0.3,-0.);
    \diagram*{
    (a) --[dashed] (b);
    (c) -- (d);
    (e) -- (f);
    (g) -- (h);
    (i) -- (j)
    };
}

\newcommand{\notintuitionfourpoint}{
    \vertex (a) at (-0.3,0.0);
    \vertex (b) at (1.3,0.0);
    \vertex (c) at (1.3,0.);
    \vertex (d) at (1.7,0.7);
    \vertex (e) at (1.7,-0.7);
    \vertex (f) at (1.3,-0.);
    \vertex (g) at (-0.3,0.);
    \vertex (h) at (-0.7,0.7);
    \vertex (i) at (-0.7,-0.7);
    \vertex (j) at (-0.3,-0.);
    \diagram*{
    (a) -- (b);
    (c) -- (d);
    (e) -- (f);
    (g) -- (h);
    (i) -- (j)
    };
}

\newcommand{\intuitionmass}{
    \vertex (a) at (-0.15,0);
    \vertex (b) at (0.5,0);
    \vertex (c) at (1.9,0);
    \vertex (d) at (2.55,0);
    \diagram*{
    (a) -- [] (b) -- [half left] (c) -- [dashed,half left] (b),
    (c) -- [] (d),
    };
}

\newcommand{\sunsetintuit}{
    \vertex (b) at (0.5,0);
    \vertex (c) at (1.8,0);
    \diagram*{
    (b) -- [half left] (c) -- [half left] (b),
    , (b) --[dashed] (c),
    };
}

\newcommand{\sunsetintuitB}{
    \vertex (b) at (0.5,0);
    \vertex (c) at (1.8,0);
    \diagram*{
    (b) -- [half left] (c) -- [dotted,thick,half left] (b),
    , (b) --[dashed] (c),
    };
}

\newcommand{\twoloopbubbleB}[1]{
    \vertex (a) at (0.5,0.5);
    \vertex[#1] (b) at (0.5,1.5) {} ;
    \draw (a) arc [start angle=-90, end angle=270, radius=0.5cm];
    \draw (b) arc [start angle=-90, end angle=270, radius=0.5cm];
}

\newcommand{\twoloopbubbleC}{
    \vertex (a) at (0.5,0.5);
    \vertex (b) at (0.5,1.5) {} ;
    \draw (a) arc [start angle=-90, end angle=270, radius=0.5cm];
    \draw[dashed] (b) arc [start angle=-90, end angle=270, radius=0.5cm];
}

\newcommand{\threeloopbubbleC}{
    \vertex (a) at (0.5,0.5);
    \vertex (b) at (0.5,1.5) {} ;
    \vertex (c) at (0.5,-0.5) {} ;
    \draw (a) arc [start angle=-90, end angle=270, radius=0.5cm];
    \draw[dashed] (b) arc [start angle=-90, end angle=270, radius=0.5cm];
    \draw[dashed] (c) arc [start angle=-90, end angle=270, radius=0.5cm];
}

\newcommand{\intuitionselfenergy}{
    \vertex (a) at (-0.15,0);
    \vertex (b) at (0.5,0);
    \vertex (c) at (1.9,0);
    \vertex (d) at (2.55,0);
    \diagram*{
    (a) -- [] (b) -- [half left,dashed] (c) -- [dashed,half left] (b),
    (c) -- [] (d),
    };
}

\newcommand{\intuitionlast}{
    \vertex (a) at (0,0);
    \vertex (b) at (0.5,0.5);
    \vertex (c) at (1,0);
    \vertex (d) at (0.5,1.3);
    \diagram*{
    (a) -- (b) -- (c);
    (b) --[dashed] (d)
    };
    \draw (d) arc [start angle=-90, end angle=270, radius=0.5cm];
}

\newcommand{\intuitzoomone}{
    \vertex (a) at (0,0);
    \vertex (b) at (0.5,0.5);
    \vertex (c) at (1.35,0.5);
    \vertex (d) at (1.85,0);
    \diagram*{
    (a) -- (b) --[dashed] (c) -- (d)
    };
    \draw (c) arc [start angle=-45, end angle=225, radius=0.6cm];
}

\newcommand{\intuitionzoomtwo}{
    \vertex (a) at (0,0);
    \vertex (b) at (0.5,0.5);
    \vertex (c) at (1,0);
    \vertex (d) at (0.5,1.1);
    \diagram*{
    (a) -- (b) -- (c);
    (b) --[dashed] (d)
    };
    \draw (d) arc [start angle=-90, end angle=270, radius=0.6cm];
}

\newcommand{\propagator}[1]{
    \vertex[] (a) at (0,0) {};
    \vertex (b) at (1.2,0);
    \diagram*{
    (a) -- [#1] (b),
    };
}

\newcommand{\selfenergyfourdot}[2]{
    \vertex (a) at (0,0);
    \vertex[#1] (b) at (0.7,0) {};
    \vertex (c) at (1.4,0);
    \diagram*{
    (a) -- (b) -- (c)
    };
    \draw[#2] (b) arc [start angle=-90, end angle=270, radius=0.5cm];
}

\newcommand{\sunsetIR}[2]{
    \vertex (a) at (0,-0.62);
    \vertex (b) at (0.5,0);
    \vertex (c) at (1.9,0);
    \vertex (d) at (2.4,-0.62);
    \vertex (e) at (1.2,-0.62);
    \diagram*{
    (a) -- [#1] (e) -- [#1] (d),
    (b) -- [#2, half left] (c) -- [#1, half left] (b),
    (c) -- [#2] (b),
    };
}

\newcommand{\selfenergyfourIRdot}[2]{
    \vertex (a) at (0,0);
    \vertex (b) at (0.7,0);
    \vertex[#1] (d) at (0.7,1.0) {};
    \vertex (c) at (1.4,0);
    \diagram*{
    (a) -- (b) -- (c)
    };
    \draw[#2] (b) arc [start angle=-90, end angle=270, radius=0.5cm];
}

\newcommand{\irdivsex}{
    \vertex (a) at (-0.15,0);
    \vertex (b) at (0.5,0);
    \vertex (c) at (1.9,0);
    \vertex (d) at (2.55,0);
    \diagram*{
    (a) -- [] (b) -- [dashed,half left] (c) -- [dashed,half left] (b),
    (c) -- [] (d),
    };
}

\newcommand{\mixingdiag}{
    \vertex (a) at (0,0);
    \vertex (b) at (1,0);
    \vertex (c) at (4,0);
    \vertex (d) at (5,0);
    \diagram*{
    (a) -- (b) -- [dashed] (c) -- (d)
    };
}